\documentclass[english,aps,prl,twocolumn,amsmath,amssymb,showpacs,superscriptaddress,notitlepage,longbibliography]{revtex4-1}
\usepackage{bm}
\usepackage{amsmath}
\usepackage{amssymb}
\usepackage{mathdots}
\usepackage{graphicx}
\usepackage{hyperref}
\usepackage{xcolor}

\begin{document}

	\title{Anomalous Coherence Length in Superconductors with Quantum Metric }
	\author{Jin-Xin Hu}
	
	\affiliation{Department of Physics, Hong Kong University of Science and Technology,
		Clear Water Bay, Hong Kong, China}
	
	\author{Shuai A. Chen}
	\email{chsh@ust.hk}
	\affiliation{Department of Physics, Hong Kong University of Science and Technology,
		Clear Water Bay, Hong Kong, China}
	
	\author{K. T. Law}
	\email{phlaw@ust.hk}
	\affiliation{Department of Physics, Hong Kong University of Science and Technology,
		Clear Water Bay, Hong Kong, China}
	
	\date{\today}
	
	\begin{abstract}
		The coherence length $\xi$ is the fundamental length scale of superconductors which governs the sizes of Cooper pairs, vortices, Andreev bound states, and more. In BCS theory, the coherence length is $\xi_\mathrm{BCS} = \hbar v_{F}/\Delta$, where $v_{F}$ is the Fermi velocity and $\Delta$ is the pairing gap. It is clear that increasing $\Delta$ will shorten $\xi_\mathrm{BCS}$. In this work, we show that the quantum metric, which is the real part of the quantum geometric tensor, gives rise to an anomalous contribution to the coherence length. Specifically, $\xi = \sqrt{\xi_\mathrm{BCS}^2 +\ell_{\mathrm{qm}}^{2}}$ for a superconductor where $\ell_{\mathrm{qm}}$ is the quantum metric contribution.  In the flat-band limit, $\xi$ does not vanish but is bound below by $\ell_{\mathrm{qm}}$. We demonstrate that under the uniform pairing condition, $\ell_{\mathrm{qm}}$ is controlled by the quantum metric of minimal trace in the flat-band limit. Physically, the Cooper pair size of a superconductor cannot be squeezed down to a size smaller than $\ell_{\mathrm{qm}}$ which is a fundamental length scale determined by the quantum geometry of the wave functions. Lastly, we compute the quantum metric contributions for the family of superconducting moir\'{e} graphene materials, demonstrating the significant role played by quantum metric effects in these narrow-band superconductors.
	\end{abstract}
	
	\maketitle
	
	\section{Introduction}
	
	Bardeen-Cooper-Schriffer (BCS) theory \cite{bardeen1957microscopic} of superconductivity stands as one of the most successful and influential theories in modern physics. It offers a mean-field, yet non-perturbative and microscopic framework for understanding superconductivity. It has been very successful in describing a large number of superconductors \cite{carbotte1990properties,sigrist1991phenomenological,blatter1994vortices}. Deviations from the BCS theory are not unusual, which are often attributed to strong interaction effects \cite{stewart1984heavy,keimer2015quantum}. Recently, the observations of superconductivity in twisted bilayer graphene~\cite{cao2018unconventional,yankowitz2019tuning,arora2020superconductivity,oh2021evidence,tian2023evidence} and related graphene family \cite{liu2020tunable,park2021tunable,park2022robust} hinted that a new theory is needed to describe superconductors with nearly flat bands. It was observed in a recent experiment~\cite{tian2023evidence} that some important physical quantities deviate greatly from BCS predictions and the microscopic origins behind them are not yet clear. 
	
	For example, the BCS superconducting coherence length $\xi_\mathrm{BCS}$ is expressed as $\hbar v_{F}/\Delta$, where $v_{F}$ is the Fermi velocity and $\Delta$ is the pairing gap. When the moir\'e band of twisted bilayer graphene is nearly flat with $v_F \approx 10^3 \text{m}/\text{s}$ and $\Delta \approx 0.2$ meV, $\xi_\mathrm{BCS}$ is estimated to be around $3$ nm which is more than one order of magnitude shorter than the values measured using upper critical field measurements~\cite{tian2023evidence}. Furthermore, the low Fermi velocity (or equivalently, large effective mass) should lead to a low superfluid stiffness. This results in an expected Berezinskii-Kosterlitz-Thouless transition temperature much lower than the transition temperature measured at optimal doping \cite{tian2023evidence}. It had been pointed out by previous works that the quantum metric \cite{hu2019geometric,julku2020superfluid} of the flat bands, which is the real part of the quantum geometric tensor, is essential in sustaining a supercurrent \cite{peotta2015superfluidity,torma2022superconductivity}. Apart from the investigations of superfluid weight \cite{julku2016geometric,liang2017band,liang2017wave,iskin2018berezinskii,iskin2018quantum,iskin2018exposing,mondaini2018pairing,
		iskin2019origin,xie2020topology,verma2021optical,herzog2022superfluid,kitamura2022superconductivity,huhtinen2022revisiting,mao2024upper,hofmann2023superconductivity,mao2023diamagnetic}, the quantum geometry affects other physical quantities such as the intrinsic nonlinear transport~\cite{gao2023quantum,wang2023quantum,kaplan2024unification} and electron-phonon coupling~\cite{yu2024non}.
	
	%%%%%%%%%%%%%%%%%%%%%
	\begin{figure}[t]
		\centering \includegraphics[width=1\linewidth]{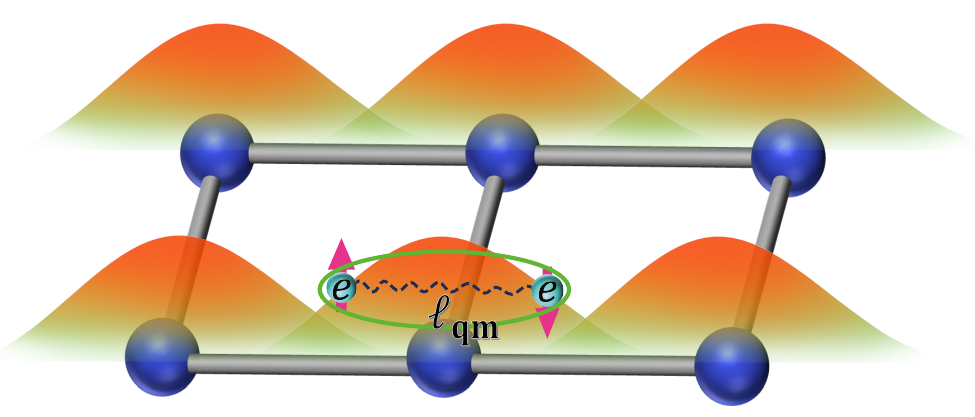} \caption{The schematic illustration of the quantum metric length $\ell_{\mathrm{qm}} $. In flat-band superconductors with quantum metric, the size (or the quadratic spread) of the optimally localized Wannier functions is $\ell_{\mathrm{qm}} $. $\ell_{\mathrm{qm}} $ is also the minimal coherence length (or the minimal size of the Cooper pairs) of the superconductor. }
		\label{fig:fig1}
	\end{figure}
	%%%%%%%%%%%%%%%%%%%%%%%%%%%%%%%%%%
	
	In a recent work, by deriving the Ginzburg-Landau theory for an exactly flat band (with zero bandwidth) \cite{chen2024ginzburg}, we pointed out that $\xi$ is determined by the quantum metric of the Bloch wave function which is independent of the interaction strength. This contradicts the intuition that stronger attractive interactions between electrons generally result in a smaller Cooper pair size and shorter coherence length, as described by the BCS theory.
	
	Explicitly, considering the Bloch states of a band represented by $|u(\bm{k})\rangle$, the quantum geometric tensor $\mathfrak{G}$ ~\cite{provost1980riemannian,berry1984quantal} is
	\begin{equation}
		\!\!\mathfrak{G}_{ab}\!=\!\langle\partial_{a}u(\bm{k})|\partial_{b}u(\bm{k})\rangle-\langle\partial_{a}u(\bm{k})|u(\bm{k})\rangle\langle u(\bm{k})|\partial_{b}u(\bm{k})\rangle.
		\label{eq:QGT}
	\end{equation}
	Here, $a$ and $b$ represent the momentum directions. The quantum geometric tensor can be decomposed into real and imaginary parts as $\mathfrak{G} = \mathcal{G} - i\mathcal F/2$, where the real part $\mathcal{G}$ is the quantum metric and the imaginary part $\mathcal F$ is the Berry curvature. Berry curvature arises from the phase difference between adjacent Bloch states and characterizes the band topology of materials~\cite{klitzing1980new,thouless1982quantized,bellissard1994noncommutative,hasan2010colloquium,qi2011topological,shapere1989geometric}. The study of the physical consequences of the Berry curvature has been one of the central topics in modern physics. On the other hand, the effect of quantum metric, which measures the distance between two quantum states~\cite{anandan1990geometry}, is much less studied. It was pointed out that the quantum metric provides the size (or the so-called quadratic spread)  $\ell_{\mathrm{qm}} $ of the optimally localized Wannier state of a band~\cite{marzari1997maximally}, where $\ell_{\mathrm{qm}}=\sqrt[4]{\det\overline{\mathcal{G}}}$. Here, $\overline{\mathcal{G}}$, defined in Eq.~\eqref{metric_average}, is the weighted average of the quantum metric of the Bloch states within a band. The mathematical definition of $\ell_{\mathrm{qm}}$, which we call the {\it quantum metric length}, measures the minimal spread of the Wannier functions, is schematically illustrated in Fig.~\ref{fig:fig1}. However,  the impact of the quantum metric length $\ell_{\mathrm{qm}}$ on physical quantities was not clear. Until very recently, the Ginzburg-Landau theory \cite{chen2024ginzburg} shown that at zero temperature, $\xi =\ell_{\mathrm{qm}}$ for an exactly flat band, which is independent of the interaction strength.
	
\begin{figure}
		\centering
		\includegraphics[width=1\linewidth]{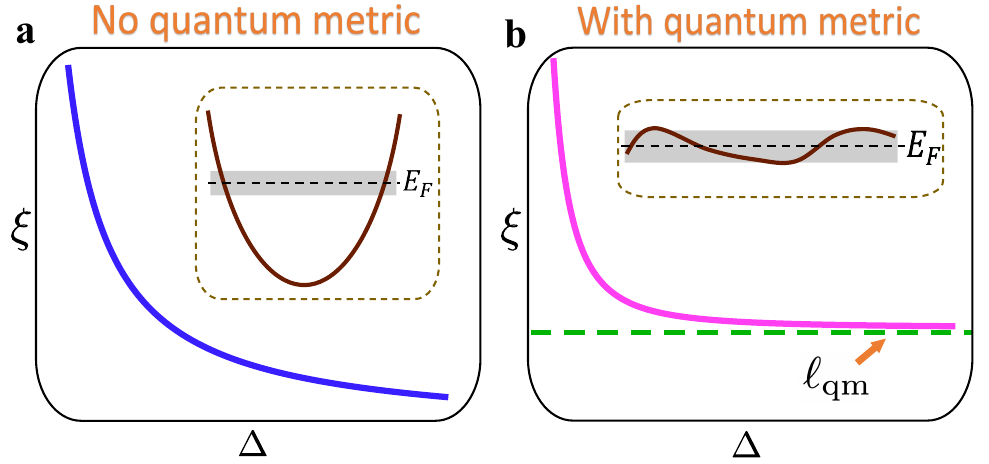}
		\caption{Bound of the coherence length by quantum metric. \textbf{a} For a conventional superconductor with a dispersive band (as illustrated by the insert) without quantum metric, the coherence length $\xi=\hbar v_F/\Delta$ decreases as $\Delta$ ($\Delta$ is the superconducting pairing gap) increases and $\xi$ is not bounded from below.  \textbf{b} In the presence of quantum metric, the superconducting coherence length $\xi$ has a lower bound of $\ell_{\mathrm{qm}}$. For a superconductor with a narrow band (as illustrated in the insert), the conventional contribution can be suppressed as $\Delta$ increases. }
		\label{fig:fig2}
	\end{figure}

	In realistic materials such as twisted bilayer graphene and related moir\'e flat-band superconductors, the bands are nearly flat, but the dispersion is still finite. One fundamental question arises: What is the interplay between the quantum metric effect and the finite dispersion of the band? In this work, we demonstrate that
	\begin{equation}
		\xi = \sqrt{ \xi_\mathrm{BCS}^{2} + \ell_{\mathrm{qm}}^{2}}~.
		\label{eq:xi}
	\end{equation}
	In other words, there is an anomalous quantum metric contribution to the superconducting coherence length (recall that $\xi_\mathrm{BCS} = \hbar v_{F}/\Delta$). In the flat-band limit with vanishing $v_{F}$, the quantum metric effect can be significant and even dominant. We show that this is indeed the case for several moir\'{e} superconductors with nearly flat bands \cite{tian2023evidence,liu2020tunable,park2021tunable,park2022robust}. Our result gives a possible explanation for why the observed superconducting coherence length in the recent experiment \cite{tian2023evidence} is much larger than expected. It is worth noting that the coherence length is lattice-geometry independent while the quantum metric is lattice-geometry dependent \cite{simon2020contrasting,huhtinen2022revisiting}. To resolve the discrepancy, we apply the uniform pairing condition when evaluating the pair correlators and then demonstrate that $\ell_\mathrm{qm}$ is related to the quantum metric of the minimal trace~\cite{huhtinen2022revisiting}. We delineate the physical picture that, in the presence of the quantum metric, increasing the attractive interaction strength between electrons can only reduce the BCS part of the coherence length and squeeze the Cooper pair size down to the quantum metric length $\ell_{\mathrm{qm}}$, but not further, as demonstrated in Fig.~\ref{fig:fig2}.

	%%%%%%%%%%%%%%%%%%%%%%%%%%%%%%%%%%%%%%%%%%
	%%%%%%%%%%%%%%%%%%%%%%%%%%%%%%%%%%%%%%%%%%
	Additionally, for a topological flat band with nontrivial (spin) Chern number, $\xi\geq a\sqrt{\vert C \vert/4\pi}$, where $a$ is the lattice constant as illustrated in Fig.~\ref{fig:fig4}. At the end of this work, we show that the quantum metric length $\ell_{\mathrm{qm}}$ is important for the superconducting moir\'e graphene family. As the new length scale $\ell_{\mathrm{qm}} $ defined by the quantum metric is a fundamental property of the band structure and its importance should be manifested beyond superconducting phenomena, we expect that $\ell_{\mathrm{qm}} $ also plays a crucial role in other interaction-driven ordered states (such as the magnetic or density-wave states~\cite{hofmann2023superconductivity,han2024quantum}) in flat-band systems.
	
	\section{Results}

	\subsection{Quantum metric and coherence length}
	\label{sec:coherence}
	We investigate the interplay between quantum metric and band dispersion in superconductors where superconductivity appears within an isolated narrow band. To begin with, we describe our formalism from a multi-orbital Hamiltonian with two components: the non-interacting part $H_0$ and the attractive interacting part $H_{\mathrm{int}}$, which read
	\begin{align}
		H_{0} & =\sum_{ij,\alpha\beta,\sigma}h_{ij,\alpha\beta}^{\sigma}a^\dagger_{i\alpha\sigma}a^{}_{j\beta\sigma},\label{smeq:H0}\\
		H_\mathrm{int} & =-\sum_{i,\alpha}U a_{i\alpha\uparrow}^{\dagger}a_{i\alpha\downarrow}^{\dagger}a^{}_{i\alpha\downarrow}a^{}_{i\alpha\uparrow}, \label{smeq:Hint}
	\end{align}
	where $h^\sigma_{ij,\alpha\beta}$ is the hopping integral and $U$ denotes the on-site attractive interaction strength. $a_{i\alpha\sigma}$ annihilates a fermion with spin $\sigma$ in the orbital $\alpha$ at the site $i$ (we may call $a_{i\alpha\sigma}$ orbital fermions). Considering an isolated band near the Fermi energy separated from other bands with a large band gap, we can have an effective one-band description. For $s$-wave superconducting phase, it is common to introduce orbital-dependent order parameters $\Delta_\alpha = - U\langle a_{i\alpha\downarrow}a_{i\alpha\uparrow}\rangle$. The mean-field ground state has been extensively investigated, particularly with regard to the superfluid weight determined by the quantum metric~\cite{peotta2015superfluidity}. It is possible to project the orbital fermion $a_{i\alpha\sigma}$ onto the fermion $c_\sigma$ of the isolated band, which is referred as the band fermion. In particular, we employ the following projection scheme
	\begin{equation}
		a_{i\alpha\sigma}\rightarrow \frac{1}{\sqrt{N}} \sum_{\bm{k}}e^{i\bm{k}\cdot(\bm{r}_i+\bm{\delta}_{\alpha})}u^*_{\alpha\sigma}(\bm{k})c_{\sigma}(\bm{k}),\label{smeq:projection}
	\end{equation}
	where we explicitly keep the orbital positions $\{\bm \delta_\alpha\}$ within a unit cell. The Bloch state $u_{\alpha\sigma}(\bm k)$ of the isolated band with energy $\epsilon_\sigma(\bm k)$ satisfies the time reversal symmetry $u_{\alpha}(\bm k)\equiv u_{\alpha\uparrow}(\bm k)=u^*_{\alpha\downarrow}(-\bm k)$. The projection in Eq.~\eqref{smeq:projection} yields an effective one-band mean-field Hamiltonian $H_\mathrm{mf}$,
	\begin{align}
		\!\!\!\!H_\mathrm{mf} =\sum_{\bm k} \epsilon_\sigma(\bm k) c^\dagger_\sigma(\bm k) c^{}_\sigma(\bm k)\!+\![\Delta c^\dagger_{\uparrow}(\bm k) c^\dagger_{\downarrow}(-\bm k)+h.c.]
	\end{align}
	with $\Delta = 1/N\sum_{\alpha\bm k}\Delta_\alpha |u_\alpha(\bm k)|^2$. 
	The projected mean-field Hamiltonian $H_\mathrm{mf}$ is independent of the choice of orbital positions $\{\bm \delta_\alpha\}$. To facilitate the theoretical analysis, we can adopt the uniform pairing condition and the minimal quantum metric~\cite{huhtinen2022revisiting}. 
	The former assumes that the pairing potentials are the same for different orbitals, and the latter is specific to orbital positions corresponding to the minimal trace of quantum metric. Then we can define the Cooper pair operator $\hat{\Delta}(\bm{q}) =\frac{1}{N} \sum_{i\alpha} e^{-i\bm q\cdot (\bm r_i+\bm \delta_\alpha)}a_{i\alpha \downarrow}a_{i\alpha \uparrow}$ which is formulated after projection  as 
	\begin{equation}
		\hat{\Delta}(\bm{q})\rightarrow \frac{1}{N}\sum_{\bm{k}}\Lambda(\bm{k}+\bm{q},\bm{k})c_{\downarrow}(-\bm{k})c_{\uparrow}(\bm{k}+\bm{q}). 
		\label{eq:proj_cooper}
	\end{equation}
	Here the form factor $\Lambda(\bm{k}+\bm{q},\bm{k})=\sum_{\alpha} u^*_{\alpha}(\bm k+\bm q)u_{\alpha}(\bm k)$ appears as the overlap between two Bloch states. Then we can evaluate the pairing correlator $\mathcal{C}(\bm{r})=\sum_{\bm q} e^{-i\bm q\cdot \bm r}  \langle 
	\hat{\Delta}(\bm{q})\hat{\Delta}^\dagger(\bm{q})\rangle$ to deduce the coherence length. The pairing correlator $\mathcal C(\bm r)$ is expected to decay exponentially as a function of $|\bm r|$ at zero temperature for an isotropic system. In other words, $\mathcal C(\bm r) \sim e^{-|\bm r|/\xi}$ and the decay length $\xi$ is the superconducting coherence length~\cite{annett2004superconductivity}. As shown in Supplementary Note 2, $\mathcal C(\bm r) \equiv\sum_{\bm{q}}e^{-i\bm{q}\cdot\bm{r}}\mathcal M(\bm q) $, where
	\begin{equation}
		\label{totallength}
		\!\mathcal M(\bm q)\! =\! \frac{T}{N} \!\sum_{n\bm{k}} \! \vert\Lambda(\bm{k}+\bm{q},\bm{k})\vert^{2}G_0(i\omega_{n},\bm{k}+\bm{q})G_0(-i\omega_{n},-\bm{k}).
	\end{equation}
	Here, $G_0(i\omega_{n},\bm{k})$ is the normal Gor'kov's Green function of the band fermions $c_{\sigma}$ and $\omega_{n}=(2n+1)\pi T$ is the Matsubara frequency, as defined in Methods section. Then we can extract the coherence length by $\xi^2 = - \frac{1}{2\mathcal M(0)}\left.\frac{d^2 \mathcal M(\bm q)}{dq^2}\right|_{q=0}$ with $q=|\bm q|$ at zero temperature. It is essential to emphasize that the validity of the expression in Eq.~\eqref{totallength} hinges on the uniform pairing condition, specifically in relation to the Bloch states of the minimal quantum metric \cite{huhtinen2022revisiting}. Without these conditions, the coherence length calculated using Eq.~\eqref{totallength} will be overestimated, and additional details can be found in the Supplementary Note 2. To see how the quantum metric affects the coherence length $\xi$, the form factor $\Lambda$ enters $\mathcal{M}(\bm q)$ such that
	\begin{equation}
		\vert\Lambda(\bm{k}+\bm{q},\bm{k})\vert^{2}=1-\sum_{ab}\mathcal{G}_{ab}(\bm{k})q_{a}q_{b}+\mathcal{O}(\bm{q}^{2})~.
	\end{equation}
	The matrix $\mathcal{G}_{ab}$ is the quantum metric of Bloch states, namely the real part of quantum geometric tensor $\mathfrak G$ in Eq.~\eqref{eq:QGT}, 
	\begin{equation}
		\mathcal{G}_{ab}(\bm{k})\!=\!\mathrm{Re}\left[\mathfrak{G}{}_{ab}(\bm{k})\right].
	\end{equation}
	By theoretically evaluating the pairing correlator, we can obtain the coherence length in Eq.~\eqref{eq:xi} as $\xi = \sqrt{\xi_\mathrm{BCS}^2 +\ell_{\mathrm{qm}}^{2}}$. In fact, the structure of the the coherence length in Eq.~\eqref{eq:xi} is general and works regardless of the uniform pairing condition and the minimal quantum metric. The anomalous coherence length is $\ell_{\mathrm{qm}} =\sqrt[4]{\det\overline{\mathcal{G}}_{ab}}$, where $\overline{\mathcal{G}}_{ab}$ is the weighted average of the quantum metric of the band, which is defined by
	\begin{equation}
		\label{metric_average}
		\overline{\mathcal{G}}_{ab}=\frac{\sum_{\bm{k}}\mathcal G_{ab}(\bm{k})/\varepsilon(\bm{k})}{\sum_{\bm{k}}1/\varepsilon(\bm{k})},
	\end{equation}
	where $\varepsilon(\bm{k})$ is the dispersion of the Bogoliubov quasiparticle. In the limit of a flat dispersion $\varepsilon(\bm k)$, the quantum metric length $\ell_{\mathrm{qm}}$ is reduced to the length scale of the minimal quantum metric. The above discussions on the coherence length in Eq.~\eqref{eq:xi} can be easily generalized to an anisotropic system with a non-circular Fermi surface where the quantum metric length becomes spatially dependent due to finite off-diagonal elements in the quantum metric.
	
	%%%%%%%%%%%%%%%%%%%%%%%%%%%%%%%%%%%%%%%%%%
	\begin{figure*}
		\centering
		\includegraphics[width=1\linewidth]{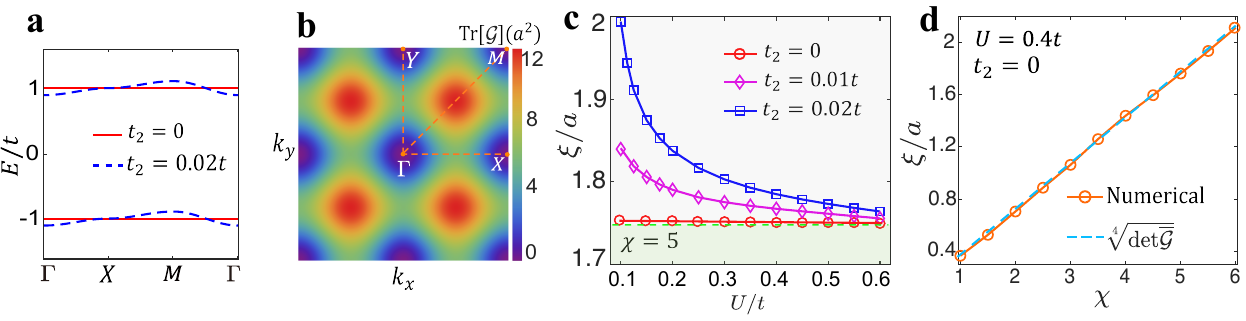}
		\caption{Quantum metric and coherence length for topological trivial flat-band model. \textbf{a} The energy spectrum of the flat-band model in Ref.~\cite{hofmann2023superconductivity}. The solid (dashed) lines denote $t_2=0$ and $t_2=0.02t$, respectively. $t_2$ denotes the nearest hopping which makes the band dispersive. \textbf{b} The profiles of the quantum metric $\mathrm{Tr} [\mathcal G]$ of the conduction band in the first Brillouin zone. The color bar denotes the magnitude of $\mathrm{Tr} [\mathcal G]$. \textbf{c} The calculated coherence length $\xi$ for $\chi=5$ as a function of the attractive interaction $U/t$. The red, purple and blue denote the cases of $t_2=0$, $t_2=0.01t$ and $t_2=0.02t$, respectively. The theoretical bound $\ell_\mathrm{qm}$ is indicated by a dashed green line which coincides with the red one.
			\textbf{d} The quantum metric dependence of $\xi$ as the parameter $\chi$ varies when $U=0.4t$. The dashed light blue line marks the length scale $\ell_\mathrm{qm}$. All calculations are conducted at $k_{B}T=0.001t$ and at half-filling $\mu=t$.}
		\label{fig:fig3}
	\end{figure*}
	%%%%%%%%%%%%%%%%%%%%%%%%%%%%%%%%%%%%%%%%%%%

	To understand the physical consequence of the anomalous coherence length, we note that for a conventional superconductor, $\xi = \xi_\mathrm{BCS}$ decreases as the interaction strength (or equivalently $\Delta$) increases, as schematically shown in Fig.~\ref{fig:fig2}{\bf a}. However, in the presence of the quantum metric, $\xi$ decreases as $\Delta$ increases, but approaches the quantum metric length $\ell_{\mathrm{qm}}$ (see Fig.~\ref{fig:fig2}{\bf b}). 
	In the flat-band limit, the coherence length (at zero temperature) is independent of the interaction strength and given by $\ell_{\mathrm{qm}}$. 
	In the presence of finite quantum metric, interactions cannot squeeze the Cooper pairs to a size smaller than $\ell_{\mathrm{qm}}$.

	\subsection{Topologically trivial flat-band model}
	To support the analytical results mentioned above, we employ the mean-field theory on a microscopic model, which features exactly flat bands without dispersion~\cite{hofmann2022heuristic,hofmann2023superconductivity}. The normal state Hamiltonian $h_s(\bm k)$ for electrons with spin index $s$ reads
	\begin{equation}
		h_s(\bm{k})=-t[\lambda_{x}\sin(\alpha_{\bm{k}})+s\lambda_{y}\cos(\alpha_{\bm{k}})].
		\label{eq:H0_flat}
	\end{equation}
	Here, $\alpha_{\bm{k}}=\chi[\cos(k_{x}a)+\cos(k_{y}a)]$ and $\lambda_i$ are the Pauli matrices in orbital basis. $s=\pm 1$ denotes the spins $\uparrow$ and $\downarrow$. The $h_s(\bm k)$ has a pair of perfectly flat bands at energies $\epsilon_{\bm{k}}=\pm t$  which are depicted in Fig.~\ref{fig:fig3}{\bf a} (solid lines) and the corresponding wave functions are $|u_{\pm}\rangle=1/\sqrt{2}(\pm 1,i s e^{is \alpha_{\bm{k}}})^{T}$ for the upper band (+) and the lower band ($-$). The flat band is topologically trivial with the Berry curvature vanishing over the whole Brillouin zone. We can tune the quantum metric by altering the parameter $\chi$ in $\alpha_{\bm k}$. It is straightforward to obtain the quantum metric for $+$ band with components $\mathcal{G}_{ab}(\bm{k})=\chi^{2}a^{2}\sin(k_{a})\sin(k_{b})/4$, which is the minimal quantum metric since the orbitals are located at high-symmetry positions. The averaged quantum metric defined by Eq.~(\ref{metric_average}) is given by $\overline{\mathcal{G}}_{ab}=\delta_{ab}\chi^{2}/8$ which is related to the quantum metric length $\ell_\mathrm{qm}=\sqrt{2}\chi/4$.  
	In Fig.~\ref{fig:fig3}{\bf b}, we plot the distribution of $\mathrm{Tr} [\mathcal G(\bm k)]$ that respects the $C_4$ symmetry and that $\mathrm{Tr} [\mathcal G(\bm k)]$ reaches its maximum at $M/2$. Since we are interested in a superconducting phase, we do not include other possible ground state ansatz. In the Nambu basis $\Psi_{\bm{k}}=(a_{A,\bm{k}\uparrow},a_{B,\bm{k}\uparrow},a^\dagger_{A,-\bm{k}\downarrow},a^\dagger_{B,-\bm{k}\downarrow})^T$ with an attractive interaction as Eq.~\eqref{smeq:Hint}, we have the mean-field Hamiltonian $H_\mathrm{mf}$ 
	\begin{equation}
		H_\mathrm{mf}=\sum_{\bm{k}}\Psi_{\bm k}^\dagger \left[
		\begin{matrix}{}
			&h_{\uparrow}(\bm{k})-\mu & \hat{\Delta}  \\
			&\hat{\Delta}^\dagger  &  -h_{\downarrow}^*(-\bm{k})+\mu
		\end{matrix}\right]\Psi_{\bm k}.
	\end{equation}
	Here, $\hat \Delta=\mathrm{diag}[\Delta_A,\Delta_B]$ is the mean-field pairing order parameters. The Fermi energy $\mu$ is chosen such that the $+$ band is half-filled. The solutions of the order parameters yield $\Delta_A=\Delta_B=U/4$, which satisfy the uniform pairing condition.
	
	Due to the absence of band dispersion, the coherence length $\xi = \sqrt{2}\chi/4$ depends solely on the quantum metric. This is illustrated in Fig.~\ref{fig:fig3}{\bf d}, where the numerical results [Eq.~\eqref{totallength}] of pair correlation functions align with $\ell_{\mathrm{qm}}$. To incorporate the finite band dispersion, one can introduce an additional nearest-hopping term $\delta h=-2t_2[\cos(k_x a)+\cos(k_y a)]\lambda_0$ to $h_s(\mathbf k)$, where $\lambda_0$ is the $2 \times 2$ identity matrix.  This term gives rise to a band dispersion as well as the conventional contribution $\xi_\mathrm{BCS}$ to the total coherence length $\xi$. In Fig.~\ref{fig:fig3}{\bf c}, the total coherence length gradually decreases for $t_2=0.01t,0.02t$ when the attractive interaction strength $U$ increases. In particular, $\xi$ approaches $\ell_{\mathrm{qm}}$ in the flat-band limit due to the suppression of $\xi_\mathrm{BCS}$, as expected from Eq.~\eqref{eq:xi}.

	%%%%%%%%%%%%%%%%%%%%%%
	\begin{figure*}[t]
		\centering
		\includegraphics[width=1\linewidth]{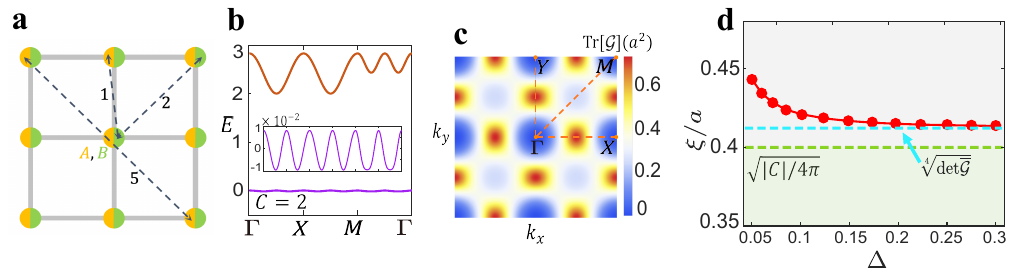}
		\caption{Quantum metric and coherence length for topological flat-band model. \textbf{a} A two-orbital square lattice with short- and long-range hoppings, \textbf{b} the electronic band structure, \textbf{c} the quantum metric distribution of the lower flat band, and \textbf{d} coherence length $\xi$ \emph{v.s.} pairing gap $\Delta$. In \textbf{a}, the inter-orbital nearest hopping, intra-orbital next-nearest-neighbor and fifth-nearest-neighbor hoppings are labeled. In \textbf{b}, the lower band (purple) has nearly zero bandwidth with the Chern number $C=2$. In \textbf{d}, the coherence length is extracted from the pair correlation function and it is bounded by a Chern number, which is guided by the dashed green line.}
		\label{fig:fig4}
	\end{figure*}
	%%%%%%%%%%%%%%%%%%%%% 
	
	\subsection{Topological bound of the coherence length \label{sec:top_bound}}
	In the previous subsection, we have demonstrated how the quantum metric gives a lower bound for the superconducting coherence length. We now consider a system which possesses topological flat bands. As pointed out previously~\cite{roy2014band,marzari1997maximally}, the quantum metric has a lower bound which is proportional to the Chern number. Therefore, we expect that there is a finite quantum metric length which serves as the lower bound of the superconducting coherence length for a superconductor with nontrivial spin Chern numbers. 
	
	Specifically, the quantum geometric tensor is a positive semidefinite matrix, and in two spatial dimensions, we have the inequality $\sqrt{\mathrm{det}\mathcal G(\bm k)}\geq  |\mathcal F_{xy}(\bm k)|/2$, which implies that a topological band must necessarily possess a finite quantum metric. According to Eq.~(\ref{eq:xi}), this indicates that there is a lower bound  on the coherence length $\xi$ which is determined by the topology of the band such that
	\begin{equation}
		\xi\geq\ell_\mathrm{qm}\geq a\sqrt{\vert C\vert/4\pi}~,
		\label{eq:lowbound}
	\end{equation} 
	where $C$ denotes the (spin) Chern number of a band with a lattice constant $a$. For demonstration, we consider a two-orbital square lattice with short- and long-range hoppings (Fig.~\ref{fig:fig4}\textbf{a}) with a finite spin Chern number~\cite{sun2011nearly,yang2012topological,mitscherling2022bound}. Under the basis $a_{\bm{k}\sigma}=(a_{A\bm{k}\sigma},a_{B\bm{k}\sigma})^T$, the non-interacting Hamiltonian is $H_0=\sum_{\bm{k},\sigma}a_{\bm{k}\sigma}^\dagger H_{\bm{k}}a_{\bm{k}\sigma}$, where $H_{\bm{k}}=\sum_{i}h_i(\bm{k})\lambda_i$. Here $h_0(\bm{k})=(\sqrt{2}-1)\cos(2k_x a)\cos(2k_y a)/2$, $h_x(\bm{k})=-\sqrt{2}[\cos(k_x a)+\cos(k_ya)]/2$, $h_y(\bm{k})=\sqrt{2}[\cos(k_xa)-\cos(k_ya)]/2$, and $h_z(\bm{k})=-\sqrt{2}\sin(k_xa)\sin(k_ya)$. The $\lambda_i$ are the Pauli matrices on the orbital basis. Importantly, the lowest band is nearly flat with a spin Chern number $C=2$ (see Fig.~\ref{fig:fig4}\textbf{b}). The bandwidth is approximately 1\% of the total band gap.
	%%%%%%%%%%%%%%%%%%%%%%
	\begin{figure*}[t]
		\centering
		\includegraphics[width=0.85\linewidth]{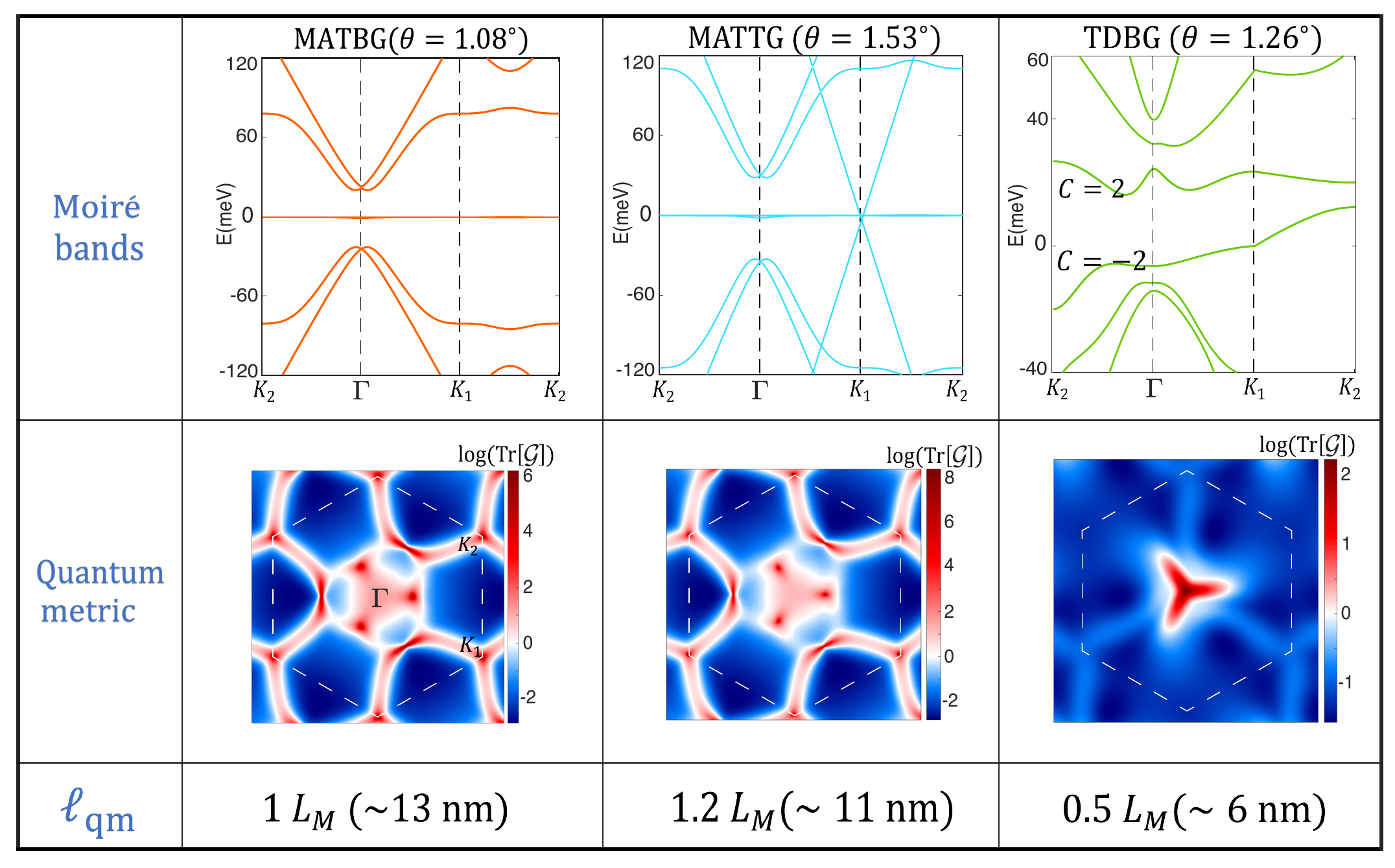}
		\caption{A list of the moir\'{e} band structures, quantum metric and geometric contributions of the coherence length $\ell_{\mathrm{qm}}$ for magic angle twisted bilayer graphene (MATBG), magic angle twisted trilayer graphene (MATTG) and twisted double bilayer graphene (TDBG). For both MATBG and MATTG, the quantum metric is plotted for the highest valence band, and it exhibits divergence near the $K$ points. In TDBG, an electric field potential of $V=40$ meV is applied, leading to flat bands near charge neutrality with Chern number $C=\pm 2$.
			The quantum metric is plotted for the lowest conduction band.
			In evaluating $\ell_\mathrm{qm}$, we ignore the band dispersion.}
		\label{fig:fig5}
	\end{figure*}
	%%%%%%%%%%%%%%%%%%%%%
	
	In Fig.~\ref{fig:fig4}\textbf{c}, we depict the distribution of $\mathrm{Tr} [\mathcal G (\bm{k})]$, which exhibits $C_4$ symmetry and has a large quantum metric at $X/2$ and points connected by symmetry. To demonstrate the effect of the nontrivial Chern number, in the mean-field calculations we assume the flat band is half-filled for simplicity. The uniform pairing condition is also satisfied as $\Delta_A=\Delta_B=\Delta$. Furthermore, we have calculated the Cooper pair correlation functions and extracted the coherence length from Eq.~\eqref{totallength}, which exhibits a decreasing trend as the band pairing potential $\Delta$ increases, as shown in Fig.~\ref{fig:fig4}\textbf{d}. Especially, in the limit of large $\Delta$, the coherence length $\xi$ converges to approximately $\sqrt[4]{\det\overline{\mathcal{G}}}$ which is larger than $\sqrt{|C|/4\pi}a$ as predicted by Eq.~(\ref{eq:lowbound}). This result clearly demonstrates how the superconducting coherence length is related to the quantum geometry (both the quantum metric and the topology) of the relevant band.
	
	\subsection{Application to Moir\'e materials}
	The graphene-based moir\'{e} systems provide the versatile platforms to explore the exotic phenomena related to the flat bands~\cite{ledwith2020fractional,da2021correlation,song2022magic,chou2023kondo,herzog2024topological,hu2023symmetric,zhang2023polynomial,hu2023kondo,
		kolavr2023anderson,kwan2021kekule}.
	In the superconducting graphene-based moir\'{e} family~\cite{park2022robust}, the quantum metric effect is indeed very crucial. Particularly, the quantum metric plays a significant role in determining the coherence length in magic angle twisted bilayer graphene (MATBG) with twisted angle $\theta\approx 1.08^{\circ}$. To provide a qualitative estimation of the impact of the quantum metric, we employ the Bistritzer-MacDonald model to elucidate the significance of the quantum metric in the context of graphene-based moir\'{e} materials~\cite{bistritzer2011moire}. We also assume the presence of an $s$-wave superconducting phase. As shown in Fig.~\ref{fig:fig5}, the quantum metric length $\ell_{\mathrm{qm}}=1.2 L_M \approx 13$ nm. Here, $L_M\approx a_0/\theta$ represents the moir\'{e} lattice constant. By employing the self-consistent mean-field study (in the Supplementary Note 4), we calculate the total coherence length using Eq.~\eqref{totallength} to take into account the band dispersion. Using the interaction strength $U=0.6$ meV, which gives $T_c \approx 1.7K$, we obtain a conventional contribution of $\sim 3$nm and $\ell_{\mathrm{qm}}\sim 13$ nm at $\theta=1.08^\circ$. Therefore, the total superconducting coherence length given by Eq.~\eqref{eq:xi}, is indeed dominated by the quantum metric contribution. 
	
	A large family of moir\'{e} systems exhibit superconductivity, such as magic-angle twisted trilayer graphene (MATTG) \cite{park2021tunable} and twisted double-bilayer graphene (TDBG) \cite{liu2020tunable}. Similar to MATBG, the quantum metric effects cannot be neglected, as shown in Fig.~\ref{fig:fig5}. For MATTG, $\ell_{\mathrm{qm}}=1.2 L_M$, and for TDBG, $\ell_{\mathrm{qm}}=0.5 L_M$. The calculations of $\ell_{\mathrm{qm}}$ in Fig.~\ref{fig:fig5} are made by averaging the quantum metric over the moir\'{e} Brillouin zone without considering the quasiparticle energy in Eq.~(\ref{metric_average}). Notably, the flat band in TDBG carries a non-zero valley Chern number $C=2$, leading to a topology-bound coherence length, as discussed previously. We focus on the quantum metric within a single band, while the generalization to multiple nearly-degenerate flat bands consists of replacing the one-band quantum metric with the non-abelian quantum metric \cite{mera2022nontrivial,herzog2022many}. The quantum metric length calculated for the moir\'{e} systems is a qualitative estimation because of the limitations of the continuum model and the simple $s$-wave pairing assumption. It will be an open question of the role that quantum metric plays in unconventional superconductivity for moir\'{e} systems.

	\section{Conclusion}
	In this work, we highlight that an intrinsic length scale, $\ell_\mathrm{qm}$, derived from the quantum metric, gives rise to an anomalous contribution of the coherence length in superconductors. Particularly in the case of flat bands, $\ell_\mathrm{qm}$ plays a dominant role in determining the length scale of physical quantities, such as the superconducting coherence length. This length scale is likely also related to the size of vortices, Andreev bound states etc.. We propose that our theory may also be applicable to quantum ordered phase in flat-band systems, since $\ell_\mathrm{qm}$ is derived from the quantum geometry of the band and is independent of the interaction-driven order parameter. Furthermore, it would also be interesting to explore potential extensions of $\ell_\mathrm{qm}$ to the physical properties of other ordered states (such as ferromagnetic and antiferromagnetic states) with flat bands and quantum metric.
	
	\section{Methods} \label{sec:method}
	\emph{Mean-field theory and Gor'kov Green function}.
	For a mean-field study, we can decouple the interaction term $H_\mathrm{int}$ in Eq.~\eqref{smeq:Hint}  with a pairing order parameters $\Delta_{\alpha}=-U\langle\hat{a}_{i\alpha\downarrow}\hat{a}_{i\alpha\uparrow}\rangle$ to yield a mean-field Hamiltonian $H_{\mathrm{mf}}$
	\begin{align}
		H_\mathrm{mf}=\sum_{\bm{k}} & \Psi_{\bm{k}}^{\dagger}(\hat{h}\otimes\tau_{z}+\mathrm{Re}\hat{\Delta}\otimes\tau_{x}+\mathrm{Im}\hat{\Delta}\otimes\tau_{y})\Psi_{\bm{k}}
	\end{align}
	where the $\Psi_{\bm{k}}$ is the spinor with components $(\Psi_{\bm{k}})_{\alpha\uparrow}=a_{\alpha\uparrow}(\bm{k})$
	and $(\Psi_{\bm{k}})_{\alpha\downarrow}=a_{\alpha\downarrow}^{\dagger}(-\bm{k})$. Here $a_{\alpha\sigma}$ is a Fermion operator on the orbital basis and $\tau_{x,y,z}$ are the Pauli matrices.
	The $\hat{h}$ is the matrix with elements $(\hat{h})_{\alpha\beta}=h_{\alpha\beta}(\bm{k})-\mu\delta_{\alpha\beta}$
	and the pairing matrix $\hat{\Delta}$ has elements $(\hat{\Delta})_{\alpha\beta}=\Delta_{\alpha}\delta_{\alpha\beta}$. Within the mean-field Hamiltonian, we can define the Green function $\hat G_{\alpha\alpha^{\prime},\sigma\sigma^{\prime}}(i\omega_{n},\bm{k})=\langle(\psi_{\bm{k}})_{\alpha\sigma}(i\omega_{n})(\psi_{\bm{k}}^{\dagger})_{\alpha^{\prime}\sigma^{\prime}}(i\omega_{n})\rangle$
	with 
	\begin{equation}
		\!\!\!\hat{G}(i\omega_{n},\bm{k})=\frac{1}{i\omega_{n}-(\hat{h}\otimes\tau_{z}+\mathrm{Re}\hat{\Delta}\otimes\tau_{x}+\mathrm{Im}\hat{\Delta}\otimes\tau_{y})}.
	\end{equation}
	where $\omega_{n}=(2n+1)\pi k_B T$ is the Matsubara frequency.
	Then one may evaluate the pairing correlation function $\mathcal C(\bm{r},\bm{r}^{\prime})$ with the Green function for a multicomponent fermion $a_{\alpha\sigma}$. For an s-wave superconductor, we expect an exponential decay behavior in $\mathcal C(\bm r,\bm r^\prime)$ as a function of $|\bm r-\bm r^\prime|$.
	
	On the other hand, we apply a mean-field theory to the effective two-band model after the projection. For a superconducting phase, we can introduce an $s$-wave pairing order parameter 
	$
	\Delta=-\frac{U}{N}\sum_{\bm{k}}\langle c_{\downarrow}(-\bm{k})c_{\uparrow}(\bm{k})\rangle,
	$
	and set $\Delta$ to be real via fixing the gauge.
	Here $c_{\sigma}$ is a Fermion operator on the flat band. Then we can have a mean field Hamiltonian
	\begin{align}
		H_\mathrm{mf} =\sum_{\bm{k}}\psi_{\bm{k}}^{\dagger}\{[\epsilon(\bm{k})-\mu]\tau_{z}+\Delta\tau_{x}\}\psi_{\bm{k}},
	\end{align}
	where $\psi_{\bm{k}}=\begin{bmatrix}c_{\uparrow}(\bm{k}),
		c_{\downarrow}^{\dagger}(-\bm{k})
	\end{bmatrix}^T$ is the Nambu spinor. 
	One can directly extract the Green's function $G$ for the band fermions as
	\begin{equation}
		G(i\omega_{n},\bm{k})=\frac{-i\omega_{n}\tau_{0}-[\epsilon(\bm{k})-\mu]\tau_{z}-\Delta\tau_{x}}{\omega_{n}^{2}+[\epsilon(\bm{k})-\mu]^{2}+\Delta^{2}},\label{smeq:greenfunction}
	\end{equation}
	In evaluating physical quantities such as the pairing correlation function, one should first project
	the observables onto an isolated band, and then apply Wick's theorem via Gor'kov's Green function. The projection helps uncover the role of quantum metric in physical quantities such as the coherence length.
	
	\section{Data Availability}
	The data generated from our codes that support the findings of this study are available from the corresponding author upon reasonable request.
	
	\section{Author Contributions}
	K.T.L. and S.C. conceived the project. J.-X.H. performed the major part of the calculations and analysis. J.-X.H., S.C. and K.T.L. wrote the manuscript with contributions from all authors. All authors are involved in the discussions.
	
	\section{Acknowledgement}
	We thank Adrian Po for illuminating discussions. K.T.L. acknowledges the support of the
	Ministry of Science and Technology, China, and the Hong Kong Research
	Grants Council through Grants No. 2020YFA0309600, No. RFS2021-6S03,
	No. C6025-19G, No. AoE/P-701/20, No. 16310520, No. 16309223 and No. 16307622.
	
	\section{Competing Interests}
	The authors declare no competing interests.
	
	%\bibliographystyle{apsrev4-1}
	%\bibliography{References}

	\onecolumngrid
	\clearpage
	\begin{center}
		{\bf Supplementary Material: Anomalous Coherence Length in Superconductors with Quantum Metric}
	\end{center}
	
	\maketitle
	\setcounter{equation}{0}
	\setcounter{figure}{0}
	\setcounter{table}{0}
	\setcounter{page}{1}
	\makeatletter
	
	\maketitle
	\setcounter{equation}{0}
	\setcounter{figure}{0}
	\setcounter{table}{0}
	\setcounter{page}{1}
	\makeatletter
	\renewcommand{\figurename}{Supplementary Fig.}
	\renewcommand{\tablename}{Supplementary Table}
	\renewcommand\thetable{1}
	
\section*{Supplementary Note 1: Mean-field theory}
The physics underlying multi-band superconductors can be elucidated by introducing a projection onto the targeted band. We aim to demonstrate the relationship between the coherence length and the Fermi velocity, pairing gap, and quantum metric for a generally isolated band. A band is considered isolated when a substantial gap exists between the conduction and valence bands, allowing us to apply the projection technique. Additionally, we assume that there is no additional degeneracy  within the isolated band.
By focusing on the isolated band, we can gain valuable insights into the behavior of the superconducting phase in multi-band systems. This approach allows us to explore the interplay between fundamental properties, such as the Fermi velocity, pairing gap, and quantum metric, which all play essential roles in determining the characteristics of the superconducting state.

\subsection*{Mean-field theory for a multi-band superconductor}
We will derive the self-consistent equations for the multi-band superconductivity. 
For the sake of completeness, we show the model Hamiltonian studied in the main text $H=H_{0}+H_{\mathrm{int}}$, which read
\begin{align}
	H_{0} & =\sum_{ij}\sum_{\alpha\beta\sigma}h_{ij\alpha\beta}^{\sigma}a_{i\alpha\sigma}^{\dagger}a_{j\beta\sigma} ,\\
	H_{\mathrm{int}} & =-\sum_{i\alpha}Ua_{i\alpha\uparrow}^{\dagger}a_{i\alpha\downarrow}^{\dagger}a_{i\alpha\downarrow}a_{i\alpha\uparrow},
\end{align}
where $h^\sigma_{ij,\alpha\beta}$ is the hopping integral and $U$ is the on-site attractive interaction strength. $a_{i\alpha\sigma}$ annihilates a fermion with spin $\sigma$ in the orbital $\alpha$ on site $i$. For the $s$-wave pairing, we can  introduce the orbital-dependent mean-field order parameters:
\begin{equation}
	\Delta_{\alpha}=-U\langle a_{i\alpha\downarrow}a_{i\alpha\uparrow}\rangle.
\end{equation}
The mean-field order parameters factorize the Hamiltonian into
\begin{align}
	H_\mathrm{mf} & =\sum_{ij}\sum_{\alpha\beta\sigma}h_{ij\alpha\beta}^{\sigma}a_{i\alpha\sigma}^{\dagger}a_{j\beta\sigma}+\sum_{i\alpha}\left[a_{i\alpha\uparrow}^{\dagger}a_{i\alpha\downarrow}^{\dagger}\Delta_{\alpha}+a_{i\alpha\downarrow}a_{i\alpha\uparrow}\Delta_{\alpha}^{*}\right]\\
	& =\sum_{\bm{k}}\sum_{\alpha\beta\sigma}h_{\alpha\beta}^\sigma(\bm{k})a_{\alpha\sigma}^{\dagger}(\bm{k})a_{\beta\sigma}(\bm{k})+\sum_{\bm{k}\alpha}\left[a_{\alpha\uparrow}^{\dagger}(\bm{k})a_{\alpha\downarrow}^{\dagger}(-\bm{k})\Delta_{\alpha}+a_{\alpha\downarrow}(\bm{k})a_{\alpha\uparrow}(-\bm{k})\Delta_{\alpha}^{*}\right],
\end{align}
where a time-reversal symmetry is assumed.
The fermion operator $a_{\alpha\sigma}^{\dagger}(\bm{k})$ 
in the momentum space depends on the orbital positions $\{\bm \delta_\alpha\}$ in a unit-cell:
\begin{align}
	a_{\alpha\sigma}^{\dagger}(\bm{k}) & =\frac{1}{\sqrt{N}}\sum_{i}e^{i(\bm{r}_{i}+\bm{\delta}_{\alpha})\cdot\bm{k}}a_{i\alpha\sigma}^{\dagger}\\
	a_{i\alpha\sigma}^{\dagger} & =\frac{1}{\sqrt{N}}\sum_{\bm{k}}e^{-i(\bm{r}_{i}+\bm{\delta}_{\alpha})\cdot\bm{k}}a_{\alpha\sigma}^{\dagger}(\bm{k}).
\end{align}
with $\bm{r}_i$ is the position of unitcells. Accordingly, the Hamiltonian $h_{\alpha\beta}^{\sigma}(\bm{k})$ has
the form as 
\begin{equation}
	h_{\alpha\beta}^{\sigma}(\bm{k})=\sum_{ij}h_{ij\alpha\beta}^{\sigma}e^{-i(\bm{r}_{i}+\bm{\delta}_{\alpha}-\bm{r}_{j}-\bm{\delta}_{\beta})\cdot\bm{k}}.
\end{equation}
For the notation simplicity, we can reformulate the mean-field theory in a matrix form,
% & =\sum_{\bm{k}}\sum_{\alpha\beta}\left[h_{\alpha\beta}(\bm{k})-\mu\delta_{\alpha\beta}\right]a_{\alpha\sigma}^{\dagger}(\bm{k})a_{\beta\sigma}(\bm{k})+\sum_{\bm{k}}\sum_{\alpha}[\Delta_{\alpha}a_{\alpha\uparrow}^{\dagger}(\bm{k})a_{\alpha\downarrow}^{\dagger}(-\bm{k})+\Delta_{\alpha}^{*}a_{\alpha\downarrow}(-\bm{k})a_{\alpha\uparrow}(\bm{k})]-\frac{1}{U}\sum_{\alpha}\vert\Delta_{\alpha}\vert^{2}\nonumber \\
%  & 
\begin{align}
	H_{\mathrm{mf}} =\sum_{\bm{k}}\psi_{\bm{k}}^{\dagger}(\hat{h}\otimes\tau_{z}+\mathrm{Re}\hat{\Delta}\otimes\tau_{x}+\mathrm{Im}\hat{\Delta}\otimes\tau_{y})\psi_{\bm{k}}+\sum_{\bm{k}}\sum_{\alpha\beta}\left[h_{\alpha\beta}(\bm{k})-\mu\delta_{\alpha\beta}\right]-\frac{1}{U}\sum_{\alpha}\vert\Delta_{\alpha}\vert^{2},
\end{align}
where the $\psi_{\bm{k}}$ is the spinor with $(\psi_{\bm{k}})_{\alpha\uparrow}=a_{\alpha\uparrow}(\bm{k})$
and $(\psi_{\bm{k}})_{\alpha\downarrow}=a_{\alpha\downarrow}^{\dagger}(-\bm{k})$.
The $\hat{h}$ is the matrix with elements $(\hat{h})_{\alpha\beta}=h^{\uparrow}_{\alpha\beta}(\bm{k})-\mu\delta_{\alpha\beta}$ when we assume a time-reversal symmetry.
The pairing matrix $\hat{\Delta}$ has the elements $(\hat{\Delta})_{\alpha\beta}=\Delta_{\alpha}\delta_{\alpha\beta}$. From the mean-field Hamiltonian, we can define the Green function $\hat G_{\alpha\alpha^{\prime},\sigma\sigma^{\prime}}(i\omega_{n},\bm{k})=\langle(\psi_{\bm{k}})_{\alpha\sigma}(i\omega_{n})(\psi_{\bm{k}}^{\dagger})_{\alpha^{\prime}\sigma^{\prime}}(i\omega_{n})\rangle$
with 
\begin{equation}
	G(i\omega_{n},\bm{k})=\frac{1}{i\omega_{n}+\hat{h}\otimes\tau_{z}+\mathrm{Re}\hat{\Delta}\otimes\tau_{x}+\mathrm{Im}\hat{\Delta}\otimes\tau_{y}}.
\end{equation}
We can determine the pairing gap 
\begin{equation}
	\hat{\Delta}_{\alpha}=\frac{T}{N}\sum_{n}\sum_{\sigma\sigma^{\prime}}\sum_{\bm k}\hat{G}_{\alpha\sigma,\alpha\sigma^{\prime}}(i\omega_{n},\bm{k})(\tau_{x})_{\sigma^{\prime}\sigma},
	\label{smeq:gap}
\end{equation}
and the number equations 
\begin{equation}
	2\nu=1+\frac{T}{N}\sum_{n}\sum_{\sigma\sigma^{\prime}}\sum_{\alpha}\sum_{\bm k}\hat G_{\alpha\sigma,\alpha\sigma^{\prime}}(i\omega_{n},\bm{k})(\tau_{z})_{\sigma\sigma^{\prime}}.
\end{equation}

With Gor'kov's Green function, we can express the Cooper pair
correlators 
% \begin{align}
	% \label{numer_corre}
	% C(\bm{r},\bm{r}^{\prime}) & =\sum_{\alpha}\langle a_{\alpha\uparrow}(\bm{r})a_{\alpha\downarrow}(\bm{r})a_{\alpha\downarrow}^{\dagger}(\bm{r}^{\prime})a_{\alpha\uparrow}^{\dagger}(\bm{r}^{\prime})\rangle\nonumber \\
	%  & =\frac{T}{N}\sum_{n}\sum_{\bm{q}}e^{-i\bm{q}\cdot(\bm{r}-\bm{r}^{\prime})}\sum_{\alpha\bm{k}}[G_0^{\alpha\alpha}(i\omega_n,\bm{k}+\bm{q})G_0^{\alpha\alpha}(-i\omega_n,-\bm{k})\\
	% &+F_0^{\alpha\alpha}(i\omega_n,\bm{k}+\bm{q})F_0^{\dagger\alpha\alpha}(-i\omega_n,-\bm{k})],
	% \end{align}
\begin{align}
	\label{numer_corre}
	C(\bm{r},\bm{r}^{\prime}) & =\sum_{\alpha}\langle a_{\alpha\uparrow}(\bm{r})a_{\alpha\downarrow}(\bm{r})a_{\alpha\downarrow}^{\dagger}(\bm{r}^{\prime})a_{\alpha\uparrow}^{\dagger}(\bm{r}^{\prime})\rangle_c\nonumber \\
	&=\frac{T}{N}\sum_{n}\sum_{\bm{q}}e^{-i\bm{q}\cdot(\bm{r}-\bm{r}^{\prime})}\sum_{\alpha\bm{k}}G_0^{\alpha\alpha}(i\omega_n,\bm{k}+\bm{q})G_0^{\alpha\alpha}(-i\omega_n,-\bm{k}),
\end{align}
with Gor’kov’s normal Green’s functions $G_0$. Here $\langle \mathcal O(\mathbf r)\mathcal O(\mathbf r^\prime) \rangle_c\equiv \langle \mathcal O(\mathbf r)\mathcal O(\mathbf r^\prime) \rangle  -\langle \mathcal O(\mathbf r)\rangle \langle \mathcal O(\mathbf r^\prime ) \rangle $ represents the correlator generated from the connected generating functional, which removes a coordinate-independent constant.
This formula can be employed to evaluate the coherence length for several lattice model, such as sawtooth and Lieb lattices.

\subsection*{Projected mean-field Hamiltonian and Green function}

We can keep the most relevant band around the Fermi energy due to the large band gap. 
Assuming the related Bloch states $u(\bm{k})$,
we introduce the projection as
\begin{equation}
	a_{i\alpha\sigma}^{\dagger}\rightarrow\frac{1}{\sqrt{N}}\sum_{\bm{k}}e^{-i(\bm{r}_{i}+\bm{\delta}_{\alpha})\cdot\bm{k}}u_{\alpha\sigma}(\bm{k})c_{\sigma}^{\dagger}(\bm{k}),
\end{equation}
which maps the mean-field Hamiltonian to
\begin{equation}
	H_{0}=\sum_{\bm{k}\sigma}\epsilon(\bm{k})c_{\bm{k}\sigma}^{\dagger}c_{\bm{k}\sigma}+\sum_{\bm{k}}\left[c_{\uparrow}^{\dagger}(\bm{k})c_{\downarrow}^{\dagger}(-\bm{k})\Delta_{0}+h.c.\right],
\end{equation}
where 
\begin{equation}
	\Delta_{0}=\frac{1}{N}\sum_{\alpha}\sum_{\bm{k}}\Delta_{\alpha}u_{\alpha}(\bm{k})u_{\alpha}^{*}(\bm{k}).
\end{equation}

We first review the basic BCS relations for a $s$-wave superconductor with uniform pairing $\Delta$. We can introduce the s-wave order parameter on the band fermion,
\begin{equation}
	\Delta=-\frac{U}{N}\sum_{\bm{k}}\langle c_{\downarrow}(-\bm{k})c_{\uparrow}(\bm{k})\rangle,
	\label{smeq:Deltaband}
\end{equation}
and by a proper gauge choice, we can set the order parameter $\Delta$
to be real-valued. Then one can derive a mean-field Hamiltonian
\begin{align}
	H_\mathrm{mf} & =\sum_{\bm{k}}\left[(\epsilon(\bm{k})-\mu)c_{\sigma}^{\dagger}(\bm{k})c_{\sigma}(\bm{k})+\Delta c_{\uparrow}^{\dagger}(\bm{k})c_{\downarrow}^{\dagger}(-\bm{k})+\Delta c_{\downarrow}(-\bm{k})c_{\uparrow}(\bm{k})\right]+\frac{1}{U}\Delta^{2}\nonumber \\
	& =\sum_{\bm{k}}\begin{bmatrix}c_{\uparrow}^{\dagger}(\bm{k}) & c_{\downarrow}(-\bm{k})\end{bmatrix}\begin{bmatrix}\epsilon(\bm{k})-\mu & \Delta\\
		\Delta & -(\epsilon(\bm{k})-\mu)
	\end{bmatrix}\begin{bmatrix}c_{\uparrow}(\bm{k})\\
		c_{\downarrow}^{\dagger}(-\bm{k})
	\end{bmatrix}+\sum_{\bm{k}}(\epsilon(\bm{k})-\mu)+\frac{1}{U}\Delta^{2}\nonumber \\
	& =\sum_{\bm{k}}\psi_{\bm{k}}^{\dagger}[(\epsilon(\bm{k})-\mu)\tau_{z}+\Delta\tau_{x}]\psi_{\bm{k}}+\sum_{\bm{k}}(\epsilon(\bm{k})-\mu)+\frac{1}{U}\Delta^{2},
	\label{smeq:hmf}
\end{align}
where $\psi_{\bm{k}}=\begin{bmatrix}c_{\uparrow}(\bm{k})\\
	c_{\downarrow}^{\dagger}(-\bm{k})
\end{bmatrix}$ is the Nambu spinor and $\tau_{x,y,z}$ are Pauli matrices. We define
the Gor'kov's Green function 
\begin{equation}
	G(\bm{r}\tau;\bm{r}^{\prime}\tau^{\prime})=\begin{bmatrix}\langle\hat{T}[c_{\uparrow}(\bm{r},\tau)c_{\uparrow}^{\dagger}(\bm{r}^{\prime},\tau^{\prime})]\rangle & \langle\hat{T}[c_{\uparrow}(\bm{r},\tau)c_{\downarrow}(\bm{r}^{\prime},\tau^{\prime})]\rangle\\
		\langle\hat{T}[c_{\downarrow}^{\dagger}(\bm{r},\tau)c_{\uparrow}^{\dagger}(\bm{r}^{\prime},\tau^{\prime})]\rangle & \langle\hat{T}[c_{\downarrow}^{\dagger}(\bm{r},\tau)c_{\downarrow}(\bm{r}^{\prime},\tau^{\prime})]\rangle
	\end{bmatrix},
\end{equation}
where $c_{\sigma}(\bm{r},\tau)=e^{H_\mathrm{mf}\tau}c_{\sigma}(\bm{r})e^{-H_\mathrm{mf}\tau}$
and $\hat{T}$ denotes the imaginary time order. In the frequency-momentum
space, one can directly extract the Green's function as 
\begin{equation}
	G(i\omega_{n},\bm{k})=\frac{-i\omega_{n}\tau_{0}-(\epsilon(\bm{k})-\mu)\tau_{z}-\Delta\tau_{x}}{\omega_{n}^{2}+(\epsilon(\bm{k})-\mu)^{2}+\Delta^{2}},\label{smeq:greenfunction}
\end{equation}
where $\omega_{n}=(2n+1)\pi T$ is the Matsubara frequency and $\tau_{0}$
is a $2\times2$ identity matrix.
With the Gor'kov's Green function in Eq.~(\ref{smeq:greenfunction}),
we can determine the order parameters by 
\begin{equation}
	\Delta=-\frac{U}{2N}T\sum_{\bm{k},n}\mathrm{Tr}G(i\omega_{n},\bm{k})\tau_{x}=\frac{U}{N}\sum_{\bm{k},n}\frac{\Delta}{\omega_{n}^{2}+(\epsilon(\bm{k})-\mu)^{2}+\Delta^{2}},
\end{equation}
or 
\begin{equation}
	1=\frac{U}{N}T\sum_{\bm{k},n}\frac{1}{\omega_{n}^{2}+(\epsilon(\bm{k})-\mu)^{2}+\Delta^{2}}.\label{smeq:gap_eq}
\end{equation}
We also need to determine the chemical potential by the number of equations
\begin{equation}
	2\nu=1+\frac{T}{N}\sum_{\bm{k},n}\mathrm{Tr}G(i\omega_{n},\bm{k})\tau_{z}.\label{smeq:number_eq}
\end{equation}
with $\nu$ being the filling factor. Combine Eqs.~(\ref{smeq:gap_eq})
and (\ref{smeq:number_eq}) and we can obtain the superconducting order parameter.

\subsection*{Interaction renormalized chemical potential $\mu$}\label{smsec:chem}

In the above discussion, we introduce two chemical potentials: the bare
one $\mu_{0}$ and interaction-renormalized one $\mu=\mu(\Delta)$.
We determine the Fermi momentum $k_{F}$ by the bare chemical potential
$\epsilon(\bm{k}_{F})=\mu_{0}$ while the latter one enters the Bogoliubov
quasiparticle dispersion together with the pairing gap. The two quantities
can be determined by the number equations with 
\begin{align}
	\nu & =\frac{1}{2}\left(1-\text{\ensuremath{\int\frac{d^{2}k}{(2\pi)^{2}}}}\frac{\epsilon(\bm{k})-\mu_{0}}{\vert\epsilon(\bm{k})-\mu_{0}\vert}\tanh\frac{\beta\vert\epsilon(\bm{k})-\mu_{0}\vert}{2}\right),\\
	\nu & =\frac{1}{2}\left(1-\int\frac{d^{2}k}{(2\pi)^{2}}\frac{\epsilon(\bm{k})-\mu}{\varepsilon(\bm{k})}\tanh\frac{\beta\varepsilon(\bm{k})}{2}\right).
\end{align}
For our purpose to illustrate the relationship between the two chemical
potentials, we can take a parabolic dispersion $\epsilon(\bm{k})=\frac{k^{2}}{2m}$
as an example. Then in the zero temperature $T=0$, we have 
\begin{align}
	\int\frac{d^{2}k}{(2\pi)^{2}}\frac{\epsilon(\bm{k})-\mu}{\varepsilon(\bm{k})}\tanh\frac{\beta\varepsilon(\bm{k})}{2} & =\int\frac{d^{2}k}{(2\pi)^{2}}\frac{\epsilon(\bm{k})-\mu}{\sqrt{[\epsilon(\bm{k})-\mu]^{2}+\Delta^{2}}}\nonumber \\
	& =\frac{2\pi}{(2\pi)^{2}}\int_{0}^{k_{F}}dk\frac{\frac{k^{2}}{2m}-\mu}{\sqrt{[\frac{k^{2}}{2m}-\mu]^{2}+\Delta^{2}}}\nonumber \\
	& =\frac{m}{2\pi}\left(\sqrt{\Delta^{2}}-\sqrt{\Delta^{2}+\mu^{2}}\right).
\end{align}
with $k_{F}=\sqrt{2m\mu}$ to regularize the integral in the second
line. Then we find the relation between $\mu_{0}$ and $\mu$
\begin{equation}
	\mu_{0}=\Delta-\sqrt{\Delta^{2}+\mu^{2}}=\frac{\mu^{2}}{\Delta+\sqrt{\Delta^{2}+\mu^{2}}},
\end{equation}
or 
\begin{equation}
	\mu=\sqrt{2\Delta\mu_{0}+\mu_{0}^{2}}.
\end{equation}
Indeed, in the regime where the superconducting gap $\Delta$ is much smaller than the bare chemical potential $\mu_0$ (i.e., $\Delta/\mu_0\ll1$), we can safely neglect the renormalization effect of the interaction on the chemical potential and simply use the bare chemical potential $\mu_{0}$.
However, in the opposite limit, when the superconducting gap $\Delta$ becomes much larger than the bare chemical potential $\mu_{0}$ (i.e., $\Delta\gg\mu_{0}$), the renormalization effect becomes significant. In this case, the chemical potential is renormalized to $\mu=\sqrt{2\Delta\mu_{0}}$, and it can deviate significantly from its bare value $\mu_{0}$. It is crucial to consider this renormalization effect in such scenarios, as it affects the low-energy physics and the behavior of the system under strong interactions.

\section*{Supplementary Note 2: Quantum geometry and coherence length}

\subsection*{Proof of the independence of coherence length from orbital positions}
With the mean-field Hamiltonian $H_\mathrm{mf}$, we deduce the coherence length from the Cooper pair correlator
\begin{align}
	C(\bm{r}_{ij}) & =\sum_{\alpha}\langle a_{i\alpha\downarrow}a_{i\alpha\uparrow}a_{j\alpha\uparrow}^{\dagger}a_{j\alpha\downarrow}^{\dagger}\rangle_c \\
	&\equiv \sum_{\alpha}\langle a_{i\alpha\downarrow}a_{i\alpha\uparrow}a_{j\alpha\uparrow}^{\dagger}a_{j\alpha\downarrow}^{\dagger}\rangle - \langle a_{i\alpha\downarrow}a_{i\alpha\uparrow}\rangle \langle a_{j\alpha\uparrow}^{\dagger}a_{j\alpha\downarrow}^{\dagger}\rangle,
	\label{smeq:c_Ij}
\end{align}
with $\bm{r}_{ij}=\bm{r}_{i}-\bm{r}_{j}$. The correlator $C(\bm{r}_{ij})$
is expected to be decay exponentially $C(\bm{r}_{ij})\sim e^{-\vert\bm{r}_{ij}\vert/\xi}$ in an isotropic system. We aim to demonstrate that the correlator in Eq.~\eqref{smeq:c_Ij} is invariant under the choice of $\bm{\delta}_{\alpha}$.
We reformulate the mean-field Hamiltonian in the band basis of $H_{0}$.
We can expand the orbital fermions in the band basis
\begin{equation}
	a_{i\alpha\sigma}^{\dagger}=\frac{1}{\sqrt{N}}\sum_{n\bm{k}}e^{-i(\bm{r}_{i}+\bm{\delta}_{\alpha})\cdot\bm{k}}u_{n\alpha\sigma}(\bm{k})c_{n\sigma}^{\dagger}(\bm{k}),
\end{equation}
with 
\begin{equation}
	\sum_{\beta}h_{\alpha\beta}^{\sigma}(\bm{k})u_{n\beta\sigma}(\bm{k})=\epsilon_{n}(\bm{k})u_{n\alpha\sigma}(\bm{k}),
\end{equation}
and then we have the mean-field Hamiltonian 
\begin{equation}
	H=\sum_{n\bm{k}\sigma}\epsilon_{n}(\bm{k})c_{n\bm{k}\sigma}^{\dagger}c_{n\bm{k}\sigma}+\sum_{\bm{k}}\sum_{nm}\left[c_{n\uparrow}^{\dagger}(\bm{k})c_{m\downarrow}^{\dagger}(-\bm{k})\Delta_{nm}+h.c.\right],
	\label{eq:hmf}
\end{equation}
where $\Delta_{nm}=1/N\sum_{\alpha\bm{k}}\Delta_{\alpha}u_{n\alpha}(\bm{k})u_{m\alpha}^{*}(\bm{k})$. The mean-field order parameters $\Delta_\alpha$ or $\Delta_{nm}$ can be solved self-consistently. Obviously, the mean-field Hamiltonian is independent of the choices of the orbital positions. Similarly, the Cooper pair correlator $C(\bm{r}_{ij})$ can be evaluated in the band basis,
\begin{align}
	\label{eq:pair_correlator}
	C(\bm{r}_{ij}) & =\sum_{\alpha}\langle a_{i\alpha\downarrow}a_{i\alpha\uparrow}a_{j\alpha\uparrow}^{\dagger}a_{j\alpha\downarrow}^{\dagger}\rangle_c\nonumber \\
	& =\frac{T}{N}\sum_{\bm{q}}e^{-i\bm{q}\cdot(\bm{r}_i-\bm{r}_j)}\sum_{nmhd}\sum_{\alpha\bm{k}}u_{n\alpha}^{*}(\bm{k}+\bm{q})u_{m\alpha}(\bm{k})u_{h\alpha}^{*}(\bm{k}^{\prime})u_{d\alpha}(\bm{k}^{\prime}+\bm{q})\langle c_{n\downarrow}(\bm{k}+\bm{q})c_{m\uparrow}(-\bm{k})c_{h\uparrow}^{\dagger}(-\bm{k}^{\prime})c_{d\downarrow}^{\dagger}(\bm{k}^{\prime}+\bm{q})\rangle_c.
\end{align}
The expression
of $u_{n\alpha}(\bm{k})$ depends on the choice of the orbital
positions $\bm{\delta}_{\alpha}$. For example, if we shift $\bm{\delta}_{\alpha}\rightarrow\bm{\delta}_{\alpha}+\bm{x}_{\alpha}$,
then we have 
\begin{align}
	h_{\alpha\beta}^{\sigma}(\bm{k}) & \rightarrow h_{\alpha\beta}^{\sigma}(\bm{k})e^{-i\bm{k}\cdot(\bm{x}_{\alpha}-\bm{x}_{\beta})},\\
	u_{\alpha}(\bm{k}) & \rightarrow e^{i\bm{k}\cdot\bm{x}_{\alpha}}u_{\alpha}(\bm{k}).
\end{align}
It is easy to check that $u_{n\alpha}^{*}(\bm{k}+\bm{q})u_{m\alpha}(\bm{k})u_{h\alpha}^{*}(\bm{k}^{\prime})u_{d\alpha}(\bm{k}^{\prime}+\bm{q})$
and $\Delta_{nm}$ are invariant under the choice of $\bm{\delta}_{\alpha}$.
Moreover, the correlator $\langle c_{n\downarrow}(\bm{k}+\bm{q})c_{m\uparrow}(-\bm{k})c_{h\uparrow}^{\dagger}(-\bm{k}^{\prime})c_{d\downarrow}^{\dagger}(\bm{k}^{\prime}+\bm{q})\rangle_c$ in Eq.~\eqref{eq:pair_correlator} can be obtained based on the mean-field Hamiltonian in Eq.~\eqref{eq:hmf},
which is also invariant under the choice of $\bm{\delta}_{\alpha}$. Therefore,
we can conclude that the Cooper correlator $C(\bm{r}_{ij})$ is variant
under the choice of $\bm{\delta}_{\alpha}$, which directly implies that
the coherence length is a quantity which is independent of the intra-cell-orbital
positions. 
% If the system we consider has an isolated flat band near Fermi energy, we can simplify Eq.~(\ref{eq:pair_correlator}) as:
% \begin{equation}
	% \label{eq:pair_corre}
	% \mathcal C(\bm{r}_{ij}) =\frac{T}{N}\sum_{\bm{q}}e^{-i\bm{q}\cdot(\bm{r}_i-\bm{r}_j)}\sum_{\omega_n,\alpha\bm{k}}\vert u_\alpha(\bm{k}-\bm{q})u^*_\alpha(\bm{k}) \vert^{2}\mathrm{Tr}[G(i\omega_{n},\bm{k})G(-i\omega_{n},-\bm{k}+\bm{q})].
	% \end{equation}
% The orbital-resolved form factor can be expanded as:
% \begin{equation}
	% \sum_{\alpha}u_\alpha(\bm{k}-\bm{q})u^*_\alpha(\bm{k})u^*_\alpha(\bm{k}-\bm{q})u_\alpha(\bm{k})=\sum_\alpha|u_\alpha(\bm{k})|^4-\sum_{\alpha,a} q_a \partial_a|u_\alpha(\bm{k})|^4-\sum_{ab}\mathcal{G}^o_{ab}(\bm{k})q_aq_b.
	% \end{equation}
% Here $\mathcal{G}^o_{ab}$ is called the orbital-resolved quantum metric, which shares the same physical dimension with quantum metric but needs to be evaluated case by case. Importantly, $\mathcal C(\bm{r}_{ij})$ does not depend on the choice of orbital positions. 

%%%%%%%%%
\subsection*{Uniform pairing condition and minimum quantum metric}

We can act the projection on the Cooper pair correlators. For the
sake of convenience, we instead investigate the quantity 
\begin{align}
	\label{eq:full_mq}
	\mathcal{M}(\bm{q}) & =\sum_{ij\alpha}e^{-i\bm{q}\cdot\bm{r}_{ij}}\langle a_{i\alpha\downarrow}a_{i\alpha\uparrow}a_{j\alpha\uparrow}^{\dagger}a_{j\alpha\downarrow}^{\dagger}\rangle_c\nonumber \\
	& =\frac{1}{N}\sum_{\bm{k}\bm{k}^{\prime}}\sum_{\alpha}\langle a_{\alpha\downarrow}(\bm{k}+\bm{q})a_{\alpha\uparrow}(-\bm{k})a_{\alpha\uparrow}^{\dagger}(-\bm{k}^{\prime})a_{\alpha\downarrow}^{\dagger}(\bm{k}^{\prime}+\bm{q})\rangle_c \nonumber \\
	& \rightarrow \frac{T}{N}\sum_{\bm{k}}\sum_{\alpha}\vert u_{\alpha}^{*}(\bm{k}+\bm{q})u_{\alpha}(\bm{k})\vert^{2} G_0(i\omega_n,\bm k+\bm{q})G_0(-i\omega_n,-\bm k).
\end{align}
where $G_0$ is the normal Gorkov's Green function in Eq.~\eqref{smeq:greenfunction}.
Obviously, $\mathcal{M}(\bm{q})$ remains invariant under the choice
of $\bm{\delta}_{\alpha}$ since both $\vert u_{\alpha}^{*}(\bm{k}+\bm{q})u_{\alpha}(\bm{k})\vert^{2}$ and $\langle c_{\downarrow}(\bm{k}+\bm{q})c_{\uparrow}(-\bm{k})c_{\uparrow}^{\dagger}(-\bm{k})c_{\downarrow}^{\dagger}(\bm{k}+\bm{q})\rangle_c$
are invariant. The quantity $\mathcal{M}(\bm{q})$ is then expected to form
as 
\begin{equation}
	\mathcal{M}(\bm{q})=\frac{\mathcal{M}(0)}{1+q^{2}\xi^{2}}~,
\end{equation}
with $q=\vert\bm{q}\vert$. One may find that the coherence length
$\xi$ can be obtained via 
\begin{equation}
	\label{eq:xi_mq}
	\xi^{2}=-\frac{1}{2\mathcal{M}(\bm{0})}\frac{\partial^{2}\mathcal{M}(\bm{q})}{\partial q^{2}}\vert_{q=0}~.
\end{equation}
Therefore, the coherence length is determined by the second derivative
around $q=0$. In order to get a compact form, we resort to the
assumption that 
\begin{align}
	\label{eq:uniform}
	\sum_{\alpha}\vert u_{\alpha}^{*}(\bm{k}+\bm{q})u_{\alpha}(\bm{k})\vert^{2} & \approx\frac{1}{N_{orb}}\sum_{\alpha\beta}u_{\alpha}^{*}(\bm{k}+\bm{q})u_{\alpha}(\bm{k})u_{\beta}(\bm{k}+\bm{q})u_{\beta}^{*}(\bm{k})\nonumber \\
	& =\frac{1}{N_{orb}}\vert\langle u(\bm{k}+\bm{q})\vert u(\bm{k})\rangle\vert^{2},
\end{align}
where $N_{orb}$ is the number of orbitals. There are two aspects of this treatment. First, it assumes that the $N_{orb}$ orbitals fulfill a condition that $1/N \sum_{\bm{k}}|u_{\alpha}(\bm{k})|^2=1/N_{orb}$, which leads to the same pairing gap of all orbitals. In Refs.~\cite{huhtinen2022revisiting,torma2018quantum}, this condition is referred to as the uniform pairing condition. Second, however, such assumption in generally fails to keep the invariance under the choice of the intra-unit cell orbital positions. Under the shift $\bm{\delta}_{\alpha}\rightarrow\bm{\delta}_{\alpha}+\bm{x}_{\alpha}$,
and $u_{\alpha}(\bm{k})\rightarrow u_{\alpha}(\bm{k})e^{i\bm{k}\cdot\bm{x}_{\alpha}}$,
one can find that 
\begin{align}
	\Gamma(\bm{k},\bm{q}) 
	= &  \sum_{\alpha\beta}u_{\alpha}^{*}(\bm{k}+\bm{q})u_{\alpha}(\bm{k})u_{\beta}(\bm{k}+\bm{q})u_{\beta}^{*}(\bm{k})\nonumber \\
	\rightarrow & \sum_{\alpha\beta}e^{-i\bm{q}\cdot(\bm{x}_{\alpha}-\bm{x}_{\beta})}u_{\alpha}^{*}(\bm{k}+\bm{q})u_{\alpha}(\bm{k})u_{\beta}(\bm{k}+\bm{q})u_{\beta}^{*}(\bm{k}) \notag \\
	= & \sum_{\alpha\beta}e^{-i\bm{q}\cdot(\bm{x}_{\alpha}-\bm{x}_{\beta})}\vert u_{\alpha}^{*}(\bm{k}+\bm{q})u_{\alpha}(\bm{k})\vert e^{i\theta_{\bm{k}\alpha}(\bm{q})}|u_{\beta}(\bm{k}+\bm{q})u_{\beta}^{*}(\bm{k})|e^{-i\theta_{\bm{k}\beta}(\bm{q})},
\end{align}
This may lead to an unphysical contribution. In the expression before
the projection, the quantity does not depend on the phase factor $\theta_{\alpha}(\bm{q})$. When the phase factor vanishes, the assumption in Eq.(\ref{eq:uniform}) is valid. For it, we can shift the orbital positions by $\bm{\delta}_{\alpha}\rightarrow\bm{\delta}_{\alpha}+\bm{x}_{\alpha}$,
then 
\begin{equation}
	\exp\left[-i(\theta_{\bm{k}\alpha}(\bm{q})-\theta_{\bm{k}\beta}(\bm{q}))\right]\rightarrow\exp\left[-i(\theta_{\bm{k}\alpha}(\bm{q})-\theta_{\bm{k}\beta}(\bm{q}))\right]e^{-i\bm{q}\cdot(\bm{x}_{\alpha}-\bm{x}_{\beta})},
\end{equation}
where $\theta_{\bm{k}\alpha\beta}=\theta_{\bm{k}\alpha}-\theta_{\bm{k}\beta}$. 
We can obtain the condition for the minimization
\begin{equation}
	\label{eq:orbital_shift}
	\frac{1}{N}\sum_{\bm{k}}\frac{d}{dq_{a}}\theta_{\bm{k}\alpha\beta}(\bm{q})|_{\bm{q}\rightarrow 0}=\bm{x}_{\alpha}^{a}-\bm{x}_{\beta}^{a}~.
\end{equation}
In this case, we will find that the quantum metric indeed becomes the minimal one. We can fix the position of the first orbital $\bm{\delta}_1=0$ and $\bm{x}_1=0$.
The shift of other orbitals can be accordingly determined by Eq.~(\ref{eq:orbital_shift}). Furthermore, due to the dependence of $\theta_{\bm{k}\alpha\beta}(\bm{q})$ on the momentum $\bm{k}$, in general it is impossible to choose a set of orbital positions such that the second term in Eq.~(\ref{eq:uniform}) vanishes. This is different from the case of superfluid weight. In fact, the best we can do is to reach the minimal quantum metric in  Eq.~(\ref{eq:uniform}). In Ref.\cite{huhtinen2022revisiting}, it was shown that the resulting $\bm{\delta}_\alpha+\bm{x}_\alpha$ for the minimal quantum metric is unique and independent of the initial positions $\bm{\delta}_\alpha$. Therefore, we can reach a conclusion that the coherence length $\xi$ for a superconductor is controlled by the minimal quantum metric under the uniform pairing condition.

One may find that the coherence length obtained under the uniform pairing condition is slightly larger than the value by directly evaluating the $\mathcal{M}(\bm{q})$ in Eq.~(\ref{eq:full_mq}). There are two origins of the errors. Suppose that we start with the orbital positions corresponding to the minimal quantum metric. The first type of the error comes from the factor $\mathcal{M}(0)$ in Eq.~(\ref{eq:xi_mq}), and the error can be evaluated from the Chebyshev's sum inequality,
\begin{equation}
	\frac{1}{N}\sum_{\bm{k}\alpha}|u_{\alpha}(\bm{k})|^4-\frac{1}{NN_{orb}}\sum_{\bm{k}}(\sum_{\alpha}|u_\alpha (\bm{k})|^2)^2=\frac{1}{NN_{orb}}\sum_{\bm{k},\alpha\beta}(|u_\alpha (\bm{k})|^2-|u_\beta (\bm{k})|^2)^2 \geq 0.
\end{equation}
Secondly, it is due to the variance $\sigma_{ab}^2$ of the $\partial_a \theta_{\bm{k}\alpha\beta}$ with $u^*_\alpha(\bm{k}+\bm{q})u_\alpha(\bm{k})=|u^*_\alpha(\bm{k}+\bm{q})u_\alpha(\bm{k})|e^{i\theta_{\bm{k}\alpha}(\bm{q})}$ which appears in $\partial^2 \mathcal{M}(\bm{q})/\partial q^2$,
\begin{equation}
	\sigma^2_{ab}=\frac{1}{2N}\sum_{\bm{k}\alpha\beta}(\partial_a \theta_{\bm{k}\alpha\beta}(0)-\overline{\partial_a\theta_{\alpha\beta}(0)})(\partial_b \theta_{\bm{k}\alpha\beta}(0)-\overline{\partial_b\theta_{\alpha\beta}(0)}),
\end{equation}
with $\overline{\partial_a\theta_{\alpha\beta}(0)}=\frac{1}{N}\sum_{\bm{k}}\partial \theta_{\bm{k}\alpha\beta}(\bm{q})/\partial q_a|_{q\rightarrow 0}$. The two derivations can lead to an overestimation. For the models we consider in the main text, the derivation is acceptable. More numerical demonstrations are shown in Sec.~III.

One may understand the difference between the role of quantum metric in the superfluid weight and the coherence length as follows. The superfluid weight can be related to gapless Goldstone mode excitations. In comparison, the coherence length is the correlation behavior of the gapped Cooper pair excitations. Thus, the coherence can depend on the local distributions of the quantum metric.

\section*{Supplementary Note 3: Coherence length for a weakly dispersive band}

\subsection*{Derivation on the total coherence length}
In this subsection, we derive the coherence length for a flat band with finite dispersion. 
In the flat band limit, where $\epsilon(\bm{k})=0$, the Bogoliubov quasiparticle remains flat, resulting in the independence of the Green function $G(i\omega_{n},\bm{k})$ on the momentum $\bm{k}$.
Our objective is to demonstrate that the long-distance behavior of the Cooper pair propagator $\mathcal C(\bm{r},\bm{r}^{\prime})$ aligns precisely with the new length scale dictated by the quantum metric. The presence of the factor $e^{-i\bm{q}\cdot(\bm{r}-\bm{r}^{\prime})}$ in the Cooper pair propagator $\mathcal C(\bm{r},\bm{r}^{\prime})$ implies that the dominant contribution arises from small $\bm{q}$ regions. These small $\bm{q}$ regions correspond to the low-energy limit, where the long-distance behavior of the Cooper pair propagator is significantly affected by the quantum metric's new length scale.
By observing for small $q$, we have 
\begin{align}
	\vert\Lambda(\bm{k}+\bm{q},\bm{k})\vert^{2} & =1-\sum_{ab}\mathcal{G}_{ab}(\bm{k})q_{a}q_{b}+\mathcal{O}(q^{3}),
\end{align}
where $\mathcal{G}_{ab}$ is nothing more than the quantum metric, 
\begin{equation}
	\label{quantum_metric}
	\mathcal{G}_{ab}(\bm{k})=\mathrm{Re}\left[\langle\partial_{a}u(\bm{k})\vert\partial_{b}u(\bm{k})\rangle-\langle\partial_{a}u(\bm{k})\vert u(\bm{k})\rangle\langle u(\bm{k})\vert\partial_{b}u(\bm{k})\rangle\right],
\end{equation}
Instead of Eq.~(\ref{eq:pair_correlator}) which only involves the normal Gor'kov's Green function, we can consider a more compacted form by combining the Cooper-pair correlator and the density-density correlator,
\begin{align}
	\mathcal C(\bm r, \bm r^\prime) &= \sum_\alpha \sum_{\mathcal O\in \mathcal S_\mathrm{op} } \langle \mathcal O(\bm r) \mathcal O^\dagger (\bm r^\prime )\rangle_c \nonumber  \\ 
	& \rightarrow \frac{T}{N}\sum_{\bm k,\bm q}|\Lambda(\bm k +\bm q,\bm k)|^2 \mathrm{Tr} G(i\omega_n,\bm k+\bm q)G(-i\omega_n, -\bm k)
\end{align}
where $\mathcal S_\mathrm{op} = \{ a_{\alpha \downarrow}a_{\alpha\uparrow}, a^\dagger_{\alpha\uparrow}a^\dagger_{\alpha \downarrow}, a^\dagger_{\alpha \downarrow}a_{\alpha \downarrow},a^\dagger_{\alpha \uparrow}a_{\alpha \uparrow}\}$ and for the notation simplification, we ignore the overall $1/N_\mathrm{orb}$ factor in this section. We make this treatment is to evaluate the coherence length analytically. In the last line, we have introduced the projection.
Since the Green function in Eq.~(\ref{eq:pair_correlator}) is independent
on the momentum in the flatband limit, we can obtain the tendency
\begin{align}
	\mathcal C(\bm{r},\bm{r}^{\prime}) & \simeq T\sum_{\bm{q}}e^{-i\bm{q}\cdot(\bm{r}-\bm{r}^{\prime})}\frac{1}{N}\sum_{\bm{k}}(1-\sum_{ab}\mathcal{G}_{ab}(\bm{k})q_{a}q_{b})\frac{\tanh\left(\frac{\beta\sqrt{\Delta^{2}+\mu^{2}}}{2}\right)}{\sqrt{\Delta^{2}+\mu^{2}}}\nonumber \\
	& =\frac{\tanh\left(\frac{\beta\sqrt{\Delta^{2}+\mu^{2}}}{2}\right)}{\sqrt{\Delta^{2}+\mu^{2}}}\sum_{\bm{q}}e^{-i\bm{q}\cdot(\bm{r}-\bm{r}^{\prime})}(1-\sum_{ab}\overline{\mathcal{G}}_{ab}q_{a}q_{b})\nonumber \\
	& =\frac{\tanh\left(\frac{\beta\sqrt{\Delta^{2}+\mu^{2}}}{2}\right)}{\sqrt{\Delta^{2}+\mu^{2}}}\sum_{\bm{q}}e^{-i\bm{q}\cdot(\bm{r}-\bm{r}^{\prime})}\frac{1}{1+\sum_{ab}\overline{\mathcal{G}}_{ab}q_{a}q_{b}},
\end{align}
where 
\begin{align}
	& T\sum_{n}\mathrm{Tr}G(i\omega_{n},\bm{k}+\bm{q})G(-i\omega_{n},-\bm{k})\nonumber \\
	= & T\sum_{n}\mathrm{Tr}\frac{-i\omega_{n}\tau_{0}+\mu\tau_{z}-\Delta\tau_{x}}{\omega_{n}^{2}+\mu{}^{2}+\Delta^{2}}\frac{i\omega_{n}\tau_{0}+\mu\tau_{z}-\Delta\tau_{x}}{\omega_{n}^{2}+\mu{}^{2}+\Delta^{2}}\nonumber \\
	= & T\sum_{n}\frac{2}{\omega_{n}^{2}+\mu{}^{2}+\Delta^{2}}=\frac{\tanh\left(\frac{\beta\sqrt{\Delta^{2}+\mu^{2}}}{2}\right)}{\sqrt{\Delta^{2}+\mu^{2}}},
\end{align}
and $\overline{\mathcal{G}}_{ab}$ is the averaged quantum metric over the first
BZ. For an isotropic system, $\overline{\mathcal{G}}_{ab}=\sqrt{\det\overline{\mathcal{G}}_{ab}}\delta_{ab}$,
yielding a much more compact form 
\begin{align}
	\mathcal C(\bm{r},\bm{r}^{\prime}) & =\frac{\tanh\left(\frac{\beta}{2}\varepsilon \right)}{\varepsilon}\int\frac{d^{2}q}{(2\pi)^{2}}\frac{e^{-i\bm{q}\cdot(\bm{r}-\bm{r}^{\prime})}}{1+q^{2}\sqrt{\det\bar{\mathcal{G}}_{ab}}}\notag\\
	& =\frac{\tanh\left(\frac{\beta}{2}\varepsilon\right)}{\varepsilon\sqrt{\det\bar{\mathcal{G}}_{ab}}}\int\frac{d^{2}q}{(2\pi)^{2}}\frac{e^{-i\bm{q}\cdot(\bm{r}-\bm{r}^{\prime})}}{q^{2}+\xi^{-2}},
\end{align}
where $\xi$ is the coherence length determined by the quantum metric.
\begin{equation}
	\xi=(\det\overline{\mathcal{G}}_{ab})^{\frac{1}{4}}.
\end{equation}
% For the orbital-resolved quantum metric, we have the similar result as $\xi=(\det\overline{\mathcal{G}}^o_{ab})^{\frac{1}{4}}$, where $\overline{\mathcal{G}}^o_{ab}=\frac{\sum_{\bm{k}}\mathcal G^o_{ab}(\bm{k})}{\sum_{\bm{k}}\Lambda_0(\bm{k})}$ with $\Lambda_0(\bm{k})=\sum_\alpha|\Lambda_\alpha(\bm{k},\bm{k})|^2$. 
For the sake of convenience, we focus on the zero-temperature case, that is, $T=0$. For a general dispersive band, the low-energy contribution to the Cooper pair propagator can be expressed as,
\begin{equation}
	\mathcal C(\bm{r},\bm{r}^{\prime})\equiv\sum_{\bm{q}}e^{-i\bm{q}\cdot(\bm{r}-\bm{r}^{\prime})}\mathcal{M}(\bm{q}),
\end{equation}
where $\mathcal{M}(\bm{q})$ has the form:
\begin{equation}
	\mathcal{M}(\bm{q})=\frac{T}{N}\sum_{n\bm{k}}\vert\Lambda(\bm{k}+\bm{q},\bm{k})\vert^{2}\mathrm{Tr}[G(i\omega_{n},\bm{k}+\bm{q})G(-i\omega_{n},-\bm{k})].
\end{equation}
The presence of two significant factors sets this apart: the quantum metric and the conventional band dispersion. Both of these elements play crucial roles in determining the behavior of the Cooper pair propagator in the low-energy regime.
Thus, we can decompose $\mathcal{M}(\mathbf{q})$,
\begin{equation}
	\mathcal{M}(\bm{q})=\mathcal{M}_{0}+\mathcal{M}_\mathrm{qm}(\bm{q})+\mathcal{M}_\mathrm{con}(\bm{q}).
\end{equation}
Here $\mathcal{M}_{0}$ is independent on the momentum $\bm{q}$
\begin{align}
	\mathcal{M}_{0} & \equiv\mathcal{M}(\bm{q}=\bm{0})=\frac{T}{N}\sum_{n\bm{k}}\mathrm{Tr}G(i\omega_{n},\bm{k})G(-i\omega_{n},-\bm{k})\nonumber \\
	& =\frac{T}{N}\sum_{n\bm{k}}\frac{2}{\omega_{n}^{2}+(\epsilon(\bm{k})-\mu)^{2}+\Delta^{2}}\nonumber \\
	& =\rho_{0}(\epsilon_{F})\left[\mathrm{arctanh}\left(\frac{W/2+\mu_{0}-\mu}{\sqrt{\Delta^{2}+(W/2+\mu_{0}-\mu)^{2}}}\right)+\mathrm{arctanh}\left(\frac{W/2-\mu_{0}+\mu}{\sqrt{\Delta^{2}+(W/2-\mu_{0}+\mu)^{2}}}\right)\right]\nonumber \\
	& \equiv \mathcal{M}_{0}(\Delta,\mu),\label{smeq:f2}
\end{align}
where $W$ denotes a regularization cutoff. 
We want to emphasize that for the narrow band case, it is necessary to consider the renormalization on the chemical potential from the interactions or the superconducting pairings. 
In $T=0$, the bare chemical potential, denoted as $\mu_{0}$, represents the Fermi energy in the absence of any interaction terms,
\begin{equation}
	\nu=\int\frac{d^{2}k}{(2\pi)^{2}}\theta(\mu_{0}-\epsilon(\bm{k})),
\end{equation}
The Fermi momentum and Fermi velocity then can be defined as  $\epsilon(\bm{k}_{F})=\mu_{0}$ and $v_{F}=k_{F}/m$. 
In the superconducting phase, the chemical potential shall get renormalized to be determined from Eq.~\eqref{smeq:number_eq}
For a highly dispersive system with a large Fermi velocity, 
we can ignore the difference between $\mu_{0}$ and $\mu$ (see Sec.~\ref{smsec:chem}). In this case, 
we can set $W$ as the Debye frequency, $W=\omega_{D}$, where $W/\mu_{0}\ll1$. 
This implies that electrons within the energy window $[-W/2,+W/2]$ dominate the low-energy physics.
In contrast, in the narrow band limit, we can relate $W$ to the bandwidth or the chemical potential $W=\mu_{0}$ (setting the band bottom at zero energy). In this case, all electrons in the Fermi sea contribute to the low-energy behaviors.
In particular, the flatband limit refers to the case $W\rightarrow0$, where $\rho_{0}(\epsilon_{F})$ diverges. However, since $\int_{-\frac{W}{2}}^{\frac{W}{2}}\rho_{0}(\epsilon)d\epsilon=1$, we can approximate the density of states at the Fermi energy as
\begin{equation}
	\rho_{0}(\epsilon_{F})=\frac{1}{W}.
\end{equation}
Then we can calculate  $\mathcal{M}_{0}(\Delta,\mu)$
in Eq.~(\ref{smeq:f2}) at zero temperature,
\begin{align}
	\lim_{W\rightarrow0}\mathcal{M}_{0}(\Delta,\mu) & =\lim_{W\rightarrow0}\rho_{0}(\epsilon_{F})\left[\mathrm{arctanh}\left(\frac{W/2+\mu_{0}-\mu}{\sqrt{\Delta^{2}+(W/2+\mu_{0}-\mu)^{2}}}\right)+\mathrm{arctanh}\left(\frac{W/2-\mu_{0}+\mu}{\sqrt{\Delta^{2}+(W/2-\mu_{0}+\mu)^{2}}}\right)\right]\nonumber \\
	& =\frac{1}{\sqrt{\Delta^{2}+\mu^{2}}}.
\end{align}

For the sake of convenience, we will focus on a parabolic band dispersion given by $\epsilon(\bm{k})=\frac{k^{2}}{2m}$ to approach a flatband scenario as the effective mass $m$ diverges.
To evaluate the coherence length, we can perform an expansion of $\mathcal{M}(\bm{q})$ up to terms of $q^{2}$.
The quantum metric part arises from the expansion of the form factor and up to
$q^{2}$, we have 
\begin{align}
	\mathcal{M}_\mathrm{qm}(\bm{q}) & =-\frac{T}{N}\sum_{n\bm{k}}\sum_{ab}q_{a}q_{b}\mathcal G_{ab}(\bm{k})\mathrm{Tr}G(i\omega_{n},\bm{k})G(-i\omega_{n},-\bm{k})\nonumber \\
	& =-\frac{T}{N}\sum_{n\bm{k}}\sum_{ab}q_{a}q_{b}\mathcal G_{ab}(\bm{k})\mathrm{Tr}G(i\omega_{n},\bm{k})G(-i\omega_{n},-\bm{k})\nonumber \\
	& =-\frac{T}{N}\sum_{\bm{k}}\sum_{ab}q_{a}q_{b}\mathcal G_{ab}(\bm{k})\frac{2}{\omega_{n}^{2}+(\epsilon(\bm{k})-\mu)^{2}+\Delta^{2}}\nonumber \\
	& =-\frac{1}{N}\sum_{\bm{k}}\sum_{ab}q_{a}q_{b}\mathcal G_{ab}(\bm{k})\frac{1}{\varepsilon(\bm{k})}\nonumber \\
	& =-q^{2}\mathcal{M}_{0}(\Delta,\mu)\sqrt{\det\overline{\mathcal G}_{ab}},
\end{align}
where $\varepsilon(\bm{k})=\sqrt{(\epsilon(\bm{k})-\mu)^{2}+\Delta^{2}}$
is the dispersion for the Bogoliubov quasiparticle and $\overline{\mathcal G}_{ab}$
is the averaged quantum metric weight by a factor from the quasiparticle
dispersion. 
\begin{equation}
	\overline{\mathcal G}_{ab}=\frac{\frac{1}{N}\sum_{\bm{k}}\mathcal G_{ab}(\bm{k})\frac{1}{\varepsilon(\bm{k})}}{\mathcal M_0(\Delta,\mu)}=\frac{\int\frac{d^{2}k}{(2\pi)^{2}}\frac{\mathcal G_{ab}(\bm{k})}{\varepsilon(\bm{k})}}{\mathcal M_0(\Delta,\mu)},\label{smeq:gamma_bar_Ef}
\end{equation}
In the narrow band limit $W\rightarrow0,m\rightarrow\infty$, $\overline{\mathcal G}_{ab}$
can be well captured by the averaged quantum metric over the first
BZ. In the flatband limit, we can recover the results with $\mathcal{M}_\mathrm{qm}(\mathbf{q})=\frac{-\sum_{ab}\overline{\mathcal G}_{ab}q_{a}q_{b}}{\sqrt{\Delta^{2}+\mu^{2}}}$.

The conventional contribution $\mathcal{M}_\mathrm{con}(\bm{q})$ from the
band dispersion can be found as
\begin{align}
	\mathcal{M}_\mathrm{con}(\bm{q}) & =\frac{T}{N}\sum_{n\bm{k}}\mathrm{Tr}G(i\omega_{n},\bm{k}+\bm{q})G(-i\omega_{n},-\bm{k})\nonumber \\
	& =-\frac{T}{N}\sum_{n\bm{k}}\frac{2\bm v_{F}\cdot \bm{q}[\epsilon(\bm{k})-\mu+\bm v_{F}\cdot\bm{q}]}{[\Delta^{2}+(\epsilon(\bm{k})-\mu)^{2}+\omega_{n}^{2}][\Delta^{2}+(\epsilon(\bm{k})-\mu+\bm v_{F}\cdot \bm{q})^{2}+\omega_{n}^{2}]},
\end{align}
Here we use $\epsilon(\bm{k+q})=\epsilon(\bm{k})+\bm v_{F}\cdot \bm{q}$ for
the isotropic system around the Fermi surface and we can further evaluate $\mathcal{M}_\mathrm{con}(\bm{q})$
as 
\begin{equation}
	\mathcal{M}_\mathrm{con}(\bm{q})=-q^{2}v_{F}^{2}f_{4}(\Delta,\mu),
\end{equation}
where 
\begin{align}
	f_{4}(\Delta,\mu) & =\frac{1}{N}\sum_{\bm{k}}\frac{2\Delta^{2}-(\epsilon(\bm{k})-\mu)^{2}}{4(\Delta^{2}+(\epsilon(\bm{k})-\mu)^{2})^{\frac{5}{2}}} \notag
	=\rho_{0}(\epsilon_{F})\int_{\mu_{0}-W/2}^{\mu_{0}+W/2}d\epsilon\frac{2\Delta^{2}-(\epsilon-\mu+\mu_{0})^{2}}{4(\Delta^{2}+(\epsilon-\mu+\mu_{0})^{2})^{\frac{5}{2}}} \notag\\
	& =\rho_{0}(\epsilon_{F})[\frac{(\mu_{0}-\mu+W/2)(2\Delta^{2}+(\mu_{0}-\mu+W/2)^{2})}{4\Delta^{2}[\Delta^{2}+(\mu_{0}-\mu+W/2)^{2}]^{\frac{3}{2}}}+\frac{(\mu-\mu_{0}+W/2)(2\Delta^{2}+(\mu-\mu_{0}+W/2)^{2})}{4\Delta^{2}[\Delta^{2}+(\mu-\mu_{0}+W/2)^{2}]^{\frac{3}{2}}}].\label{smeq:f4}
\end{align}
Collect all terms together and we extract the coherence length by
observing 
\begin{align}
	\mathcal{M}(\bm{q}) & =\mathcal{M}_{0}+\mathcal{M}_\mathrm{qm}(\bm{q})+\mathcal{M}_\mathrm{con}(\bm{q})\nonumber  =\mathcal{M}_{0}(\Delta,\mu)(1-\sum_{ab}\overline{\mathcal G}_{ab}q_{a}q_{b})-v_{F}^{2}q^{2}f_{4}(\Delta,\mu)\nonumber \\
	& \simeq\frac{\mathcal M_0(\Delta,\mu)}{1+\left(\sqrt{\det\overline{ \mathcal G}_{ab}}+v_{F}^{2}\frac{f_{4}(\Delta,\mu)}{f_{2}(\Delta,\mu)}\right)q^{2}}\nonumber  \equiv\frac{\mathcal M_0(\Delta,\mu)}{1+\xi^{2}q^{2}},
\end{align}
with 
\begin{equation}
	\xi=\sqrt{\sqrt{\det\overline{\mathcal G}_{ab}}+v_{F}^{2}\frac{f_{4}(\Delta,\mu)}{\mathcal M_0(\Delta,\mu)}},
\end{equation}
where the factor $\mathcal M_0(\Delta,\mu)$ and $f_{4}(\Delta,\mu)$ are
defined in Eqs.~(\ref{smeq:f2}) and (\ref{smeq:f4}).

We then consider two limits: the narrow band $W\sim\mu_{0}$ and the
broad band $W/\mu\ll1$. For the narrow band limit, we can expand
$\frac{f_{4}(\Delta,\mu)}{\mathcal M_0(\Delta,\mu)}$ around $W/\mu_{0}=1$,
which induces 
\begin{equation}
	\frac{f_{4}(\Delta,\mu)}{\mathcal M_0(\Delta,\mu)}=\frac{1}{4\Delta^{2}}\frac{\frac{(\mu_{0}-\mu+W/2)(2\Delta^{2}+(\mu_{0}-\mu+W/2)^{2})}{[\Delta^{2}+(\mu_{0}-\mu+W/2)^{2}]^{\frac{3}{2}}}+\frac{(\mu-\mu_{0}+W/2)(2\Delta^{2}+(\mu-\mu_{0}+W/2)^{2})}{[\Delta^{2}+(\mu-\mu_{0}+W/2)^{2}]^{\frac{3}{2}}}}{\mathrm{arctanh}\left(\frac{W/2+\mu_{0}-\mu}{\sqrt{\Delta^{2}+(W/2+\mu_{0}-\mu)^{2}}}\right)+\mathrm{arctanh}\left(\frac{W/2-\mu_{0}+\mu}{\sqrt{\Delta^{2}+(W/2-\mu_{0}+\mu)^{2}}}\right)}.
\end{equation}
If the bandwidth $W$ is much smaller than the attractive interaction,
that is, $\Delta\gg\mu_{0}$, which corresponds to the correction
to a flatband, we have 
\begin{equation}
	\frac{f_{4}(\Delta,\mu)}{\mathcal M_0(\Delta,\mu)}=\frac{1}{\Delta^{2}}
\end{equation}
and the coherence length can be formulated as $\xi=\sqrt{\sqrt{\det\overline{\mathcal G}_{ab}}+\frac{\text{\ensuremath{v_{F}^{2}}}}{\Delta^{2}}}$.
If the pairing gap is much smaller than the bandwidth $\Delta/W,\Delta/\mu_{0}\ll1$,
we can approximate the chemical potential $\text{\ensuremath{\mu=\mu_{0}}}$, yielding
\begin{equation}
	\frac{f_{4}(\Delta,\mu)}{\mathcal M_0(\Delta,\mu)}=\frac{1}{4\Delta^{2}}\frac{1}{\mathrm{arctanh}(1-2\Delta^{2}/\mu_{0}^{2})},
\end{equation}
which diverges as $-\frac{1}{4\Delta^{2}\log(\Delta^{2}/\mu_{0}^{2})}$.
For a broad band case $W/\mu_{0}\ll1$, we have the expansion,
\begin{equation}
	\frac{f_{4}(\Delta,\mu)}{\mathcal M_0(\Delta,\mu)}=\frac{1}{\Delta^{2}}+\mathcal{O}\left(\frac{W}{\mu_{0}}\right),
\end{equation}
where we ignore the renormalization on the chemical potential $\mu_{0}=\mu$. And
we still have $\xi=\sqrt{\sqrt{\det\overline{\mathcal G}_{ab}}+\frac{v_{F}^{2}}{\Delta^{2}}}$.

We can summarize all conditions for the coherence length $\xi=\sqrt{\sqrt{\det\overline{\mathcal G}_{ab}}+v_{F}^{2}\frac{f_{4}(\Delta,\mu)}{\mathcal M_0(\Delta,\mu)}}$
with 
\begin{equation}
	v_{F}^{2}\frac{f_{4}(\Delta,\mu)}{\mathcal M_0(\Delta,\mu)}=\begin{cases}
		\frac{\text{\ensuremath{v_{F}^{2}}}}{\Delta^{2}} & \frac{W}{\mu_{0}}\rightarrow1,\frac{\Delta}{\mu_{0}}\gg1 \\
		\frac{v_{F}^{2}}{\Delta^{2}}\frac{1}{4\mathrm{arctanh}(1-2\Delta^{2}/\mu_{0}^{2})} & \frac{W}{\mu_{0}}\rightarrow1,\frac{\Delta}{\mu_{0}}\ll1\\
		\frac{v_{F}^{2}}{\Delta^{2}} & \frac{W}{\mu_{0}}\ll1
	\end{cases}
\end{equation}
The conventional BCS superconductor corresponds to the case $\frac{W}{\mu_{0}}\ll1$,
where $\sqrt{\det\overline{\mathcal G}_{ab}}$ can be suppressed from Eq.~(\ref{smeq:gamma_bar_Ef}). 
% The weighted average of the orbital-resolved quantum metric is defined by
% \begin{equation}
	% \label{orbital_metric_average}
	% \overline{\mathcal{G}}^o_{ab}=\frac{\sum_{\bm{k}}\mathcal G^o_{ab}(\bm{k})/\varepsilon(\bm{k})}{\sum_{\bm{k}}\Lambda_0(\bm{k})/\varepsilon(\bm{k})},
	% \end{equation}
% where we have similar result as $\ell_{\mathrm{qm}} =\sqrt[4]{\det\overline{\mathcal{G}}^o_{ab}}$.

\subsection*{Angle dependence of coherence length}
In this subsection we discuss the case for a $s$-wave superconductor with non-circular Fermi surface. It is worth noting that the Fermi surface is ill-defined in the flat-band limit when $v_F=0$. Thus, we conduct the discussion for a superconductor with finite dispersion. In two dimensions, we can parameterize the anisotropic coherence length via $\xi=\xi(\theta)$. In other words, the Cooper pair correlator now becomes anisotropic with 
\begin{equation}
	C(\bm{r}_{ij})\propto e^{-|\bm{r}_{ij}|/\xi(\theta)}
\end{equation}
with $\bm{r}_{ij}=|\bm{r}_{ij}|(\cos\theta,\sin\theta)$. The total coherence length is still contributed by the band dispersion and quantum metric $\xi(\theta)=\sqrt{\xi_{BCS}^2(\theta)+\ell^2_{qm}(\theta)}$. The conventional term $\xi_{BCS}^2(\theta)$ has the same expression as predicted by BCS theory. For the quantum metric contribution, we have the expression:
\begin{equation}
	\ell_{qm}(\theta)=\sqrt{\cos^2\theta \overline{\mathcal{G}}_{xx}+\sin^2\theta \overline{\mathcal{G}}_{yy}+\sin 2\theta \overline{\mathcal{G}}_{xy}}
\end{equation}
One may find that the $\ell_{qm}(\theta)$ can be anisotropic due to the finite off-diagonal element $\overline{\mathcal{G}}_{xy}$.

\section*{Supplementary Note 4: Flat-band lattice model}
In this section, we provide detailed information on several flat-band models. Specifically, we calculate the coherence length for the sawtooth lattice and Lieb lattice models, demonstrating that the coherence length is independent of the orbital positions. Additionally, we highlight the discrepancy between the exact numerical results and the minimal metric. We also offer more comprehensive details on the flat-band models discussed in the main text.

\subsection*{Sawtooth lattice model}
In this subsection we study the 1D sawtooth lattice model that host flat band as illustrated in Supplementary Fig.\ref{fig:figS1}~(a). The unit cell has $A$ and $B$ sites with the hopping terms $V_1$ and $V_2$. To realize the flat band, $V_1=1$ and $V_2=\sqrt{2}$. The Hamiltonian in momentum space can be written as
\begin{equation}
	\mathcal{H}_{\bm{k}}=\left(
	\begin{matrix}{}
		-2V_1\cos ka &  -V_2[e^{-ik x}+e^{ik(1-x)}]  \\
		-V_2[e^{ik x}+e^{-ik(1-x)}]  & 0
	\end{matrix}\right)~.
\end{equation}
From the Hamiltonian, we can get the quantum metric as
\begin{equation}
	\mathcal{G}_{xx}=\frac{a^2+2x(x-1)(1+\cos ka)}{2(2+\cos k)^2}~,
\end{equation}
which reaches minimal at $x=0.5a$. The band structure is shown in Supplementary Fig.\ref{fig:figS1}~(b) with the exact flat band at $E=2$. In the mean-field calculations, we use the attractive interaction $U=0.4$, and we have $\Delta_A=0.118$ and $\Delta_B=0.088$. The calculated coherence length $\xi$ from Eq.~\eqref{numer_corre} is shown in  Supplementary Fig.\ref{fig:figS1}~(c) which is independent of the orbital position. The quantum metric $\sqrt{\overline{\mathcal{G}}_{xx}}$ depends on the orbital position which is parameterized by $x$. $\sqrt{\overline{\mathcal{G}}_{xx}}$ reaches its minimum at $x=0.5$, which is the minimal quantum metric. The discrepancy between the numerical result and the minimal quantum metric can be attributed to the approximation in Eq.~(\ref{eq:uniform}). One can expand the orbital-resolved form factor as $\sum_{\bm{k}\alpha}|u_{\alpha}^*(\bm{k}+\bm{q})u_{\alpha}^*(\bm{k})|^2=\sum_{\bm{k}\alpha}|u_{\alpha}^*(\bm{k})|^4-\sum_{\bm{k}}\mathcal{G}_{ab}^o q_a q_b$, where 
$\mathcal{G}_{ab}^o$ is a quantity related to quantum metric but not equal to it. The numerical value of the coherence length is consistent with $\ell_{\mathrm{qm}}=\sqrt[4]{\det\overline{\mathcal{G}}^o}$, where $\overline{\mathcal{G}}_{ab}^o=\sum_{\bm{k}}\mathcal{G}_{ab}^o(\bm{k})/\sum_{\bm{k}\alpha}|u_{\alpha}^*(\bm{k})|^4$. In the following, we will provide the calculations for 2D Lieb lattice, which has similar results.

%%%%%%%%%%%%%%%%%%%%%%
\begin{figure}[t]
	\centering 
	\includegraphics[width=1\linewidth]{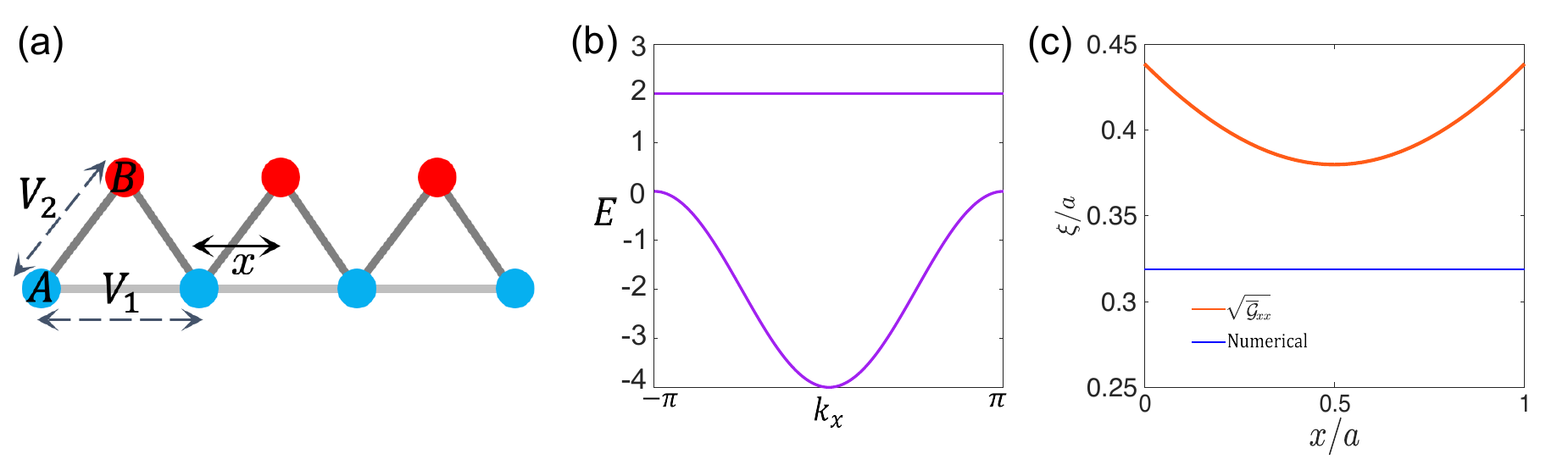}
	\caption{(a) The sawtooth lattice with $A$ and $B$ atoms forming a unit cell. The $AA$ and $AB$ hoppings are denoted as $V_1$ and $V_2$, respectively. $x$ is the position of $B$ atoms. (b) The band structure of sawtooth lattice for $V_1=1$ and $V_2=\sqrt{2}$. (c) The calculated coherence length as a function of $x$ (blue line) and the quantum metric $\sqrt{\overline{\mathcal{G}}_{xx}}$ (orange line). We set the on-site attractive interaction $U=0.4$.}
	\label{fig:figS1} 
\end{figure}
%%%%%%%%%%%%%%%%%%%%%%%
\subsection*{Lieb lattice}
In this subsection we study the coherence length for Lieb lattice model. The model Hamiltonian of Lieb lattice in the momentum space can be written as $H_0=\sum_{\bm{k},\sigma}a_{\bm{k}\sigma}^\dagger \mathcal{H}_{\bm{k}}a_{\bm{k}\sigma}$, where
\begin{equation}
	\mathcal{H}_{\bm{k}}=\left(
	\begin{matrix}{}
		&0 & f_{x} &f_2  \\
		&f^*_{x}  & 0 &f_{y}\\
		&f^*_2 &f^*_{y} &0
	\end{matrix}\right).
	\label{eq:liebh}
\end{equation}
Here $f_x=2J(\cos k_x a/2 +i\eta\sin k_x a/2 )$, $f_y=2J(\cos k_y a/2 +i\eta\sin k_y a/2)$ and $f_2=2t_2[\cos (k_x+k_y)a/2 +\cos (k_x-k_y)a/2 ]$. By changing the orbital positions, the Hamiltonian will transform as $\mathcal{H}_{\bm{k}} \rightarrow \mathcal{U}^\dagger(\bm{k})\mathcal{H}_{\bm{k}}\mathcal{U}(\bm{k})$ with $[\mathcal{U}(\bm{k})]_{\alpha\beta}=e^{i\bm{k}\cdot\bm{\delta}_{\alpha}}\delta_{\alpha\beta}$.

To support the analytical results as discussed above, we utilized self-consistent mean-field calculations on the Lieb lattice model, illustrated in Supplementary Fig.\ref{fig:figS2}~(a), exhibiting a perfectly flat band at $E=0$, as shown in Supplementary Fig.\ref{fig:figS2}~(b). Each unit cell comprises three atomic orbitals identified as $A$, $B$, and $C$. The staggered hoppings between $A$ and $B$ orbitals are denoted as $(1+\eta)J$ and $(1-\eta)J$, respectively. The quantum metric and band gap can be tuned by varying $\eta$. The next nearest-neighbor hopping is parameterized as $t_2$, which induces band dispersion. In Supplementary Fig.\ref{fig:figS2}~(c), the distribution of $\mathrm{Tr} [\mathcal G(\bm k)]$ at $\eta=0.3$ and $t_2=0$ is plotted, reaching a maximum near $M$. We then proceed with fully self-consistent solutions of the on-site order parameter $\hat{\Delta} = \mathrm{diag}(\Delta_A, \Delta_B, \Delta_C)$. The stable solutions indicate $\Delta_B \ll \Delta_A = \Delta_C$ due to the fact that the wave function of the flat band has zero orbital weight on the $B$ site. The intra-unit-cell positions of the orbitals can by varied by $x$, which is the distance between $B$ and $A$/$C$ orbitals.

We numerically evaluate the coherence length $\xi$ by using the multi-orbital correlation function in Eq.~\eqref{numer_corre}. In Supplementary Fig.~\ref{fig:figS2}(d) we plot $\xi$ as a function of $x$, which reveals that $\xi$ does not depend on the orbital position choice. However, the quantum metric of the flat band is lattice geometry dependent, which reaches the minimum at some specific $x$ corresponding to the minimal quantum metric~\cite{huhtinen2022revisiting}. The discrepancy between the numerical result and the minimal quantum metric can be attributed to the approximation in Eq.~(\ref{eq:uniform}). Due to the absence of the dispersion of the bands, the coherence length $\xi$ depends solely on the quantum metric by tuning $\eta$. This is illustrated in Supplementary Fig.~\ref{fig:figS2}(e), where the numerical results of pair correlation functions align with the theoretical results obtained from minimal quantum metric. 

To incorporate finite band dispersion, one can introduce an additional nearest-hopping $t_2$. This term gives rise to a band dispersion as well as the conventional contribution $\xi_\mathrm{BCS}$ to the total coherence length $\xi$. In Supplementary Supplementary Fig.~\ref{fig:figS2}(f), the total coherence length gradually decreases for $t_2=0.01J,0.02J$ when the attractive interaction strength $U$ increases. Remarkably, the total coherence length $\xi$ approaches the flat-band limit due to the suppression of the conventional contribution $\xi_\mathrm{BCS}$.

%%%%%%%%%%%%%%%%%%%%%%%%%%%%%%%%%%%%%%%%%%
\begin{figure*}[t]
	\centering
	\includegraphics[width=0.8\linewidth]{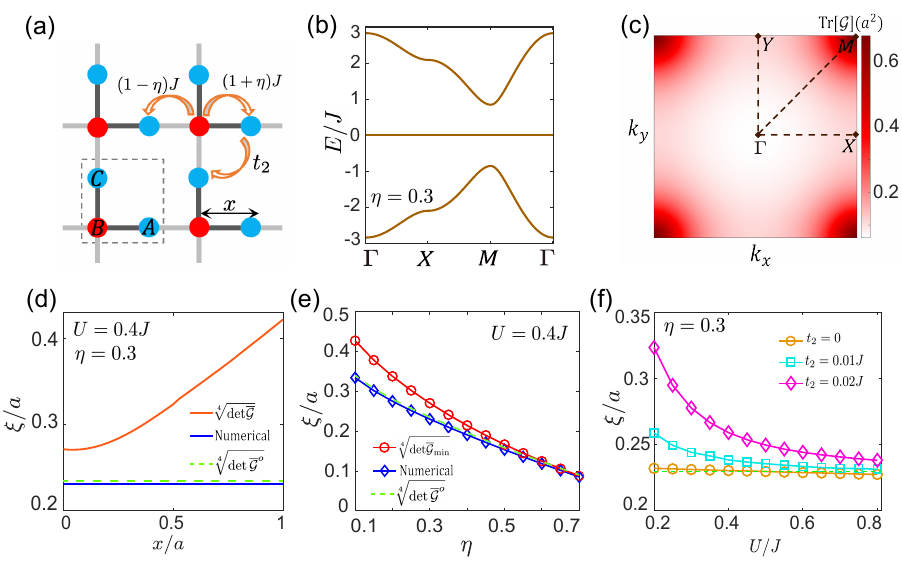}
	\caption{(a) A schematic plot of the Lieb lattice with $A$, $B$ and $C$ orbitals forming a unit cell. The thick and thin lines denote the nearest hoppings with $(1\pm\eta)J$, respectively. A next-nearest hopping $t_2$ can make the band dispersive. (b) The calculated band structure of the lattice model at $\eta=0.3$ and $t_2=0$ The flat band is separated from the other bands by a gap $\Delta_{\mathrm{gap}}=\sqrt{8}\eta J$. (c) The profiles of quantum metric $\mathrm{Tr} [\mathcal G]$ of the flat band in the first Brillouin zone. (d) The calculated coherence length $\xi$ as a function of the orbital position $x$.
		% The theoretical value $\ell_{\mathrm{qm}} =\sqrt[4]{\det\overline{\mathcal{G}}^o}$ is indicated by a dashed green line which coincides with the numerical one (blue).
		(e) The quantum metric dependence of $\xi$ as the parameter $\eta$ varies when $U=0.4J$. 
		% The dashed green line marks the length scale $\ell_{\mathrm{qm}} =\sqrt[4]{\det\overline{\mathcal{G}}^o}$. 
		The red circled line denotes the value from the minimal quantum metric. (f) The calculated $\xi$ at $\eta=0.3$ as a function of the attractive interaction $U/J$ for $t_2=0,0.01J,0.02J$. All calculations are conducted at $k_{B}T=0.005J$ and at half-filling $\mu=0$.}
	\label{fig:figS2}
\end{figure*}
%%%%%%%%%%%%%%%%%%%%%%%%%%%%%%%%%%%%%%%%%%%

\subsection*{Flat-band model with tunable quantum metric}
In this subsection we would like to study an exact flat-band model with tunable quantum metric and zero Berry curvature. The topologically trivial flat-band model Hamiltonian in Ref.~\cite{hofmann2023superconductivity} is given by 
\begin{equation}
	h_s(\bm{k})=-t [\sin(\alpha_{\bm{k}})\lambda_x+ s_z \cos(\alpha_{\bm{k}})\lambda_y]+[-2t_2(\cos k_x+\cos k_y)-\mu]\lambda_0~,
\end{equation}
where $\alpha(\bm{k})=\chi[\cos(k_x a)+\cos(k_y a)]$, $t_2$ is the nearest hopping which makes the band dispersive and $\mu$ is the chemical potential. $s=\{ \uparrow,\downarrow \}$ is the spin index with the eigenvalue $s_z=\pm 1$. This Hamiltonian preserves the time-reversal symmetry with $h_{\uparrow}(\bm{k})=h^{*}_{\downarrow}(-\bm{k})$. The energy dispersion is obtained as $\varepsilon_{\pm}(\bm{k})=\pm t+2t_2(\cos k_x+\cos k_y)-\mu$. The wave function for the upper (+) and lower ($-$) band can be written as
\begin{equation}
	\vert u_{+}\rangle =\frac{1}{\sqrt{2}}\begin{pmatrix}
		1 \\
		i s_ze^{is_z \alpha_{\bm{k}}}
	\end{pmatrix},\\\\\vert u_{-}\rangle =\frac{1}{\sqrt{2}}\begin{pmatrix}
		-1 \\
		i s_ze^{is_z \alpha_{\bm{k}}}
	\end{pmatrix}~.
\end{equation}
From the definition in Eq.~(1) of the main text,
% \begin{equation}
	% \mathfrak G_{ab}=\langle \partial_a u(\bm{k})|\left[1-|u(\bm{k})\rangle\langle u(\bm{k})|\right]|\partial_b u(\bm{k})\rangle.
	% \end{equation}
we can calculate the quantum geometric tensor for the ($+$) band,
\begin{equation}
	\mathfrak G_{ab}=\frac{1}{4}\partial_a \alpha_{\bm{k}}\partial_b \alpha_{\bm{k}}.
\end{equation}
Thus the quantum metric can be obtained as 
\begin{equation}
	\mathcal{G}_{ab}=a^2\chi^2 \sin(k_a)\sin(k_b)/4,
\end{equation}
and the Berry curvature vanishes identically.
\subsection*{Topological flat-band model with $C=2$}
% %%%%%%%%%%%%%%%%%%%%%%
% \begin{figure}[t]
	% \centering 
	% \includegraphics[width=1\linewidth]{supple_fig3}
	% \caption{(a) two-orbital square lattice with short and long-range hoppings. The inter-orbital nearest hopping, intra-orbital next-nearest-neighbor and fifth-nearest-neighbor hoppings are labeled. (b) The self-consistent pairing strength $\Delta_A$ and $\Delta_B$ as a function of $U$. (c) The coherence length is extracted from the pair correlation function. The bound from orbital-resolved quantum metric is guided by the dashed green line. }
	% \label{fig:figS3} 
	% \end{figure}
% %%%%%%%%%%%%%%%%%%%%%%%
In the main text, we study a system exhibiting topological flat band. Here we show the details of the model Hamiltonian as well as the mean-field calculations. In the tight binding model, the inter-orbital nearest hopping, intra-orbital next-nearest-neighbor and fifth-nearest-neighbor hoppings are involved. The Hamiltonian reads $\mathcal{H}_{0}=\sum_{\bm{k}}\hat{c}_{\bm{k}}^{\dagger}h(\bm{k})\hat{c}_{\bm{k}}$ with the fermion operators $\hat{c}_{\bm{k}}=(\hat{c}_{A\bm{k}},\hat{c}_{B\bm{k}})^{T}$, with $A$ and $B$ being the orbital index:
\begin{align}
	h(\bm{k})=&-\frac{1-\sqrt{2}}{4}[\cos(2(k_x+k_y))+\cos(2(k_x-k_y))]\lambda_0-\sqrt{2}\sin(k_x)\sin(k_y)\lambda_z \notag \\
	&-\frac{\sqrt{2}}{2}[\cos(k_x)+\cos(k_y)]\lambda_x-\frac{\sqrt{2}}{2}[-\cos(k_x)+\cos(k_y)]\lambda_y+\delta_E \lambda_0.
\end{align}
Here $\lambda$ is the Pauli matrix acting on the sublattice basis. We set $\delta_E =1.215$ to make the lowest flat band lying on $E=0$. In a compact form we can write $h(\bm{k})=\sum_i h_i (\bm{k}) \sigma_i$. The quantum metric can be obtained as $\mathcal{G}_{ab}=1/4 \partial_a \hat{\bf{h}}\cdot \partial_b \hat{\bf{h}}$ where $\hat{\bf{h}}=(h_x,h_y,h_z)$. We can write the quantum metric for the flat band as
\begin{eqnarray}
	\mathcal{G}_{xx} &=& \frac{2\sin^2 k_y(1+\cos^2 k_x \cos^2 k_y)+\cos^2 k_y \sin^2 k_x}{(3+\cos 2k_x \cos 2k_y)^2},\\
	\mathcal{G}_{yy} &=& \frac{2\sin^2 k_x(1+\cos^2 k_x \cos^2 k_y)+\cos^2 k_x \sin^2 k_y}{(3+\cos 2k_x \cos 2k_y)^2},\\
	\mathcal{G}_{xy} &=& \frac{3 \sin 2k_x \sin 2k_y}{4(3+\cos 2k_x \cos 2k_y)^2}.
\end{eqnarray}

The Berry curvature of the flat band can be obtained with $\mathcal{F}_{xy}=-1/2 \hat{\bf{h}}\cdot (\partial_x \hat{\bf{h}} \times \partial_y \hat{\bf{h}})$, which reads 
\begin{equation}
	\mathcal{F}_{xy}=\frac{2(\cos^2 k_y \sin^2 k_x + \sin^2 k_y)}{(3+\cos 2k_x \cos 2k_y)^{3/2}}.
\end{equation}
One can verify that $\sqrt{\mathrm{det}\mathcal G(\bm k)}=\mathcal F_{xy}(\bm k)/2$ and $\mathcal F_{xy}(\bm k)>0$ over the whole Brillouin zone. The Chern number is given by $C=\frac{1}{2\pi}\int \mathcal{F}_{xy}(\bm k) dk_x dk_y =2$.

In the main text, we have calculated the coherence length by using the projected model Hamiltonian and shown that the coherence length is bound by the quantum metric. We assume the pairing gap $\Delta_0$ on the flat band. Here we show that this model obeys the uniform pairing condition. Since the system has time-reversal symmetry, the Hamiltonians of the spin $\uparrow$ and spin $\downarrow$ can be written as $h_{\uparrow}(\bm{k})=h(\bm{k})$ and $h_{\downarrow}(\bm{k})=h^*(-\bm{k})$, respectively. The full BdG Hamiltonian can be written in terms of the Nambu basis $\Psi_{\bm{k}}=(\hat{c}_{A,\bm{k}\uparrow},\hat{c}_{B,\bm{k}\uparrow},\hat{c}^\dagger_{A,-\bm{k}\downarrow},\hat{c}^\dagger_{B,-\bm{k}\downarrow})^T$:
\begin{equation}
	\mathcal{H}_\mathrm{BdG}=\sum_{\bm{k}}\Psi_{\bm{k}}^\dagger\left[
	\begin{matrix}{}
		&h_{\uparrow}(\bm{k})-\mu & \text{diag}(\Delta_A,\Delta_B)   \\
		&\text{diag}(\Delta_A^*,\Delta_B^*)  &  -h_{\downarrow}(-\bm{k})+\mu
	\end{matrix}\right]\Psi_{\bm{k}}.
\end{equation}
Here $A (B)$ denotes the orbital index. In the mean-field calculation, we set $\mu=0$ in the half-filling region. With self-consistent calculations of the two intra-orbital pairing order parameters $\Delta_{A}$ and $\Delta_{B}$, we have identified stable solutions with $\Delta_{A}=\Delta_{B}=\Delta_{0}$ and the uniform pairing conduction is satisfied. This is because the charge densities of $A$ and $B$ orbitals are same with $\sum_{\bm{k}}|u_{A\bm{k}}|^2=\sum_{\bm{k}}|u_{B\bm{k}}|^2=0.5$.
% By varying the attractive interaction $U$, 
% $\Delta_{0}$ is approximately linear with $U$ as shown in Fig.~\ref{fig:figS3}(b) with the blue circled line. 

% By implementing a mean-field theory (see Supplementary Information III),  we have calculated the Cooper pair correlation functions and extracted the coherence length, which exhibits a decreasing trend as the pairing gap $\Delta$ increases, as depicted in Fig.~\ref{fig:figS3}~(c). The coherence length $\xi$ converges to approximately $\ell_{\mathrm{qm}} =\sqrt[4]{\det\overline{\mathcal{G}}^o}$, which is the orbital-resolved quantum metric. It is clear that quantitatively the discrepancy comes from the approximation in Eq.~\ref{eq:uniform}. Thus, the topological bound of the coherence length in the main text is under the approximation in Eq.~(\ref{eq:uniform}). In general cases, the coherence length calculated from the quantum metric can give a relatively qualitatively result. 

We summarize the flat-band models studied in our work, focusing on the interplay between the quantum metric and band dispersion effects. We provide a general expression for the coherence length as $\xi = \sqrt{\xi_\mathrm{BCS}^2 +\ell_{\mathrm{qm}}^{2}}$ is directly related to the minimal quantum metric. However, we observe a slight discrepancy between the minimal quantum metric and the numerical results from self-consistent mean-field calculations in certain lattice models. This discrepancy is attributed to the uniform pairing assumption in Eq.\eqref{eq:uniform}. In the main text, we use the band-projection method to study the interplay between the quantum metric and band dispersion effects, and in both cases, the quantum metric is the minimal one. Generally, the minimal quantum metric is a reliable measure for evaluating the coherence length contributed by the quantum metric effect, as seen in twisted moir\'{e} materials.

\section*{Supplementary Note 5: Continuum model of moir\'{e} graphene family}
\subsection*{Continuum model of twisted bilayer graphene}
In this section, we provide the calculations of the band structure as well as the quantum metric for twisted bilayer graphene (TBG), twisted trilayer graphene (TTG), and twisted double bilayer graphene (TDBG) in the main text. TBG is formed by two layers with a twist angle $-\theta/2$ and $\theta/2$ relative to $x$ axis respectively (see Supplementary Fig.~\ref{fig:figS4}). In the continuum limit by neglecting the intervalley mixing, the Hamiltonian reads:
\begin{equation}
	H=H_1+H_2+H_{int}.
\end{equation}
Here, $H_{1/2}$ denotes the Hamiltonian of the top/bottom layer, and $H_{int}$ denotes the interlayer coupling. $H_{1/2}$ reads:
\begin{align}
	H_{1} & =\sum_{\xi,\bm{k}}a^\dagger_{1,\xi,\bm{k}}\hbar v_F \hat{R}_{\theta/2}(\bm{k}-\bm{K_1})\cdot(\xi \sigma_x,\sigma_y)a_{1,\xi,\bm{k}}, \\ 
	H_{2} & =\sum_{\xi,\bm{k}}a^\dagger_{2,\xi,\bm{k}}\hbar v_F \hat{R}_{-\theta/2}(\bm{k}-\bm{K_2})\cdot(\xi \sigma_x,\sigma_y)a_{2,\xi,\bm{k}},
\end{align}
where $\hat R_{\theta} = \begin{pmatrix} 
	\cos \theta & -\sin \theta \\
	\sin \theta & \cos \theta 
\end{pmatrix}$ is the rotation operator. $\sigma$ acts on the spinor of sublattices A and B in graphene. $\xi=\pm$ denotes the $K$ and $K'$ valleys. The interlayer coupling can be written as 
\begin{equation}
	H_{int}=\sum_{\xi}\int_{\bm{r}}\psi^\dagger_{1,\xi}(\bm{r})T(\bm{r})\psi_{2,\xi}(\bm{r})+h.c.~,
\end{equation}
where 
\begin{equation}
	T(\bm{r})=\left(
	\begin{array}{cc}
		w_0 & w_1 \\
		w_1 & w_0 \\
	\end{array}
	\right)
	+ \left(
	\begin{array}{cc}
		w_0 & w_1e^{-i\frac{2\pi}{3}} \\
		w_1e^{i\frac{2\pi}{3}} & w_0 \\
	\end{array}
	\right)e^{-i\bm{G}_1^M\cdot \bm{r}}
	+   \left(
	\begin{array}{cc}
		w_0 & w_1e^{i\frac{2\pi}{3}} \\
		w_1e^{-i\frac{2\pi}{3}} & w_0 \\
	\end{array}
	\right)e^{-i(\bm{G}_1^M+\bm{G}_2^M)\cdot \bm{r}}   ~.
\end{equation}
With the relation $\psi_{i,\xi}(\bm{r})=\sum_{\bm{k}}a_{i,\xi,\bm{k}}e^{i\bm{k}\cdot\bm{r}}$, the interlayer coupling can be expressed as 
\begin{equation}
	H_{int}=\sum_{\xi}a^\dagger_{1,\xi,\bm{k}}[\left(
	\begin{array}{cc}
		w_0 & w_1 \\
		w_1 & w_0 \\
	\end{array}
	\right)\delta_{\bm{k},\bm{k}'}+\left(
	\begin{array}{cc}
		w_0 & w_1e^{-i\frac{2\pi}{3}} \\
		w_1e^{i\frac{2\pi}{3}} & w_0 \\
	\end{array}
	\right)\delta_{\bm{k+\bm{G}_1^M},\bm{k}'}+ \left(
	\begin{array}{cc}
		w_0 & w_1e^{i\frac{2\pi}{3}} \\
		w_1e^{-i\frac{2\pi}{3}} & w_0 \\
	\end{array}
	\right)\delta_{\bm{k}+\bm{G}_1^M+\bm{G}_2^M,\bm{k}'}]a_{2,\xi,\bm{k}'}+h.c.~.
\end{equation}

%Generally speaking, the moir\'{e} graphene are formed by two sectors S1 and S2 with a twist angle $-\theta/2$ and $\theta/2$ relative to $x$ axis respectively (see Fig.\ref{fig:fig2}). For TBG, the sector S1 is layer 1 and sector S2 is layer 2. For TTG,  the sector S1 is layer 1,3 and sector S2 is layer 2. For TDBG, the sector S1 is layer 1,2 and sector S2 is layer 3,4. The moir\'{e} energy bands originating from the $K_{+}$ or $K_{-}$ valleys can be described by the continuum Hamiltonian $H=\sum_{v}\int d\bm{r}\psi_{v\bm{k}}^{\dagger}(\bm{r})\mathcal{\hat{H}}_{v}(\bm{k},\bm{r})\psi_{v\bm{k}}(\bm{r})$, where $v=\pm$ denotes the valley index. The eigenstate can be written as $\psi_{v\bm{k}}(\bm{r})=\sum_{G}C_{v\bm{k}}^X(\bm{G})e^{i(\bm{k}+\bm{G})\cdot \bm{r}}$, where $X$ is the sublattice index and $\bm{G}=m_1 G_1^M+m_1 G_2^M$ and $m_1$ and $m_2$ are integers. The continuum model can be spanned with the plane-wave basis, and we choose a cutoff of 37 reciprocal lattice vectors.
%%%%%%%%%%%%%%%%%%%%%%
\begin{figure}[t]
	\centering 
	\includegraphics[width=0.5\linewidth]{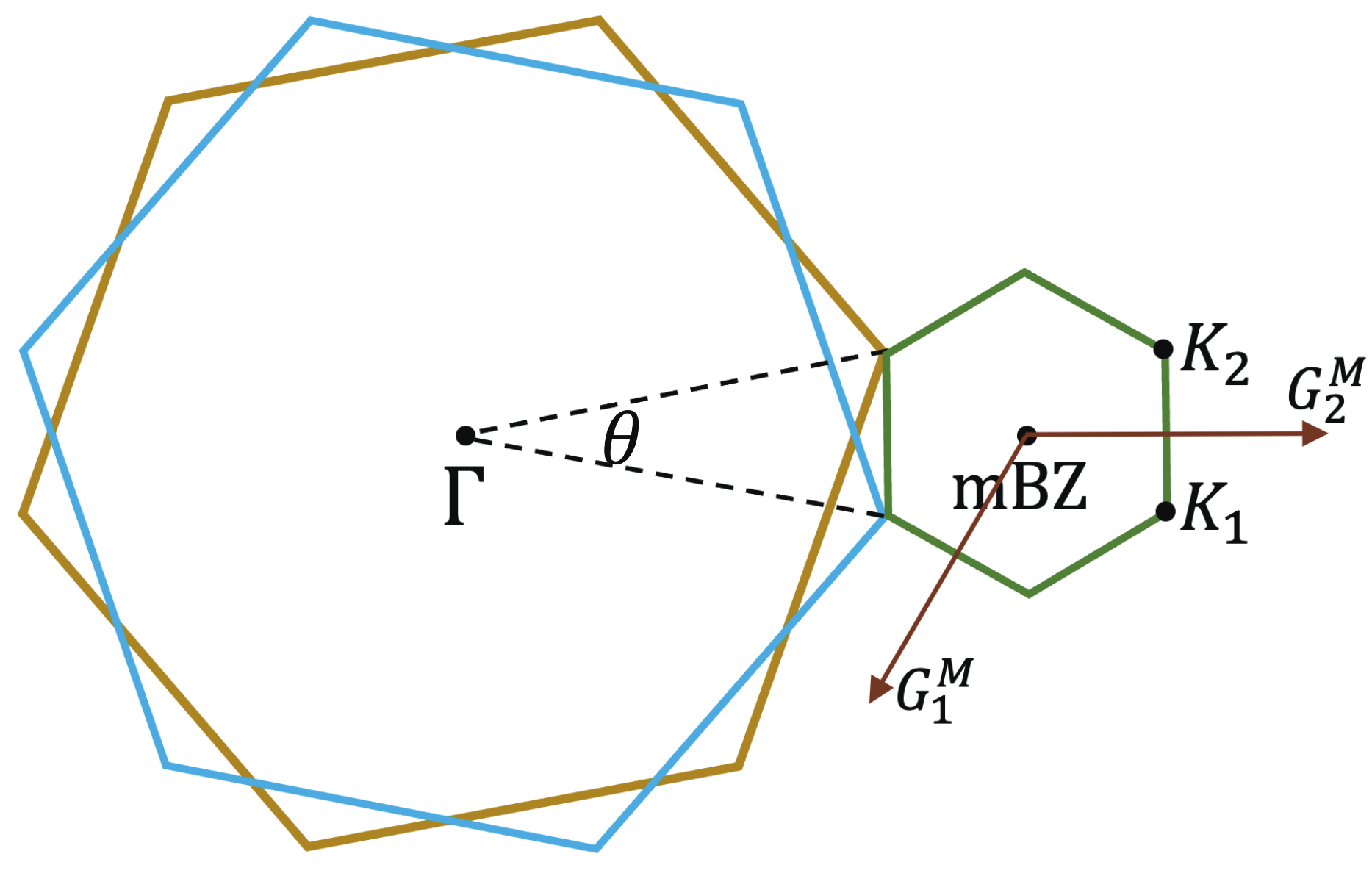}
	\caption{The moir\'{e} Brillouin zone (mBZ) of moir\'{e} graphene family. The original Brillouin zones from the two sectors are plotted as blue and brown colors, respectively. $G_1^M$ and $G_2^M$ are reciprocal lattice vectors.}
	\label{fig:figS4} 
\end{figure}
%%%%%%%%%%%%%%%%%%%%%%%

We adopt the parameters with $\hbar v_F=5252.44$ meV$\cdot\AA$, $w_0=79.7$ meV and $w_1=97.5$ meV \cite{2018PhRvX...8c1087K}.
\subsection*{Continuum model of twisted trilayer graphene}
For TTG, the moir\'{e} pattern is formed by the rotation of the top and bottom layer with angle $-\theta/2$ and the middle layer with angle $\theta/2$ relative to $x$ axis respectively. The moir\'{e} Brillouin zone is the same as Supplementary Fig.~\ref{fig:figS2}. For TTG, the Hamiltonian reads
\begin{equation}
	H=H_t+H_m+H_b+H_{int}^{tm}+H_{int}^{bm}.
\end{equation}
Here $H_{t,m,b}$ denote the Hamiltonian of the top, middle, and bottom layers, respectively. $H_{int}^{tm}$ is the interlayer coupling of the top and middle layers and $H_{int}^{bm}$ is the interlayer coupling of the bottom and middle layers. $H_{t,m,b}$ reads:
\begin{align}
	H_{t}&=\sum_{\xi,\bm{k}}a^\dagger_{t,\xi,\bm{k}}\hbar v_F \hat{R}_{\theta/2}(\bm{k}-\bm{K_1})\cdot(\xi \sigma_x,\sigma_y)a_{t,\xi,\bm{k}} ~,\\
	H_{m}&=\sum_{\xi,\bm{k}}a^\dagger_{m,\xi,\bm{k}}\hbar v_F \hat{R}_{-\theta/2}(\bm{k}-\bm{K_2})\cdot(\xi \sigma_x,\sigma_y)a_{m,\xi,\bm{k}}~, \\
	H_{b}&=\sum_{\xi,\bm{k}}a^\dagger_{b,\xi,\bm{k}}\hbar v_F \hat{R}_{\theta/2}(\bm{k}-\bm{K_1})\cdot(\xi \sigma_x,\sigma_y)a_{b,\xi,\bm{k}},
\end{align}
and
\begin{equation}
	H_{int}^{tm}=\sum_{\xi}a^\dagger_{t,\xi,\bm{k}}\left[\left(
	\begin{array}{cc}
		w_0 & w_1 \\
		w_1 & w_0 \\
	\end{array}
	\right)\delta_{\bm{k},\bm{k}'}+\left(
	\begin{array}{cc}
		w_0 & w_1e^{-i\frac{2\pi}{3}} \\
		w_1e^{i\frac{2\pi}{3}} & w_0 \\
	\end{array}
	\right)\delta_{\bm{k+\bm{G}_1^M},\bm{k}'}+ \left(
	\begin{array}{cc}
		w_0 & w_1e^{i\frac{2\pi}{3}} \\
		w_1e^{-i\frac{2\pi}{3}} & w_0 \\
	\end{array}
	\right)\delta_{\bm{k}+\bm{G}_1^M+\bm{G}_2^M,\bm{k}'}\right]a_{m,\xi,\bm{k}'}+h.c.
\end{equation}
\begin{equation}
	H_{int}^{bm}=\sum_{\xi}a^\dagger_{b,\xi,\bm{k}}\left[\left(
	\begin{array}{cc}
		w_0 & w_1 \\
		w_1 & w_0 \\
	\end{array}
	\right)\delta_{\bm{k},\bm{k}'}+\left(
	\begin{array}{cc}
		w_0 & w_1e^{-i\frac{2\pi}{3}} \\
		w_1e^{i\frac{2\pi}{3}} & w_0 \\
	\end{array}
	\right)\delta_{\bm{k+\bm{G}_1^M},\bm{k}'}+ \left(
	\begin{array}{cc}
		w_0 & w_1e^{i\frac{2\pi}{3}} \\
		w_1e^{-i\frac{2\pi}{3}} & w_0 \\
	\end{array}
	\right)\delta_{\bm{k}+\bm{G}_1^M+\bm{G}_2^M,\bm{k}'}\right]a_{m,\xi,\bm{k}'}+h.c.
\end{equation}
The parameters are the same as TBG.

\subsection*{Continuum model of twisted double bilayer graphene}
The TDBG is consisted of two Bernal stacked bilayer graphenes, twisted with respect to one another. We denote the Hamiltonian of up (down) Bernal stacked bilayer graphene $H_u$ and $H_d$, which read
\begin{eqnarray}
	H_u=\sum_{\xi,\bm{k}} a^\dagger_{u,\xi,,\bm{k}}\left(
	\begin{array}{cc}
		H_1 & U  \\
		U^\dagger & H_2
	\end{array}\right)a_{u,\xi,\bm{k}} ~, \\
	H_d=\sum_{\xi,\bm{k}} a^\dagger_{d,\xi,\bm{k}}\left(
	\begin{array}{cc}
		H_3 & U  \\
		U^\dagger & H_4
	\end{array}\right)a_{d,\xi,\bm{k}}  ~.
\end{eqnarray}
The Hamiltonian of each layer reads
\begin{align}
	H_{1} &=\hbar v_0 \hat{R}_{\theta/2}(\bm{k}-\bm{K_1})\cdot(\xi\sigma_x,\sigma_y)+\frac{V}{2}\sigma_0 ~,\\
	H_{2} &=\hbar v_0 \hat{R}_{\theta/2}(\bm{k}-\bm{K_1})\cdot(\xi\sigma_x,\sigma_y)+\frac{V}{6}\sigma_0  ~,\\
	H_{3} &=\hbar v_0 \hat{R}_{-\theta/2}(\bm{k}-\bm{K_2})\cdot(\xi\sigma_x,\sigma_y)-\frac{V}{6}\sigma_0~, \\
	H_{4} &=\hbar v_0 \hat{R}_{-\theta/2}(\bm{k}-\bm{K_2})\cdot(\xi\sigma_x,\sigma_y)-\frac{V}{2}\sigma_0~.
\end{align}
and $U$ describe the inter-layer tunneling between the Bernal-stacked bilayer graphene:
\begin{equation}
	U=\left(
	\begin{array}{cc}
		-v_4(\xi k_x-ik_y) & v_3(\xi k_x+ik_y) \\
		\gamma_1 & -v_4(\xi k_x-ik_y) \\
	\end{array}
	\right)~.
\end{equation}
The interlayer coupling is written as
\begin{equation}
	H_{int}=\sum_{\xi}\int_{\bm{r}}\psi^\dagger_{u,\xi}(\bm{r})  
	\left(
	\begin{array}{cc}
		0 & 0 \\
		T(\bm{r}) & 0 \\
	\end{array}
	\right) \psi_{d,\xi}(\bm{r})+h.c.~.
\end{equation}
The full Hamiltonian is $H=H_u+H_d+H_{int}$.

We adopt the parameters with $(\gamma_0, \gamma_1,\gamma_2,\gamma_3,\gamma_4)=(2610,361,283,140)$ meV, and $v_i=\frac{\sqrt{3}}{2\hbar}a_0 \gamma_i$ \cite{2019NatCo..10.5333L}. The interlayer coupling is $(w_0,w_1)=(88,100)$ meV. The effect of an out-of-plane electric field is introduced by $V=40$ meV. Once the Bloch wavefunction is obtained, the quantum metric $\mathcal{G}_{ab}$ can be evaluated by Eq.~\ref{quantum_metric}. Because of the time-reversal symmetry, we have $\mathcal{G}^{\xi=+}_{ab}(\bm{k})=\mathcal{G}^{\xi=-}_{ab}(-\bm{k})$. For TBG and TTG, since the quantum metric is divergent at the Dirac points, in our calculations we add a small sublattice potential $\Delta_p \sigma_z$ with $\Delta_p=0.01$ to the bottom layer of graphene. This is reasonable because the superconductivity occurs near half filling of the flat band, not near the Dirac points. 

%%%%%%%%%%%%%%%%%%%%%%
\begin{figure}[t]
	\centering 
	\includegraphics[width=1\linewidth]{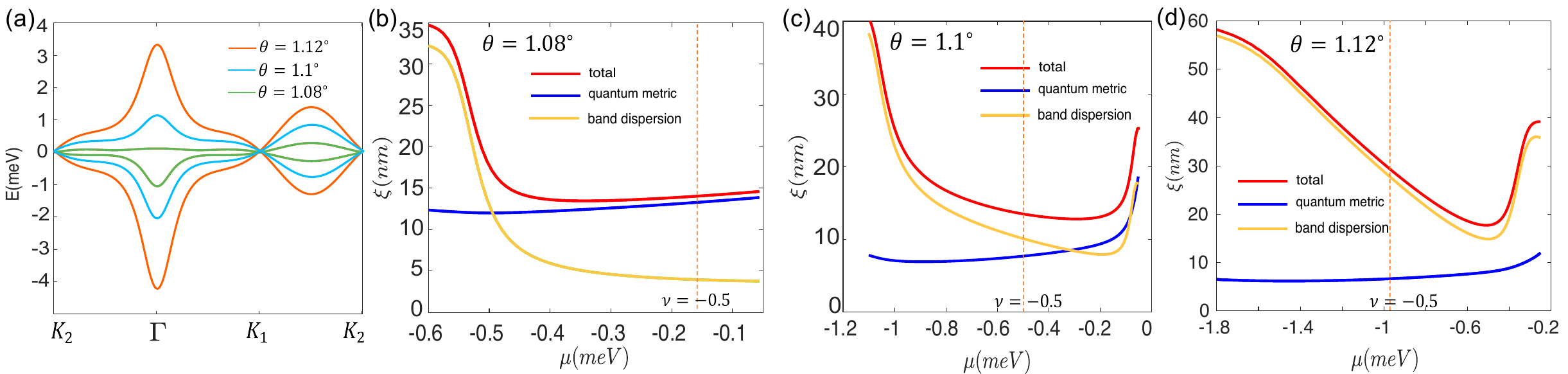}
	\caption{(a) The band structure of TBG for $\theta=1.08^\circ,1.1^\circ,1.12^\circ$. At $\theta=1.08^\circ$, the bandwith $W< 1$ meV. (b)-(d) The calculation of the superconducting coherence length $\xi$ as a function of $\mu$ for $\theta=1.08^\circ,1.1^\circ,1.12^\circ$. In (b), $\xi$ from band dispersion $\leq$ 2nm and the quantum metric contribution dominates. When the twist angle is larger, the Fermi velocity changes dramatically, leading to the enhanced contribution from band dispersion in (c) and (d). In (c) and (d), $\xi$ get divergent when $\mu$ is away from half-filling because of the weak pairing gap. The half-filling ($\nu=-0.5$) is labeled by the orange dashed line. The attractive interaction $U_0=0.6$ meV.}
	\label{fig:figS5} 
\end{figure}
%%%%%%%%%%%%%%%%%%%%%%%
%%%%%%%%%%%%%%%%%%%%%%
\begin{figure}[ht]
	\centering 
	\includegraphics[width=1\linewidth]{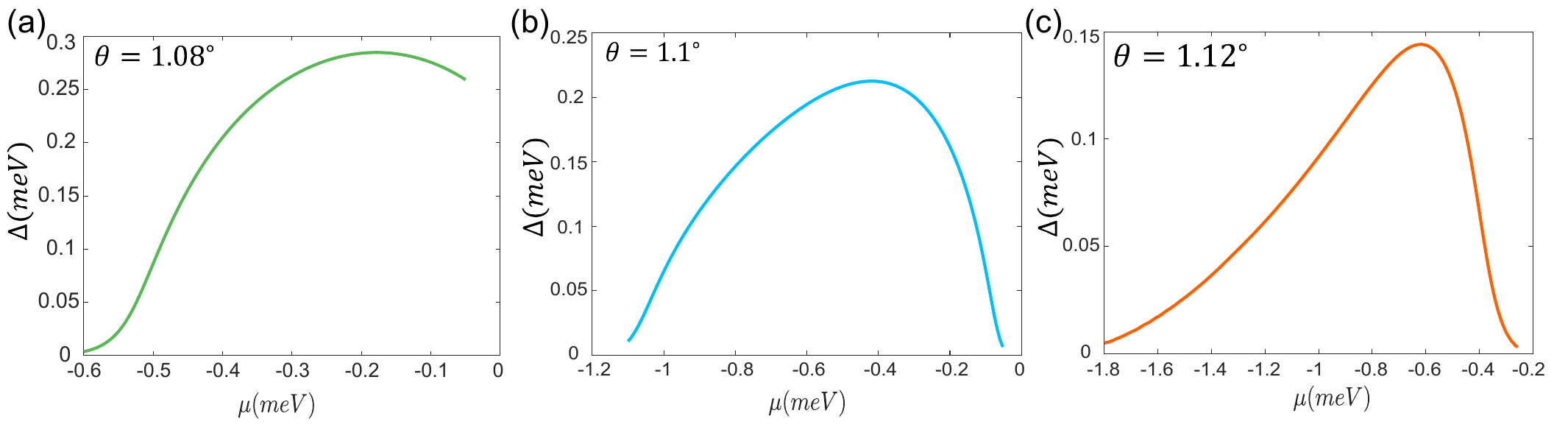}
	\caption{The self-consistent calculation of the pairing gap $\Delta$ as a function of $\mu$ for different $\theta$. The attractive interaction $U_0=0.6$ meV.}
	\label{fig:figS6} 
\end{figure}
%%%%%%%%%%%%%%%%%%%%%%%
\subsection*{Coherence length of TBG: Mean-field study}
In this section, we provide the calculation of the superconducting coherence length of TBG in detail. Considering the superconductivity occurs on the highest valence band, the mean-field BdG Hamiltonian of TBG can be written as
\begin{equation}
	H_{\mathrm{BdG}}=\left(
	\begin{array}{cccc}
		E_{+\uparrow,\bm{k}}-\mu &0 &0 & \Delta\\
		0 &E_{-\downarrow,\bm{k}}-\mu &-\Delta &0\\
		0 &-\Delta & -E_{+\uparrow,-\bm{k}}+\mu &0 \\
		\Delta &0 & 0 & -E_{-\downarrow,-\bm{k}}+\mu
	\end{array}
	\right).
\end{equation}
Here $+(-)$ and $\uparrow (\downarrow)$ are the valley and spin index, respectively. $E$ is the energy spectrum of the highest valence band of TBG. Due to the time-reversal symmetry, we have $E_{+\uparrow,\bm{k}}=E_{-\downarrow,-\bm{k}}$ and $E_{-\downarrow,\bm{k}}=E_{+\uparrow,-\bm{k}}$. The zero-temperature pairing gap $\Delta$ can be solved self-consistently by $1=U_0\sum_k 1/[2\sqrt{(E_{+\uparrow,\bm{k}}-\mu)^2+\Delta^2}]$, with $U_0$ being the interaction strength.

We summarize the calculation of the zero temperature superconducting coherence length $\xi$ in Supplementary Fig.~\ref{fig:figS5}. We calculate the $\xi$ by using Eq.(22) and (23). In our calculation, we use $U_0=0.6$ meV, yielding a critical temperature $T_c\approx 1.7$K at half filling for $\theta=1.08^\circ$. We use the notation $\xi_{bd},\xi_{qm},\xi_{t}$ to denote the coherence length from band dispersion, quantum metric, and total coherence length, respectively.

Near the magic angle ($\theta=1.08^{\circ}$), the bandwidth of TBG changes dramatically as shown in Supplementary Fig.~\ref{fig:figS5}{\bf a}. At $\theta=1.08^\circ$, the density of states near the Dirac point is almost unchanged and $\Delta\approx 0.25$ meV. Near half filling ($\nu=-0.5$), $\xi_{bd}\leq 2$nm due to the narrow band case and $\xi_{qm}$ is about $15$ nm as shown in Supplementary Fig.~\ref{fig:figS5}{\bf b}. When $\mu$ is away from half filling ($\mu<0.5$meV), $\xi$ from band dispersion is largely enhanced because the pairing is weak in this regime (see Supplementary Fig.~\ref{fig:figS6}{\bf a}). 

When the twist angle is slightly away from $1.08^\circ$, the band structure and the Fermi velocity $v_F$ change dramatically. As shown in Supplementary Fig.~\ref{fig:figS5}(b)-(d), $\xi_{bd}$ is gradually enhanced. For example, at $\theta=1.1^\circ$, near half filling ($\nu=-0.5$), $\xi_{bd}\approx 10$ nm, which is comparable to $\xi_{qm}$.  At $\theta=1.12^\circ$, $\xi_{bd}$ further get enhanced.

%\bibliographystyle{apsrev4-2}
%\bibliography{Reference_supple}

\begin{thebibliography}{70}%
		\makeatletter
		\providecommand \@ifxundefined [1]{%
			\@ifx{#1\undefined}
		}%
		\providecommand \@ifnum [1]{%
			\ifnum #1\expandafter \@firstoftwo
			\else \expandafter \@secondoftwo
			\fi
		}%
		\providecommand \@ifx [1]{%
			\ifx #1\expandafter \@firstoftwo
			\else \expandafter \@secondoftwo
			\fi
		}%
		\providecommand \natexlab [1]{#1}%
		\providecommand \enquote  [1]{``#1''}%
		\providecommand \bibnamefont  [1]{#1}%
		\providecommand \bibfnamefont [1]{#1}%
		\providecommand \citenamefont [1]{#1}%
		\providecommand \href@noop [0]{\@secondoftwo}%
		\providecommand \href [0]{\begingroup \@sanitize@url \@href}%
		\providecommand \@href[1]{\@@startlink{#1}\@@href}%
		\providecommand \@@href[1]{\endgroup#1\@@endlink}%
		\providecommand \@sanitize@url [0]{\catcode `\\12\catcode `\$12\catcode
			`\&12\catcode `\#12\catcode `\^12\catcode `\_12\catcode `\%12\relax}%
		\providecommand \@@startlink[1]{}%
		\providecommand \@@endlink[0]{}%
		\providecommand \url  [0]{\begingroup\@sanitize@url \@url }%
		\providecommand \@url [1]{\endgroup\@href {#1}{\urlprefix }}%
		\providecommand \urlprefix  [0]{URL }%
		\providecommand \Eprint [0]{\href }%
		\providecommand \doibase [0]{http://dx.doi.org/}%
		\providecommand \selectlanguage [0]{\@gobble}%
		\providecommand \bibinfo  [0]{\@secondoftwo}%
		\providecommand \bibfield  [0]{\@secondoftwo}%
		\providecommand \translation [1]{[#1]}%
		\providecommand \BibitemOpen [0]{}%
		\providecommand \bibitemStop [0]{}%
		\providecommand \bibitemNoStop [0]{.\EOS\space}%
		\providecommand \EOS [0]{\spacefactor3000\relax}%
		\providecommand \BibitemShut  [1]{\csname bibitem#1\endcsname}%
		\let\auto@bib@innerbib\@empty
		%</preamble>
		\bibitem [{\citenamefont {Bardeen}\ \emph {et~al.}(1957)\citenamefont
			{Bardeen}, \citenamefont {Cooper},\ and\ \citenamefont
			{Schrieffer}}]{bardeen1957microscopic}%
		\BibitemOpen
		\bibfield  {author} {\bibinfo {author} {\bibfnamefont {John}\ \bibnamefont
				{Bardeen}}, \bibinfo {author} {\bibfnamefont {Leon~N}\ \bibnamefont
				{Cooper}}, \ and\ \bibinfo {author} {\bibfnamefont {J~Robert}\ \bibnamefont
				{Schrieffer}},\ }\bibfield  {title} {\enquote {\bibinfo {title} {Microscopic
					theory of superconductivity},}\ }\href@noop {} {\bibfield  {journal}
			{\bibinfo  {journal} {Physical Review}\ }\textbf {\bibinfo {volume} {106}},\
			\bibinfo {pages} {162} (\bibinfo {year} {1957})}\BibitemShut {NoStop}%
		\bibitem [{\citenamefont {Carbotte}(1990)}]{carbotte1990properties}%
		\BibitemOpen
		\bibfield  {author} {\bibinfo {author} {\bibfnamefont {JP}~\bibnamefont
				{Carbotte}},\ }\bibfield  {title} {\enquote {\bibinfo {title} {Properties of
					boson-exchange superconductors},}\ }\href@noop {} {\bibfield  {journal}
			{\bibinfo  {journal} {Reviews of Modern Physics}\ }\textbf {\bibinfo {volume}
				{62}},\ \bibinfo {pages} {1027} (\bibinfo {year} {1990})}\BibitemShut
		{NoStop}%
		\bibitem [{\citenamefont {Sigrist}\ and\ \citenamefont
			{Ueda}(1991)}]{sigrist1991phenomenological}%
		\BibitemOpen
		\bibfield  {author} {\bibinfo {author} {\bibfnamefont {Manfred}\ \bibnamefont
				{Sigrist}}\ and\ \bibinfo {author} {\bibfnamefont {Kazuo}\ \bibnamefont
				{Ueda}},\ }\bibfield  {title} {\enquote {\bibinfo {title} {Phenomenological
					theory of unconventional superconductivity},}\ }\href@noop {} {\bibfield
			{journal} {\bibinfo  {journal} {Reviews of Modern physics}\ }\textbf
			{\bibinfo {volume} {63}},\ \bibinfo {pages} {239} (\bibinfo {year}
			{1991})}\BibitemShut {NoStop}%
		\bibitem [{\citenamefont {Blatter}\ \emph {et~al.}(1994)\citenamefont
			{Blatter}, \citenamefont {Feigel'man}, \citenamefont {Geshkenbein},
			\citenamefont {Larkin},\ and\ \citenamefont {Vinokur}}]{blatter1994vortices}%
		\BibitemOpen
		\bibfield  {author} {\bibinfo {author} {\bibfnamefont {Gianni}\ \bibnamefont
				{Blatter}}, \bibinfo {author} {\bibfnamefont {Mikhail~V}\ \bibnamefont
				{Feigel'man}}, \bibinfo {author} {\bibfnamefont {Vadim~B}\ \bibnamefont
				{Geshkenbein}}, \bibinfo {author} {\bibfnamefont {Anatoly~I}\ \bibnamefont
				{Larkin}}, \ and\ \bibinfo {author} {\bibfnamefont {Valerii~M}\ \bibnamefont
				{Vinokur}},\ }\bibfield  {title} {\enquote {\bibinfo {title} {Vortices in
					high-temperature superconductors},}\ }\href@noop {} {\bibfield  {journal}
			{\bibinfo  {journal} {Reviews of modern physics}\ }\textbf {\bibinfo {volume}
				{66}},\ \bibinfo {pages} {1125} (\bibinfo {year} {1994})}\BibitemShut
		{NoStop}%
		\bibitem [{\citenamefont {Stewart}(1984)}]{stewart1984heavy}%
		\BibitemOpen
		\bibfield  {author} {\bibinfo {author} {\bibfnamefont {See~GR}\ \bibnamefont
				{Stewart}},\ }\bibfield  {title} {\enquote {\bibinfo {title} {Heavy-fermion
					systems},}\ }\href@noop {} {\bibfield  {journal} {\bibinfo  {journal}
				{Reviews of Modern Physics}\ }\textbf {\bibinfo {volume} {56}},\ \bibinfo
			{pages} {755} (\bibinfo {year} {1984})}\BibitemShut {NoStop}%
		\bibitem [{\citenamefont {Keimer}\ \emph {et~al.}(2015)\citenamefont {Keimer},
			\citenamefont {Kivelson}, \citenamefont {Norman}, \citenamefont {Uchida},\
			and\ \citenamefont {Zaanen}}]{keimer2015quantum}%
		\BibitemOpen
		\bibfield  {author} {\bibinfo {author} {\bibfnamefont {Bernhard}\
				\bibnamefont {Keimer}}, \bibinfo {author} {\bibfnamefont {Steven~A}\
				\bibnamefont {Kivelson}}, \bibinfo {author} {\bibfnamefont {Michael~R}\
				\bibnamefont {Norman}}, \bibinfo {author} {\bibfnamefont {Shinichi}\
				\bibnamefont {Uchida}}, \ and\ \bibinfo {author} {\bibfnamefont
				{J}~\bibnamefont {Zaanen}},\ }\bibfield  {title} {\enquote {\bibinfo {title}
				{From quantum matter to high-temperature superconductivity in copper
					oxides},}\ }\href@noop {} {\bibfield  {journal} {\bibinfo  {journal}
				{Nature}\ }\textbf {\bibinfo {volume} {518}},\ \bibinfo {pages} {179--186}
			(\bibinfo {year} {2015})}\BibitemShut {NoStop}%
		\bibitem [{\citenamefont {Cao}\ \emph {et~al.}(2018)\citenamefont {Cao},
			\citenamefont {Fatemi}, \citenamefont {Fang}, \citenamefont {Watanabe},
			\citenamefont {Taniguchi}, \citenamefont {Kaxiras},\ and\ \citenamefont
			{Jarillo-Herrero}}]{cao2018unconventional}%
		\BibitemOpen
		\bibfield  {author} {\bibinfo {author} {\bibfnamefont {Yuan}\ \bibnamefont
				{Cao}}, \bibinfo {author} {\bibfnamefont {Valla}\ \bibnamefont {Fatemi}},
			\bibinfo {author} {\bibfnamefont {Shiang}\ \bibnamefont {Fang}}, \bibinfo
			{author} {\bibfnamefont {Kenji}\ \bibnamefont {Watanabe}}, \bibinfo {author}
			{\bibfnamefont {Takashi}\ \bibnamefont {Taniguchi}}, \bibinfo {author}
			{\bibfnamefont {Efthimios}\ \bibnamefont {Kaxiras}}, \ and\ \bibinfo {author}
			{\bibfnamefont {Pablo}\ \bibnamefont {Jarillo-Herrero}},\ }\bibfield  {title}
		{\enquote {\bibinfo {title} {Unconventional superconductivity in magic-angle
					graphene superlattices},}\ }\href@noop {} {\bibfield  {journal} {\bibinfo
				{journal} {Nature}\ }\textbf {\bibinfo {volume} {556}},\ \bibinfo {pages}
			{43--50} (\bibinfo {year} {2018})}\BibitemShut {NoStop}%
		\bibitem [{\citenamefont {Yankowitz}\ \emph {et~al.}(2019)\citenamefont
			{Yankowitz}, \citenamefont {Chen}, \citenamefont {Polshyn}, \citenamefont
			{Zhang}, \citenamefont {Watanabe}, \citenamefont {Taniguchi}, \citenamefont
			{Graf}, \citenamefont {Young},\ and\ \citenamefont
			{Dean}}]{yankowitz2019tuning}%
		\BibitemOpen
		\bibfield  {author} {\bibinfo {author} {\bibfnamefont {Matthew}\ \bibnamefont
				{Yankowitz}}, \bibinfo {author} {\bibfnamefont {Shaowen}\ \bibnamefont
				{Chen}}, \bibinfo {author} {\bibfnamefont {Hryhoriy}\ \bibnamefont
				{Polshyn}}, \bibinfo {author} {\bibfnamefont {Yuxuan}\ \bibnamefont {Zhang}},
			\bibinfo {author} {\bibfnamefont {K}~\bibnamefont {Watanabe}}, \bibinfo
			{author} {\bibfnamefont {T}~\bibnamefont {Taniguchi}}, \bibinfo {author}
			{\bibfnamefont {David}\ \bibnamefont {Graf}}, \bibinfo {author}
			{\bibfnamefont {Andrea~F}\ \bibnamefont {Young}}, \ and\ \bibinfo {author}
			{\bibfnamefont {Cory~R}\ \bibnamefont {Dean}},\ }\bibfield  {title} {\enquote
			{\bibinfo {title} {Tuning superconductivity in twisted bilayer graphene},}\
		}\href@noop {} {\bibfield  {journal} {\bibinfo  {journal} {Science}\ }\textbf
			{\bibinfo {volume} {363}},\ \bibinfo {pages} {1059--1064} (\bibinfo {year}
			{2019})}\BibitemShut {NoStop}%
		\bibitem [{\citenamefont {Arora}\ \emph {et~al.}(2020)\citenamefont {Arora},
			\citenamefont {Polski}, \citenamefont {Zhang}, \citenamefont {Thomson},
			\citenamefont {Choi}, \citenamefont {Kim}, \citenamefont {Lin}, \citenamefont
			{Wilson}, \citenamefont {Xu}, \citenamefont {Chu} \emph
			{et~al.}}]{arora2020superconductivity}%
		\BibitemOpen
		\bibfield  {author} {\bibinfo {author} {\bibfnamefont {Harpreet~Singh}\
				\bibnamefont {Arora}}, \bibinfo {author} {\bibfnamefont {Robert}\
				\bibnamefont {Polski}}, \bibinfo {author} {\bibfnamefont {Yiran}\
				\bibnamefont {Zhang}}, \bibinfo {author} {\bibfnamefont {Alex}\ \bibnamefont
				{Thomson}}, \bibinfo {author} {\bibfnamefont {Youngjoon}\ \bibnamefont
				{Choi}}, \bibinfo {author} {\bibfnamefont {Hyunjin}\ \bibnamefont {Kim}},
			\bibinfo {author} {\bibfnamefont {Zhong}\ \bibnamefont {Lin}}, \bibinfo
			{author} {\bibfnamefont {Ilham~Zaky}\ \bibnamefont {Wilson}}, \bibinfo
			{author} {\bibfnamefont {Xiaodong}\ \bibnamefont {Xu}}, \bibinfo {author}
			{\bibfnamefont {Jiun-Haw}\ \bibnamefont {Chu}},  \emph {et~al.},\ }\bibfield
		{title} {\enquote {\bibinfo {title} {Superconductivity in metallic twisted
					bilayer graphene stabilized by wse2},}\ }\href@noop {} {\bibfield  {journal}
			{\bibinfo  {journal} {Nature}\ }\textbf {\bibinfo {volume} {583}},\ \bibinfo
			{pages} {379--384} (\bibinfo {year} {2020})}\BibitemShut {NoStop}%
		\bibitem [{\citenamefont {Oh}\ \emph {et~al.}(2021)\citenamefont {Oh},
			\citenamefont {Nuckolls}, \citenamefont {Wong}, \citenamefont {Lee},
			\citenamefont {Liu}, \citenamefont {Watanabe}, \citenamefont {Taniguchi},\
			and\ \citenamefont {Yazdani}}]{oh2021evidence}%
		\BibitemOpen
		\bibfield  {author} {\bibinfo {author} {\bibfnamefont {Myungchul}\
				\bibnamefont {Oh}}, \bibinfo {author} {\bibfnamefont {Kevin~P}\ \bibnamefont
				{Nuckolls}}, \bibinfo {author} {\bibfnamefont {Dillon}\ \bibnamefont {Wong}},
			\bibinfo {author} {\bibfnamefont {Ryan~L}\ \bibnamefont {Lee}}, \bibinfo
			{author} {\bibfnamefont {Xiaomeng}\ \bibnamefont {Liu}}, \bibinfo {author}
			{\bibfnamefont {Kenji}\ \bibnamefont {Watanabe}}, \bibinfo {author}
			{\bibfnamefont {Takashi}\ \bibnamefont {Taniguchi}}, \ and\ \bibinfo {author}
			{\bibfnamefont {Ali}\ \bibnamefont {Yazdani}},\ }\bibfield  {title} {\enquote
			{\bibinfo {title} {Evidence for unconventional superconductivity in twisted
					bilayer graphene},}\ }\href@noop {} {\bibfield  {journal} {\bibinfo
				{journal} {Nature}\ }\textbf {\bibinfo {volume} {600}},\ \bibinfo {pages}
			{240--245} (\bibinfo {year} {2021})}\BibitemShut {NoStop}%
		\bibitem [{\citenamefont {Tian}\ \emph {et~al.}(2023)\citenamefont {Tian},
			\citenamefont {Gao}, \citenamefont {Zhang}, \citenamefont {Che},
			\citenamefont {Xu}, \citenamefont {Cheung}, \citenamefont {Watanabe},
			\citenamefont {Taniguchi}, \citenamefont {Randeria}, \citenamefont {Zhang}
			\emph {et~al.}}]{tian2023evidence}%
		\BibitemOpen
		\bibfield  {author} {\bibinfo {author} {\bibfnamefont {Haidong}\ \bibnamefont
				{Tian}}, \bibinfo {author} {\bibfnamefont {Xueshi}\ \bibnamefont {Gao}},
			\bibinfo {author} {\bibfnamefont {Yuxin}\ \bibnamefont {Zhang}}, \bibinfo
			{author} {\bibfnamefont {Shi}\ \bibnamefont {Che}}, \bibinfo {author}
			{\bibfnamefont {Tianyi}\ \bibnamefont {Xu}}, \bibinfo {author} {\bibfnamefont
				{Patrick}\ \bibnamefont {Cheung}}, \bibinfo {author} {\bibfnamefont {Kenji}\
				\bibnamefont {Watanabe}}, \bibinfo {author} {\bibfnamefont {Takashi}\
				\bibnamefont {Taniguchi}}, \bibinfo {author} {\bibfnamefont {Mohit}\
				\bibnamefont {Randeria}}, \bibinfo {author} {\bibfnamefont {Fan}\
				\bibnamefont {Zhang}},  \emph {et~al.},\ }\bibfield  {title} {\enquote
			{\bibinfo {title} {Evidence for dirac flat band superconductivity enabled by
					quantum geometry},}\ }\href@noop {} {\bibfield  {journal} {\bibinfo
				{journal} {Nature}\ }\textbf {\bibinfo {volume} {614}},\ \bibinfo {pages}
			{440--444} (\bibinfo {year} {2023})}\BibitemShut {NoStop}%
		\bibitem [{\citenamefont {Liu}\ \emph {et~al.}(2020)\citenamefont {Liu},
			\citenamefont {Hao}, \citenamefont {Khalaf}, \citenamefont {Lee},
			\citenamefont {Ronen}, \citenamefont {Yoo}, \citenamefont {Haei~Najafabadi},
			\citenamefont {Watanabe}, \citenamefont {Taniguchi}, \citenamefont
			{Vishwanath} \emph {et~al.}}]{liu2020tunable}%
		\BibitemOpen
		\bibfield  {author} {\bibinfo {author} {\bibfnamefont {Xiaomeng}\
				\bibnamefont {Liu}}, \bibinfo {author} {\bibfnamefont {Zeyu}\ \bibnamefont
				{Hao}}, \bibinfo {author} {\bibfnamefont {Eslam}\ \bibnamefont {Khalaf}},
			\bibinfo {author} {\bibfnamefont {Jong~Yeon}\ \bibnamefont {Lee}}, \bibinfo
			{author} {\bibfnamefont {Yuval}\ \bibnamefont {Ronen}}, \bibinfo {author}
			{\bibfnamefont {Hyobin}\ \bibnamefont {Yoo}}, \bibinfo {author}
			{\bibfnamefont {Danial}\ \bibnamefont {Haei~Najafabadi}}, \bibinfo {author}
			{\bibfnamefont {Kenji}\ \bibnamefont {Watanabe}}, \bibinfo {author}
			{\bibfnamefont {Takashi}\ \bibnamefont {Taniguchi}}, \bibinfo {author}
			{\bibfnamefont {Ashvin}\ \bibnamefont {Vishwanath}},  \emph {et~al.},\
		}\bibfield  {title} {\enquote {\bibinfo {title} {Tunable spin-polarized
					correlated states in twisted double bilayer graphene},}\ }\href@noop {}
		{\bibfield  {journal} {\bibinfo  {journal} {Nature}\ }\textbf {\bibinfo
				{volume} {583}},\ \bibinfo {pages} {221--225} (\bibinfo {year}
			{2020})}\BibitemShut {NoStop}%
		\bibitem [{\citenamefont {Park}\ \emph {et~al.}(2021)\citenamefont {Park},
			\citenamefont {Cao}, \citenamefont {Watanabe}, \citenamefont {Taniguchi},\
			and\ \citenamefont {Jarillo-Herrero}}]{park2021tunable}%
		\BibitemOpen
		\bibfield  {author} {\bibinfo {author} {\bibfnamefont {Jeong~Min}\
				\bibnamefont {Park}}, \bibinfo {author} {\bibfnamefont {Yuan}\ \bibnamefont
				{Cao}}, \bibinfo {author} {\bibfnamefont {Kenji}\ \bibnamefont {Watanabe}},
			\bibinfo {author} {\bibfnamefont {Takashi}\ \bibnamefont {Taniguchi}}, \ and\
			\bibinfo {author} {\bibfnamefont {Pablo}\ \bibnamefont {Jarillo-Herrero}},\
		}\bibfield  {title} {\enquote {\bibinfo {title} {Tunable strongly coupled
					superconductivity in magic-angle twisted trilayer graphene},}\ }\href@noop {}
		{\bibfield  {journal} {\bibinfo  {journal} {Nature}\ }\textbf {\bibinfo
				{volume} {590}},\ \bibinfo {pages} {249--255} (\bibinfo {year}
			{2021})}\BibitemShut {NoStop}%
		\bibitem [{\citenamefont {Park}\ \emph {et~al.}(2022)\citenamefont {Park},
			\citenamefont {Cao}, \citenamefont {Xia}, \citenamefont {Sun}, \citenamefont
			{Watanabe}, \citenamefont {Taniguchi},\ and\ \citenamefont
			{Jarillo-Herrero}}]{park2022robust}%
		\BibitemOpen
		\bibfield  {author} {\bibinfo {author} {\bibfnamefont {Jeong~Min}\
				\bibnamefont {Park}}, \bibinfo {author} {\bibfnamefont {Yuan}\ \bibnamefont
				{Cao}}, \bibinfo {author} {\bibfnamefont {Li-Qiao}\ \bibnamefont {Xia}},
			\bibinfo {author} {\bibfnamefont {Shuwen}\ \bibnamefont {Sun}}, \bibinfo
			{author} {\bibfnamefont {Kenji}\ \bibnamefont {Watanabe}}, \bibinfo {author}
			{\bibfnamefont {Takashi}\ \bibnamefont {Taniguchi}}, \ and\ \bibinfo {author}
			{\bibfnamefont {Pablo}\ \bibnamefont {Jarillo-Herrero}},\ }\bibfield  {title}
		{\enquote {\bibinfo {title} {Robust superconductivity in magic-angle
					multilayer graphene family},}\ }\href@noop {} {\bibfield  {journal} {\bibinfo
				{journal} {Nature Materials}\ }\textbf {\bibinfo {volume} {21}},\ \bibinfo
			{pages} {877--883} (\bibinfo {year} {2022})}\BibitemShut {NoStop}%
		\bibitem [{\citenamefont {Hu}\ \emph {et~al.}(2019)\citenamefont {Hu},
			\citenamefont {Hyart}, \citenamefont {Pikulin},\ and\ \citenamefont
			{Rossi}}]{hu2019geometric}%
		\BibitemOpen
		\bibfield  {author} {\bibinfo {author} {\bibfnamefont {Xiang}\ \bibnamefont
				{Hu}}, \bibinfo {author} {\bibfnamefont {Timo}\ \bibnamefont {Hyart}},
			\bibinfo {author} {\bibfnamefont {Dmitry~I}\ \bibnamefont {Pikulin}}, \ and\
			\bibinfo {author} {\bibfnamefont {Enrico}\ \bibnamefont {Rossi}},\ }\bibfield
		{title} {\enquote {\bibinfo {title} {Geometric and conventional contribution
					to the superfluid weight in twisted bilayer graphene},}\ }\href@noop {}
		{\bibfield  {journal} {\bibinfo  {journal} {Physical Review Letters}\
			}\textbf {\bibinfo {volume} {123}},\ \bibinfo {pages} {237002} (\bibinfo
			{year} {2019})}\BibitemShut {NoStop}%
		\bibitem [{\citenamefont {Julku}\ \emph {et~al.}(2020)\citenamefont {Julku},
			\citenamefont {Peltonen}, \citenamefont {Liang}, \citenamefont
			{Heikkil{\"a}},\ and\ \citenamefont {T{\"o}rm{\"a}}}]{julku2020superfluid}%
		\BibitemOpen
		\bibfield  {author} {\bibinfo {author} {\bibfnamefont {Aleksi}\ \bibnamefont
				{Julku}}, \bibinfo {author} {\bibfnamefont {Teemu~J}\ \bibnamefont
				{Peltonen}}, \bibinfo {author} {\bibfnamefont {Long}\ \bibnamefont {Liang}},
			\bibinfo {author} {\bibfnamefont {Tero~T}\ \bibnamefont {Heikkil{\"a}}}, \
			and\ \bibinfo {author} {\bibfnamefont {P{\"a}ivi}\ \bibnamefont
				{T{\"o}rm{\"a}}},\ }\bibfield  {title} {\enquote {\bibinfo {title}
				{Superfluid weight and berezinskii-kosterlitz-thouless transition temperature
					of twisted bilayer graphene},}\ }\href@noop {} {\bibfield  {journal}
			{\bibinfo  {journal} {Physical Review B}\ }\textbf {\bibinfo {volume}
				{101}},\ \bibinfo {pages} {060505} (\bibinfo {year} {2020})}\BibitemShut
		{NoStop}%
		\bibitem [{\citenamefont {Peotta}\ and\ \citenamefont
			{T{\"o}rm{\"a}}(2015)}]{peotta2015superfluidity}%
		\BibitemOpen
		\bibfield  {author} {\bibinfo {author} {\bibfnamefont {Sebastiano}\
				\bibnamefont {Peotta}}\ and\ \bibinfo {author} {\bibfnamefont {P{\"a}ivi}\
				\bibnamefont {T{\"o}rm{\"a}}},\ }\bibfield  {title} {\enquote {\bibinfo
				{title} {Superfluidity in topologically nontrivial flat bands},}\ }\href@noop
		{} {\bibfield  {journal} {\bibinfo  {journal} {Nature communications}\
			}\textbf {\bibinfo {volume} {6}},\ \bibinfo {pages} {8944} (\bibinfo {year}
			{2015})}\BibitemShut {NoStop}%
		\bibitem [{\citenamefont {T{\"o}rm{\"a}}\ \emph {et~al.}(2022)\citenamefont
			{T{\"o}rm{\"a}}, \citenamefont {Peotta},\ and\ \citenamefont
			{Bernevig}}]{torma2022superconductivity}%
		\BibitemOpen
		\bibfield  {author} {\bibinfo {author} {\bibfnamefont {P{\"a}ivi}\
				\bibnamefont {T{\"o}rm{\"a}}}, \bibinfo {author} {\bibfnamefont {Sebastiano}\
				\bibnamefont {Peotta}}, \ and\ \bibinfo {author} {\bibfnamefont {Bogdan~A}\
				\bibnamefont {Bernevig}},\ }\bibfield  {title} {\enquote {\bibinfo {title}
				{Superconductivity, superfluidity and quantum geometry in twisted multilayer
					systems},}\ }\href@noop {} {\bibfield  {journal} {\bibinfo  {journal} {Nature
					Reviews Physics}\ }\textbf {\bibinfo {volume} {4}},\ \bibinfo {pages}
			{528--542} (\bibinfo {year} {2022})}\BibitemShut {NoStop}%
		\bibitem [{\citenamefont {Julku}\ \emph {et~al.}(2016)\citenamefont {Julku},
			\citenamefont {Peotta}, \citenamefont {Vanhala}, \citenamefont {Kim},\ and\
			\citenamefont {T{\"o}rm{\"a}}}]{julku2016geometric}%
		\BibitemOpen
		\bibfield  {author} {\bibinfo {author} {\bibfnamefont {Aleksi}\ \bibnamefont
				{Julku}}, \bibinfo {author} {\bibfnamefont {Sebastiano}\ \bibnamefont
				{Peotta}}, \bibinfo {author} {\bibfnamefont {Tuomas~I}\ \bibnamefont
				{Vanhala}}, \bibinfo {author} {\bibfnamefont {Dong-Hee}\ \bibnamefont {Kim}},
			\ and\ \bibinfo {author} {\bibfnamefont {P{\"a}ivi}\ \bibnamefont
				{T{\"o}rm{\"a}}},\ }\bibfield  {title} {\enquote {\bibinfo {title} {Geometric
					origin of superfluidity in the lieb-lattice flat band},}\ }\href@noop {}
		{\bibfield  {journal} {\bibinfo  {journal} {Physical review letters}\
			}\textbf {\bibinfo {volume} {117}},\ \bibinfo {pages} {045303} (\bibinfo
			{year} {2016})}\BibitemShut {NoStop}%
		\bibitem [{\citenamefont {Liang}\ \emph
			{et~al.}(2017{\natexlab{a}})\citenamefont {Liang}, \citenamefont {Vanhala},
			\citenamefont {Peotta}, \citenamefont {Siro}, \citenamefont {Harju},\ and\
			\citenamefont {T{\"o}rm{\"a}}}]{liang2017band}%
		\BibitemOpen
		\bibfield  {author} {\bibinfo {author} {\bibfnamefont {Long}\ \bibnamefont
				{Liang}}, \bibinfo {author} {\bibfnamefont {Tuomas~I}\ \bibnamefont
				{Vanhala}}, \bibinfo {author} {\bibfnamefont {Sebastiano}\ \bibnamefont
				{Peotta}}, \bibinfo {author} {\bibfnamefont {Topi}\ \bibnamefont {Siro}},
			\bibinfo {author} {\bibfnamefont {Ari}\ \bibnamefont {Harju}}, \ and\
			\bibinfo {author} {\bibfnamefont {P{\"a}ivi}\ \bibnamefont {T{\"o}rm{\"a}}},\
		}\bibfield  {title} {\enquote {\bibinfo {title} {Band geometry, berry
					curvature, and superfluid weight},}\ }\href@noop {} {\bibfield  {journal}
			{\bibinfo  {journal} {Physical Review B}\ }\textbf {\bibinfo {volume} {95}},\
			\bibinfo {pages} {024515} (\bibinfo {year} {2017}{\natexlab{a}})}\BibitemShut
		{NoStop}%
		\bibitem [{\citenamefont {Liang}\ \emph
			{et~al.}(2017{\natexlab{b}})\citenamefont {Liang}, \citenamefont {Peotta},
			\citenamefont {Harju},\ and\ \citenamefont {T{\"o}rm{\"a}}}]{liang2017wave}%
		\BibitemOpen
		\bibfield  {author} {\bibinfo {author} {\bibfnamefont {Long}\ \bibnamefont
				{Liang}}, \bibinfo {author} {\bibfnamefont {Sebastiano}\ \bibnamefont
				{Peotta}}, \bibinfo {author} {\bibfnamefont {Ari}\ \bibnamefont {Harju}}, \
			and\ \bibinfo {author} {\bibfnamefont {P{\"a}ivi}\ \bibnamefont
				{T{\"o}rm{\"a}}},\ }\bibfield  {title} {\enquote {\bibinfo {title}
				{Wave-packet dynamics of bogoliubov quasiparticles: Quantum metric
					effects},}\ }\href@noop {} {\bibfield  {journal} {\bibinfo  {journal}
				{Physical Review B}\ }\textbf {\bibinfo {volume} {96}},\ \bibinfo {pages}
			{064511} (\bibinfo {year} {2017}{\natexlab{b}})}\BibitemShut {NoStop}%
		\bibitem [{\citenamefont {Iskin}(2018{\natexlab{a}})}]{iskin2018berezinskii}%
		\BibitemOpen
		\bibfield  {author} {\bibinfo {author} {\bibfnamefont {M}~\bibnamefont
				{Iskin}},\ }\bibfield  {title} {\enquote {\bibinfo {title}
				{Berezinskii-kosterlitz-thouless transition in the time-reversal-symmetric
					hofstadter-hubbard model},}\ }\href@noop {} {\bibfield  {journal} {\bibinfo
				{journal} {Physical Review A}\ }\textbf {\bibinfo {volume} {97}},\ \bibinfo
			{pages} {013618} (\bibinfo {year} {2018}{\natexlab{a}})}\BibitemShut
		{NoStop}%
		\bibitem [{\citenamefont {Iskin}(2018{\natexlab{b}})}]{iskin2018quantum}%
		\BibitemOpen
		\bibfield  {author} {\bibinfo {author} {\bibfnamefont {M}~\bibnamefont
				{Iskin}},\ }\bibfield  {title} {\enquote {\bibinfo {title} {Quantum-metric
					contribution to the pair mass in spin-orbit-coupled fermi superfluids},}\
		}\href@noop {} {\bibfield  {journal} {\bibinfo  {journal} {Physical Review
					A}\ }\textbf {\bibinfo {volume} {97}},\ \bibinfo {pages} {033625} (\bibinfo
			{year} {2018}{\natexlab{b}})}\BibitemShut {NoStop}%
		\bibitem [{\citenamefont {Iskin}(2018{\natexlab{c}})}]{iskin2018exposing}%
		\BibitemOpen
		\bibfield  {author} {\bibinfo {author} {\bibfnamefont {Menderes}\
				\bibnamefont {Iskin}},\ }\bibfield  {title} {\enquote {\bibinfo {title}
				{Exposing the quantum geometry of spin-orbit-coupled fermi superfluids},}\
		}\href@noop {} {\bibfield  {journal} {\bibinfo  {journal} {Physical Review
					A}\ }\textbf {\bibinfo {volume} {97}},\ \bibinfo {pages} {063625} (\bibinfo
			{year} {2018}{\natexlab{c}})}\BibitemShut {NoStop}%
		\bibitem [{\citenamefont {Mondaini}\ \emph {et~al.}(2018)\citenamefont
			{Mondaini}, \citenamefont {Batrouni},\ and\ \citenamefont
			{Gr{\'e}maud}}]{mondaini2018pairing}%
		\BibitemOpen
		\bibfield  {author} {\bibinfo {author} {\bibfnamefont {Rubem}\ \bibnamefont
				{Mondaini}}, \bibinfo {author} {\bibfnamefont {G~George}\ \bibnamefont
				{Batrouni}}, \ and\ \bibinfo {author} {\bibfnamefont {Beno{\^\i}t}\
				\bibnamefont {Gr{\'e}maud}},\ }\bibfield  {title} {\enquote {\bibinfo {title}
				{Pairing and superconductivity in the flat band: Creutz lattice},}\
		}\href@noop {} {\bibfield  {journal} {\bibinfo  {journal} {Physical Review
					B}\ }\textbf {\bibinfo {volume} {98}},\ \bibinfo {pages} {155142} (\bibinfo
			{year} {2018})}\BibitemShut {NoStop}%
		\bibitem [{\citenamefont {Iskin}(2019)}]{iskin2019origin}%
		\BibitemOpen
		\bibfield  {author} {\bibinfo {author} {\bibfnamefont {M}~\bibnamefont
				{Iskin}},\ }\bibfield  {title} {\enquote {\bibinfo {title} {Origin of
					flat-band superfluidity on the mielke checkerboard lattice},}\ }\href@noop {}
		{\bibfield  {journal} {\bibinfo  {journal} {Physical Review A}\ }\textbf
			{\bibinfo {volume} {99}},\ \bibinfo {pages} {053608} (\bibinfo {year}
			{2019})}\BibitemShut {NoStop}%
		\bibitem [{\citenamefont {Xie}\ \emph {et~al.}(2020)\citenamefont {Xie},
			\citenamefont {Song}, \citenamefont {Lian},\ and\ \citenamefont
			{Bernevig}}]{xie2020topology}%
		\BibitemOpen
		\bibfield  {author} {\bibinfo {author} {\bibfnamefont {Fang}\ \bibnamefont
				{Xie}}, \bibinfo {author} {\bibfnamefont {Zhida}\ \bibnamefont {Song}},
			\bibinfo {author} {\bibfnamefont {Biao}\ \bibnamefont {Lian}}, \ and\
			\bibinfo {author} {\bibfnamefont {B~Andrei}\ \bibnamefont {Bernevig}},\
		}\bibfield  {title} {\enquote {\bibinfo {title} {Topology-bounded superfluid
					weight in twisted bilayer graphene},}\ }\href@noop {} {\bibfield  {journal}
			{\bibinfo  {journal} {Physical review letters}\ }\textbf {\bibinfo {volume}
				{124}},\ \bibinfo {pages} {167002} (\bibinfo {year} {2020})}\BibitemShut
		{NoStop}%
		\bibitem [{\citenamefont {Verma}\ \emph {et~al.}(2021)\citenamefont {Verma},
			\citenamefont {Hazra},\ and\ \citenamefont {Randeria}}]{verma2021optical}%
		\BibitemOpen
		\bibfield  {author} {\bibinfo {author} {\bibfnamefont {Nishchhal}\
				\bibnamefont {Verma}}, \bibinfo {author} {\bibfnamefont {Tamaghna}\
				\bibnamefont {Hazra}}, \ and\ \bibinfo {author} {\bibfnamefont {Mohit}\
				\bibnamefont {Randeria}},\ }\bibfield  {title} {\enquote {\bibinfo {title}
				{Optical spectral weight, phase stiffness, and t c bounds for trivial and
					topological flat band superconductors},}\ }\href@noop {} {\bibfield
			{journal} {\bibinfo  {journal} {Proceedings of the National Academy of
					Sciences}\ }\textbf {\bibinfo {volume} {118}},\ \bibinfo {pages}
			{e2106744118} (\bibinfo {year} {2021})}\BibitemShut {NoStop}%
		\bibitem [{\citenamefont {Herzog-Arbeitman}\ \emph
			{et~al.}(2022{\natexlab{a}})\citenamefont {Herzog-Arbeitman}, \citenamefont
			{Peri}, \citenamefont {Schindler}, \citenamefont {Huber},\ and\ \citenamefont
			{Bernevig}}]{herzog2022superfluid}%
		\BibitemOpen
		\bibfield  {author} {\bibinfo {author} {\bibfnamefont {Jonah}\ \bibnamefont
				{Herzog-Arbeitman}}, \bibinfo {author} {\bibfnamefont {Valerio}\ \bibnamefont
				{Peri}}, \bibinfo {author} {\bibfnamefont {Frank}\ \bibnamefont {Schindler}},
			\bibinfo {author} {\bibfnamefont {Sebastian~D}\ \bibnamefont {Huber}}, \ and\
			\bibinfo {author} {\bibfnamefont {B~Andrei}\ \bibnamefont {Bernevig}},\
		}\bibfield  {title} {\enquote {\bibinfo {title} {Superfluid weight bounds
					from symmetry and quantum geometry in flat bands},}\ }\href@noop {}
		{\bibfield  {journal} {\bibinfo  {journal} {Physical review letters}\
			}\textbf {\bibinfo {volume} {128}},\ \bibinfo {pages} {087002} (\bibinfo
			{year} {2022}{\natexlab{a}})}\BibitemShut {NoStop}%
		\bibitem [{\citenamefont {Kitamura}\ \emph {et~al.}(2022)\citenamefont
			{Kitamura}, \citenamefont {Yamashita}, \citenamefont {Ishizuka},
			\citenamefont {Daido},\ and\ \citenamefont
			{Yanase}}]{kitamura2022superconductivity}%
		\BibitemOpen
		\bibfield  {author} {\bibinfo {author} {\bibfnamefont {Taisei}\ \bibnamefont
				{Kitamura}}, \bibinfo {author} {\bibfnamefont {Tatsuya}\ \bibnamefont
				{Yamashita}}, \bibinfo {author} {\bibfnamefont {Jun}\ \bibnamefont
				{Ishizuka}}, \bibinfo {author} {\bibfnamefont {Akito}\ \bibnamefont {Daido}},
			\ and\ \bibinfo {author} {\bibfnamefont {Youichi}\ \bibnamefont {Yanase}},\
		}\bibfield  {title} {\enquote {\bibinfo {title} {Superconductivity in
					monolayer fese enhanced by quantum geometry},}\ }\href@noop {} {\bibfield
			{journal} {\bibinfo  {journal} {Physical Review Research}\ }\textbf {\bibinfo
				{volume} {4}},\ \bibinfo {pages} {023232} (\bibinfo {year}
			{2022})}\BibitemShut {NoStop}%
		\bibitem [{\citenamefont {Huhtinen}\ \emph {et~al.}(2022)\citenamefont
			{Huhtinen}, \citenamefont {Herzog-Arbeitman}, \citenamefont {Chew},
			\citenamefont {Bernevig},\ and\ \citenamefont
			{T{\"o}rm{\"a}}}]{huhtinen2022revisiting}%
		\BibitemOpen
		\bibfield  {author} {\bibinfo {author} {\bibfnamefont {Kukka-Emilia}\
				\bibnamefont {Huhtinen}}, \bibinfo {author} {\bibfnamefont {Jonah}\
				\bibnamefont {Herzog-Arbeitman}}, \bibinfo {author} {\bibfnamefont {Aaron}\
				\bibnamefont {Chew}}, \bibinfo {author} {\bibfnamefont {Bogdan~A}\
				\bibnamefont {Bernevig}}, \ and\ \bibinfo {author} {\bibfnamefont
				{P{\"a}ivi}\ \bibnamefont {T{\"o}rm{\"a}}},\ }\bibfield  {title} {\enquote
			{\bibinfo {title} {Revisiting flat band superconductivity: Dependence on
					minimal quantum metric and band touchings},}\ }\href@noop {} {\bibfield
			{journal} {\bibinfo  {journal} {Physical Review B}\ }\textbf {\bibinfo
				{volume} {106}},\ \bibinfo {pages} {014518} (\bibinfo {year}
			{2022})}\BibitemShut {NoStop}%
		\bibitem [{\citenamefont {Mao}\ and\ \citenamefont
			{Chowdhury}(2024)}]{mao2024upper}%
		\BibitemOpen
		\bibfield  {author} {\bibinfo {author} {\bibfnamefont {Dan}\ \bibnamefont
				{Mao}}\ and\ \bibinfo {author} {\bibfnamefont {Debanjan}\ \bibnamefont
				{Chowdhury}},\ }\bibfield  {title} {\enquote {\bibinfo {title} {Upper bounds
					on superconducting and excitonic phase stiffness for interacting isolated
					narrow bands},}\ }\href@noop {} {\bibfield  {journal} {\bibinfo  {journal}
				{Physical Review B}\ }\textbf {\bibinfo {volume} {109}},\ \bibinfo {pages}
			{024507} (\bibinfo {year} {2024})}\BibitemShut {NoStop}%
		\bibitem [{\citenamefont {Hofmann}\ \emph {et~al.}(2023)\citenamefont
			{Hofmann}, \citenamefont {Berg},\ and\ \citenamefont
			{Chowdhury}}]{hofmann2023superconductivity}%
		\BibitemOpen
		\bibfield  {author} {\bibinfo {author} {\bibfnamefont {Johannes~S}\
				\bibnamefont {Hofmann}}, \bibinfo {author} {\bibfnamefont {Erez}\
				\bibnamefont {Berg}}, \ and\ \bibinfo {author} {\bibfnamefont {Debanjan}\
				\bibnamefont {Chowdhury}},\ }\bibfield  {title} {\enquote {\bibinfo {title}
				{Superconductivity, charge density wave, and supersolidity in flat bands with
					a tunable quantum metric},}\ }\href@noop {} {\bibfield  {journal} {\bibinfo
				{journal} {Physical review letters}\ }\textbf {\bibinfo {volume} {130}},\
			\bibinfo {pages} {226001} (\bibinfo {year} {2023})}\BibitemShut {NoStop}%
		\bibitem [{\citenamefont {Mao}\ and\ \citenamefont
			{Chowdhury}(2023)}]{mao2023diamagnetic}%
		\BibitemOpen
		\bibfield  {author} {\bibinfo {author} {\bibfnamefont {Dan}\ \bibnamefont
				{Mao}}\ and\ \bibinfo {author} {\bibfnamefont {Debanjan}\ \bibnamefont
				{Chowdhury}},\ }\bibfield  {title} {\enquote {\bibinfo {title} {Diamagnetic
					response and phase stiffness for interacting isolated narrow bands},}\
		}\href@noop {} {\bibfield  {journal} {\bibinfo  {journal} {Proceedings of the
					National Academy of Sciences}\ }\textbf {\bibinfo {volume} {120}},\ \bibinfo
			{pages} {e2217816120} (\bibinfo {year} {2023})}\BibitemShut {NoStop}%
		\bibitem [{\citenamefont {Gao}\ \emph {et~al.}(2023)\citenamefont {Gao},
			\citenamefont {Liu}, \citenamefont {Qiu}, \citenamefont {Ghosh},
			\citenamefont {V.~Trevisan}, \citenamefont {Onishi}, \citenamefont {Hu},
			\citenamefont {Qian}, \citenamefont {Tien}, \citenamefont {Chen} \emph
			{et~al.}}]{gao2023quantum}%
		\BibitemOpen
		\bibfield  {author} {\bibinfo {author} {\bibfnamefont {Anyuan}\ \bibnamefont
				{Gao}}, \bibinfo {author} {\bibfnamefont {Yu-Fei}\ \bibnamefont {Liu}},
			\bibinfo {author} {\bibfnamefont {Jian-Xiang}\ \bibnamefont {Qiu}}, \bibinfo
			{author} {\bibfnamefont {Barun}\ \bibnamefont {Ghosh}}, \bibinfo {author}
			{\bibfnamefont {Tha{\'\i}s}\ \bibnamefont {V.~Trevisan}}, \bibinfo {author}
			{\bibfnamefont {Yugo}\ \bibnamefont {Onishi}}, \bibinfo {author}
			{\bibfnamefont {Chaowei}\ \bibnamefont {Hu}}, \bibinfo {author}
			{\bibfnamefont {Tiema}\ \bibnamefont {Qian}}, \bibinfo {author}
			{\bibfnamefont {Hung-Ju}\ \bibnamefont {Tien}}, \bibinfo {author}
			{\bibfnamefont {Shao-Wen}\ \bibnamefont {Chen}},  \emph {et~al.},\ }\bibfield
		{title} {\enquote {\bibinfo {title} {Quantum metric nonlinear hall effect in
					a topological antiferromagnetic heterostructure},}\ }\href@noop {} {\bibfield
			{journal} {\bibinfo  {journal} {Science}\ }\textbf {\bibinfo {volume}
				{381}},\ \bibinfo {pages} {181--186} (\bibinfo {year} {2023})}\BibitemShut
		{NoStop}%
		\bibitem [{\citenamefont {Wang}\ \emph {et~al.}(2023)\citenamefont {Wang},
			\citenamefont {Kaplan}, \citenamefont {Zhang}, \citenamefont {Holder},
			\citenamefont {Cao}, \citenamefont {Wang}, \citenamefont {Zhou},
			\citenamefont {Zhou}, \citenamefont {Jiang}, \citenamefont {Zhang} \emph
			{et~al.}}]{wang2023quantum}%
		\BibitemOpen
		\bibfield  {author} {\bibinfo {author} {\bibfnamefont {Naizhou}\ \bibnamefont
				{Wang}}, \bibinfo {author} {\bibfnamefont {Daniel}\ \bibnamefont {Kaplan}},
			\bibinfo {author} {\bibfnamefont {Zhaowei}\ \bibnamefont {Zhang}}, \bibinfo
			{author} {\bibfnamefont {Tobias}\ \bibnamefont {Holder}}, \bibinfo {author}
			{\bibfnamefont {Ning}\ \bibnamefont {Cao}}, \bibinfo {author} {\bibfnamefont
				{Aifeng}\ \bibnamefont {Wang}}, \bibinfo {author} {\bibfnamefont {Xiaoyuan}\
				\bibnamefont {Zhou}}, \bibinfo {author} {\bibfnamefont {Feifei}\ \bibnamefont
				{Zhou}}, \bibinfo {author} {\bibfnamefont {Zhengzhi}\ \bibnamefont {Jiang}},
			\bibinfo {author} {\bibfnamefont {Chusheng}\ \bibnamefont {Zhang}},  \emph
			{et~al.},\ }\bibfield  {title} {\enquote {\bibinfo {title}
				{Quantum-metric-induced nonlinear transport in a topological
					antiferromagnet},}\ }\href@noop {} {\bibfield  {journal} {\bibinfo  {journal}
				{Nature}\ }\textbf {\bibinfo {volume} {621}},\ \bibinfo {pages} {487--492}
			(\bibinfo {year} {2023})}\BibitemShut {NoStop}%
		\bibitem [{\citenamefont {Kaplan}\ \emph {et~al.}(2024)\citenamefont {Kaplan},
			\citenamefont {Holder},\ and\ \citenamefont {Yan}}]{kaplan2024unification}%
		\BibitemOpen
		\bibfield  {author} {\bibinfo {author} {\bibfnamefont {Daniel}\ \bibnamefont
				{Kaplan}}, \bibinfo {author} {\bibfnamefont {Tobias}\ \bibnamefont {Holder}},
			\ and\ \bibinfo {author} {\bibfnamefont {Binghai}\ \bibnamefont {Yan}},\
		}\bibfield  {title} {\enquote {\bibinfo {title} {Unification of nonlinear
					anomalous hall effect and nonreciprocal magnetoresistance in metals by the
					quantum geometry},}\ }\href@noop {} {\bibfield  {journal} {\bibinfo
				{journal} {Physical review letters}\ }\textbf {\bibinfo {volume} {132}},\
			\bibinfo {pages} {026301} (\bibinfo {year} {2024})}\BibitemShut {NoStop}%
		\bibitem [{\citenamefont {Yu}\ \emph {et~al.}(2024)\citenamefont {Yu},
			\citenamefont {Ciccarino}, \citenamefont {Bianco}, \citenamefont {Errea},
			\citenamefont {Narang},\ and\ \citenamefont {Bernevig}}]{yu2024non}%
		\BibitemOpen
		\bibfield  {author} {\bibinfo {author} {\bibfnamefont {Jiabin}\ \bibnamefont
				{Yu}}, \bibinfo {author} {\bibfnamefont {Christopher~J}\ \bibnamefont
				{Ciccarino}}, \bibinfo {author} {\bibfnamefont {Raffaello}\ \bibnamefont
				{Bianco}}, \bibinfo {author} {\bibfnamefont {Ion}\ \bibnamefont {Errea}},
			\bibinfo {author} {\bibfnamefont {Prineha}\ \bibnamefont {Narang}}, \ and\
			\bibinfo {author} {\bibfnamefont {B~Andrei}\ \bibnamefont {Bernevig}},\
		}\bibfield  {title} {\enquote {\bibinfo {title} {Non-trivial quantum geometry
					and the strength of electron--phonon coupling},}\ }\href@noop {} {\bibfield
			{journal} {\bibinfo  {journal} {Nature Physics}\ ,\ \bibinfo {pages} {1--7}}
			(\bibinfo {year} {2024})}\BibitemShut {NoStop}%
		\bibitem [{\citenamefont {Chen}\ and\ \citenamefont
			{Law}(2024)}]{chen2024ginzburg}%
		\BibitemOpen
		\bibfield  {author} {\bibinfo {author} {\bibfnamefont {Shuai~A}\ \bibnamefont
				{Chen}}\ and\ \bibinfo {author} {\bibfnamefont {KT}~\bibnamefont {Law}},\
		}\bibfield  {title} {\enquote {\bibinfo {title} {Ginzburg-landau theory of
					flat-band superconductors with quantum metric},}\ }\href@noop {} {\bibfield
			{journal} {\bibinfo  {journal} {Physical Review Letters}\ }\textbf {\bibinfo
				{volume} {132}},\ \bibinfo {pages} {026002} (\bibinfo {year}
			{2024})}\BibitemShut {NoStop}%
		\bibitem [{\citenamefont {Provost}\ and\ \citenamefont
			{Vallee}(1980)}]{provost1980riemannian}%
		\BibitemOpen
		\bibfield  {author} {\bibinfo {author} {\bibfnamefont {JP}~\bibnamefont
				{Provost}}\ and\ \bibinfo {author} {\bibfnamefont {G}~\bibnamefont
				{Vallee}},\ }\bibfield  {title} {\enquote {\bibinfo {title} {Riemannian
					structure on manifolds of quantum states},}\ }\href@noop {} {\bibfield
			{journal} {\bibinfo  {journal} {Communications in Mathematical Physics}\
			}\textbf {\bibinfo {volume} {76}},\ \bibinfo {pages} {289--301} (\bibinfo
			{year} {1980})}\BibitemShut {NoStop}%
		\bibitem [{\citenamefont {Berry}(1984)}]{berry1984quantal}%
		\BibitemOpen
		\bibfield  {author} {\bibinfo {author} {\bibfnamefont {Michael~Victor}\
				\bibnamefont {Berry}},\ }\bibfield  {title} {\enquote {\bibinfo {title}
				{Quantal phase factors accompanying adiabatic changes},}\ }\href@noop {}
		{\bibfield  {journal} {\bibinfo  {journal} {Proceedings of the Royal Society
					of London. A. Mathematical and Physical Sciences}\ }\textbf {\bibinfo
				{volume} {392}},\ \bibinfo {pages} {45--57} (\bibinfo {year}
			{1984})}\BibitemShut {NoStop}%
		\bibitem [{\citenamefont {Klitzing}\ \emph {et~al.}(1980)\citenamefont
			{Klitzing}, \citenamefont {Dorda},\ and\ \citenamefont
			{Pepper}}]{klitzing1980new}%
		\BibitemOpen
		\bibfield  {author} {\bibinfo {author} {\bibfnamefont {K~v}\ \bibnamefont
				{Klitzing}}, \bibinfo {author} {\bibfnamefont {Gerhard}\ \bibnamefont
				{Dorda}}, \ and\ \bibinfo {author} {\bibfnamefont {Michael}\ \bibnamefont
				{Pepper}},\ }\bibfield  {title} {\enquote {\bibinfo {title} {New method for
					high-accuracy determination of the fine-structure constant based on quantized
					hall resistance},}\ }\href@noop {} {\bibfield  {journal} {\bibinfo  {journal}
				{Physical review letters}\ }\textbf {\bibinfo {volume} {45}},\ \bibinfo
			{pages} {494} (\bibinfo {year} {1980})}\BibitemShut {NoStop}%
		\bibitem [{\citenamefont {Thouless}\ \emph {et~al.}(1982)\citenamefont
			{Thouless}, \citenamefont {Kohmoto}, \citenamefont {Nightingale},\ and\
			\citenamefont {den Nijs}}]{thouless1982quantized}%
		\BibitemOpen
		\bibfield  {author} {\bibinfo {author} {\bibfnamefont {David~J}\ \bibnamefont
				{Thouless}}, \bibinfo {author} {\bibfnamefont {Mahito}\ \bibnamefont
				{Kohmoto}}, \bibinfo {author} {\bibfnamefont {M~Peter}\ \bibnamefont
				{Nightingale}}, \ and\ \bibinfo {author} {\bibfnamefont {Marcel}\
				\bibnamefont {den Nijs}},\ }\bibfield  {title} {\enquote {\bibinfo {title}
				{Quantized hall conductance in a two-dimensional periodic potential},}\
		}\href@noop {} {\bibfield  {journal} {\bibinfo  {journal} {Physical review
					letters}\ }\textbf {\bibinfo {volume} {49}},\ \bibinfo {pages} {405}
			(\bibinfo {year} {1982})}\BibitemShut {NoStop}%
		\bibitem [{\citenamefont {Bellissard}\ \emph {et~al.}(1994)\citenamefont
			{Bellissard}, \citenamefont {van Elst},\ and\ \citenamefont
			{Schulz-Baldes}}]{bellissard1994noncommutative}%
		\BibitemOpen
		\bibfield  {author} {\bibinfo {author} {\bibfnamefont {Jean}\ \bibnamefont
				{Bellissard}}, \bibinfo {author} {\bibfnamefont {Andreas}\ \bibnamefont {van
					Elst}}, \ and\ \bibinfo {author} {\bibfnamefont {Hermann}\ \bibnamefont
				{Schulz-Baldes}},\ }\bibfield  {title} {\enquote {\bibinfo {title} {The
					noncommutative geometry of the quantum hall effect},}\ }\href@noop {}
		{\bibfield  {journal} {\bibinfo  {journal} {Journal of Mathematical Physics}\
			}\textbf {\bibinfo {volume} {35}},\ \bibinfo {pages} {5373--5451} (\bibinfo
			{year} {1994})}\BibitemShut {NoStop}%
		\bibitem [{\citenamefont {Hasan}\ and\ \citenamefont
			{Kane}(2010)}]{hasan2010colloquium}%
		\BibitemOpen
		\bibfield  {author} {\bibinfo {author} {\bibfnamefont {M~Zahid}\ \bibnamefont
				{Hasan}}\ and\ \bibinfo {author} {\bibfnamefont {Charles~L}\ \bibnamefont
				{Kane}},\ }\bibfield  {title} {\enquote {\bibinfo {title} {Colloquium:
					topological insulators},}\ }\href@noop {} {\bibfield  {journal} {\bibinfo
				{journal} {Reviews of modern physics}\ }\textbf {\bibinfo {volume} {82}},\
			\bibinfo {pages} {3045--3067} (\bibinfo {year} {2010})}\BibitemShut {NoStop}%
		\bibitem [{\citenamefont {Qi}\ and\ \citenamefont
			{Zhang}(2011)}]{qi2011topological}%
		\BibitemOpen
		\bibfield  {author} {\bibinfo {author} {\bibfnamefont {Xiao-Liang}\
				\bibnamefont {Qi}}\ and\ \bibinfo {author} {\bibfnamefont {Shou-Cheng}\
				\bibnamefont {Zhang}},\ }\bibfield  {title} {\enquote {\bibinfo {title}
				{Topological insulators and superconductors},}\ }\href@noop {} {\bibfield
			{journal} {\bibinfo  {journal} {Reviews of modern physics}\ }\textbf
			{\bibinfo {volume} {83}},\ \bibinfo {pages} {1057--1110} (\bibinfo {year}
			{2011})}\BibitemShut {NoStop}%
		\bibitem [{\citenamefont {Shapere}\ and\ \citenamefont
			{Wilczek}(1989)}]{shapere1989geometric}%
		\BibitemOpen
		\bibfield  {author} {\bibinfo {author} {\bibfnamefont {Alfred}\ \bibnamefont
				{Shapere}}\ and\ \bibinfo {author} {\bibfnamefont {Frank}\ \bibnamefont
				{Wilczek}},\ }\href@noop {} {\emph {\bibinfo {title} {Geometric phases in
					physics}}},\ Vol.~\bibinfo {volume} {5}\ (\bibinfo  {publisher} {World
			scientific},\ \bibinfo {year} {1989})\BibitemShut {NoStop}%
		\bibitem [{\citenamefont {Anandan}\ and\ \citenamefont
			{Aharonov}(1990)}]{anandan1990geometry}%
		\BibitemOpen
		\bibfield  {author} {\bibinfo {author} {\bibfnamefont {Jeeva}\ \bibnamefont
				{Anandan}}\ and\ \bibinfo {author} {\bibfnamefont {Yakir}\ \bibnamefont
				{Aharonov}},\ }\bibfield  {title} {\enquote {\bibinfo {title} {Geometry of
					quantum evolution},}\ }\href@noop {} {\bibfield  {journal} {\bibinfo
				{journal} {Physical review letters}\ }\textbf {\bibinfo {volume} {65}},\
			\bibinfo {pages} {1697} (\bibinfo {year} {1990})}\BibitemShut {NoStop}%
		\bibitem [{\citenamefont {Marzari}\ and\ \citenamefont
			{Vanderbilt}(1997)}]{marzari1997maximally}%
		\BibitemOpen
		\bibfield  {author} {\bibinfo {author} {\bibfnamefont {Nicola}\ \bibnamefont
				{Marzari}}\ and\ \bibinfo {author} {\bibfnamefont {David}\ \bibnamefont
				{Vanderbilt}},\ }\bibfield  {title} {\enquote {\bibinfo {title} {Maximally
					localized generalized wannier functions for composite energy bands},}\
		}\href@noop {} {\bibfield  {journal} {\bibinfo  {journal} {Physical review
					B}\ }\textbf {\bibinfo {volume} {56}},\ \bibinfo {pages} {12847} (\bibinfo
			{year} {1997})}\BibitemShut {NoStop}%
		\bibitem [{\citenamefont {Simon}\ and\ \citenamefont
			{Rudner}(2020)}]{simon2020contrasting}%
		\BibitemOpen
		\bibfield  {author} {\bibinfo {author} {\bibfnamefont {Steven~H}\
				\bibnamefont {Simon}}\ and\ \bibinfo {author} {\bibfnamefont {Mark~S}\
				\bibnamefont {Rudner}},\ }\bibfield  {title} {\enquote {\bibinfo {title}
				{Contrasting lattice geometry dependent versus independent quantities:
					Ramifications for berry curvature, energy gaps, and dynamics},}\ }\href@noop
		{} {\bibfield  {journal} {\bibinfo  {journal} {Physical Review B}\ }\textbf
			{\bibinfo {volume} {102}},\ \bibinfo {pages} {165148} (\bibinfo {year}
			{2020})}\BibitemShut {NoStop}%
		\bibitem [{\citenamefont {Han}\ \emph {et~al.}(2024)\citenamefont {Han},
			\citenamefont {Herzog-Arbeitman}, \citenamefont {Bernevig},\ and\
			\citenamefont {Kivelson}}]{han2024quantum}%
		\BibitemOpen
		\bibfield  {author} {\bibinfo {author} {\bibfnamefont {Zhaoyu}\ \bibnamefont
				{Han}}, \bibinfo {author} {\bibfnamefont {Jonah}\ \bibnamefont
				{Herzog-Arbeitman}}, \bibinfo {author} {\bibfnamefont {B~Andrei}\
				\bibnamefont {Bernevig}}, \ and\ \bibinfo {author} {\bibfnamefont {Steven~A}\
				\bibnamefont {Kivelson}},\ }\bibfield  {title} {\enquote {\bibinfo {title}
				{“quantum geometric nesting” and solvable model flat-band systems},}\
		}\href@noop {} {\bibfield  {journal} {\bibinfo  {journal} {Physical Review
					X}\ }\textbf {\bibinfo {volume} {14}},\ \bibinfo {pages} {041004} (\bibinfo
			{year} {2024})}\BibitemShut {NoStop}%
		\bibitem [{\citenamefont {Annett}(2004)}]{annett2004superconductivity}%
		\BibitemOpen
		\bibfield  {author} {\bibinfo {author} {\bibfnamefont {James~F}\ \bibnamefont
				{Annett}},\ }\href@noop {} {\emph {\bibinfo {title} {Superconductivity,
					superfluids and condensates}}},\ Vol.~\bibinfo {volume} {5}\ (\bibinfo
		{publisher} {Oxford University Press},\ \bibinfo {year} {2004})\BibitemShut
		{NoStop}%
		\bibitem [{\citenamefont {Hofmann}\ \emph {et~al.}(2022)\citenamefont
			{Hofmann}, \citenamefont {Chowdhury}, \citenamefont {Kivelson},\ and\
			\citenamefont {Berg}}]{hofmann2022heuristic}%
		\BibitemOpen
		\bibfield  {author} {\bibinfo {author} {\bibfnamefont {Johannes~S}\
				\bibnamefont {Hofmann}}, \bibinfo {author} {\bibfnamefont {Debanjan}\
				\bibnamefont {Chowdhury}}, \bibinfo {author} {\bibfnamefont {Steven~A}\
				\bibnamefont {Kivelson}}, \ and\ \bibinfo {author} {\bibfnamefont {Erez}\
				\bibnamefont {Berg}},\ }\bibfield  {title} {\enquote {\bibinfo {title}
				{Heuristic bounds on superconductivity and how to exceed them},}\ }\href@noop
		{} {\bibfield  {journal} {\bibinfo  {journal} {npj quantum materials}\
			}\textbf {\bibinfo {volume} {7}},\ \bibinfo {pages} {83} (\bibinfo {year}
			{2022})}\BibitemShut {NoStop}%
		\bibitem [{\citenamefont {Roy}(2014)}]{roy2014band}%
		\BibitemOpen
		\bibfield  {author} {\bibinfo {author} {\bibfnamefont {Rahul}\ \bibnamefont
				{Roy}},\ }\bibfield  {title} {\enquote {\bibinfo {title} {Band geometry of
					fractional topological insulators},}\ }\href@noop {} {\bibfield  {journal}
			{\bibinfo  {journal} {Physical Review B}\ }\textbf {\bibinfo {volume} {90}},\
			\bibinfo {pages} {165139} (\bibinfo {year} {2014})}\BibitemShut {NoStop}%
		\bibitem [{\citenamefont {Sun}\ \emph {et~al.}(2011)\citenamefont {Sun},
			\citenamefont {Gu}, \citenamefont {Katsura},\ and\ \citenamefont
			{Das~Sarma}}]{sun2011nearly}%
		\BibitemOpen
		\bibfield  {author} {\bibinfo {author} {\bibfnamefont {Kai}\ \bibnamefont
				{Sun}}, \bibinfo {author} {\bibfnamefont {Zhengcheng}\ \bibnamefont {Gu}},
			\bibinfo {author} {\bibfnamefont {Hosho}\ \bibnamefont {Katsura}}, \ and\
			\bibinfo {author} {\bibfnamefont {S}~\bibnamefont {Das~Sarma}},\ }\bibfield
		{title} {\enquote {\bibinfo {title} {Nearly flatbands with nontrivial
					topology},}\ }\href@noop {} {\bibfield  {journal} {\bibinfo  {journal}
				{Physical review letters}\ }\textbf {\bibinfo {volume} {106}},\ \bibinfo
			{pages} {236803} (\bibinfo {year} {2011})}\BibitemShut {NoStop}%
		\bibitem [{\citenamefont {Yang}\ \emph {et~al.}(2012)\citenamefont {Yang},
			\citenamefont {Gu}, \citenamefont {Sun},\ and\ \citenamefont
			{Das~Sarma}}]{yang2012topological}%
		\BibitemOpen
		\bibfield  {author} {\bibinfo {author} {\bibfnamefont {Shuo}\ \bibnamefont
				{Yang}}, \bibinfo {author} {\bibfnamefont {Zheng-Cheng}\ \bibnamefont {Gu}},
			\bibinfo {author} {\bibfnamefont {Kai}\ \bibnamefont {Sun}}, \ and\ \bibinfo
			{author} {\bibfnamefont {S}~\bibnamefont {Das~Sarma}},\ }\bibfield  {title}
		{\enquote {\bibinfo {title} {Topological flat band models with arbitrary
					chern numbers},}\ }\href@noop {} {\bibfield  {journal} {\bibinfo  {journal}
				{Physical Review B—Condensed Matter and Materials Physics}\ }\textbf
			{\bibinfo {volume} {86}},\ \bibinfo {pages} {241112} (\bibinfo {year}
			{2012})}\BibitemShut {NoStop}%
		\bibitem [{\citenamefont {Mitscherling}\ and\ \citenamefont
			{Holder}(2022)}]{mitscherling2022bound}%
		\BibitemOpen
		\bibfield  {author} {\bibinfo {author} {\bibfnamefont {Johannes}\
				\bibnamefont {Mitscherling}}\ and\ \bibinfo {author} {\bibfnamefont {Tobias}\
				\bibnamefont {Holder}},\ }\bibfield  {title} {\enquote {\bibinfo {title}
				{Bound on resistivity in flat-band materials due to the quantum metric},}\
		}\href@noop {} {\bibfield  {journal} {\bibinfo  {journal} {Physical Review
					B}\ }\textbf {\bibinfo {volume} {105}},\ \bibinfo {pages} {085154} (\bibinfo
			{year} {2022})}\BibitemShut {NoStop}%
		\bibitem [{\citenamefont {Ledwith}\ \emph {et~al.}(2020)\citenamefont
			{Ledwith}, \citenamefont {Tarnopolsky}, \citenamefont {Khalaf},\ and\
			\citenamefont {Vishwanath}}]{ledwith2020fractional}%
		\BibitemOpen
		\bibfield  {author} {\bibinfo {author} {\bibfnamefont {Patrick~J}\
				\bibnamefont {Ledwith}}, \bibinfo {author} {\bibfnamefont {Grigory}\
				\bibnamefont {Tarnopolsky}}, \bibinfo {author} {\bibfnamefont {Eslam}\
				\bibnamefont {Khalaf}}, \ and\ \bibinfo {author} {\bibfnamefont {Ashvin}\
				\bibnamefont {Vishwanath}},\ }\bibfield  {title} {\enquote {\bibinfo {title}
				{Fractional chern insulator states in twisted bilayer graphene: An analytical
					approach},}\ }\href@noop {} {\bibfield  {journal} {\bibinfo  {journal}
				{Physical Review Research}\ }\textbf {\bibinfo {volume} {2}},\ \bibinfo
			{pages} {023237} (\bibinfo {year} {2020})}\BibitemShut {NoStop}%
		\bibitem [{\citenamefont {Da~Liao}\ \emph {et~al.}(2021)\citenamefont
			{Da~Liao}, \citenamefont {Kang}, \citenamefont {Brei{\o}}, \citenamefont
			{Xu}, \citenamefont {Wu}, \citenamefont {Andersen}, \citenamefont
			{Fernandes},\ and\ \citenamefont {Meng}}]{da2021correlation}%
		\BibitemOpen
		\bibfield  {author} {\bibinfo {author} {\bibfnamefont {Yuan}\ \bibnamefont
				{Da~Liao}}, \bibinfo {author} {\bibfnamefont {Jian}\ \bibnamefont {Kang}},
			\bibinfo {author} {\bibfnamefont {Clara~N}\ \bibnamefont {Brei{\o}}},
			\bibinfo {author} {\bibfnamefont {Xiao~Yan}\ \bibnamefont {Xu}}, \bibinfo
			{author} {\bibfnamefont {Han-Qing}\ \bibnamefont {Wu}}, \bibinfo {author}
			{\bibfnamefont {Brian~M}\ \bibnamefont {Andersen}}, \bibinfo {author}
			{\bibfnamefont {Rafael~M}\ \bibnamefont {Fernandes}}, \ and\ \bibinfo
			{author} {\bibfnamefont {Zi~Yang}\ \bibnamefont {Meng}},\ }\bibfield  {title}
		{\enquote {\bibinfo {title} {Correlation-induced insulating topological
					phases at charge neutrality in twisted bilayer graphene},}\ }\href@noop {}
		{\bibfield  {journal} {\bibinfo  {journal} {Physical Review X}\ }\textbf
			{\bibinfo {volume} {11}},\ \bibinfo {pages} {011014} (\bibinfo {year}
			{2021})}\BibitemShut {NoStop}%
		\bibitem [{\citenamefont {Song}\ and\ \citenamefont
			{Bernevig}(2022)}]{song2022magic}%
		\BibitemOpen
		\bibfield  {author} {\bibinfo {author} {\bibfnamefont {Zhi-Da}\ \bibnamefont
				{Song}}\ and\ \bibinfo {author} {\bibfnamefont {B~Andrei}\ \bibnamefont
				{Bernevig}},\ }\bibfield  {title} {\enquote {\bibinfo {title} {Magic-angle
					twisted bilayer graphene as a topological heavy fermion problem},}\
		}\href@noop {} {\bibfield  {journal} {\bibinfo  {journal} {Physical review
					letters}\ }\textbf {\bibinfo {volume} {129}},\ \bibinfo {pages} {047601}
			(\bibinfo {year} {2022})}\BibitemShut {NoStop}%
		\bibitem [{\citenamefont {Chou}\ and\ \citenamefont
			{Das~Sarma}(2023)}]{chou2023kondo}%
		\BibitemOpen
		\bibfield  {author} {\bibinfo {author} {\bibfnamefont {Yang-Zhi}\
				\bibnamefont {Chou}}\ and\ \bibinfo {author} {\bibfnamefont {Sankar}\
				\bibnamefont {Das~Sarma}},\ }\bibfield  {title} {\enquote {\bibinfo {title}
				{Kondo lattice model in magic-angle twisted bilayer graphene},}\ }\href@noop
		{} {\bibfield  {journal} {\bibinfo  {journal} {Physical Review Letters}\
			}\textbf {\bibinfo {volume} {131}},\ \bibinfo {pages} {026501} (\bibinfo
			{year} {2023})}\BibitemShut {NoStop}%
		\bibitem [{\citenamefont {Herzog-Arbeitman}\ \emph {et~al.}(2024)\citenamefont
			{Herzog-Arbeitman}, \citenamefont {Yu}, \citenamefont {C{\u{a}}lug{\u{a}}ru},
			\citenamefont {Hu}, \citenamefont {Regnault}, \citenamefont {Liu},
			\citenamefont {Vafek}, \citenamefont {Coleman}, \citenamefont {Tsvelik},
			\citenamefont {Song} \emph {et~al.}}]{herzog2024topological}%
		\BibitemOpen
		\bibfield  {author} {\bibinfo {author} {\bibfnamefont {Jonah}\ \bibnamefont
				{Herzog-Arbeitman}}, \bibinfo {author} {\bibfnamefont {Jiabin}\ \bibnamefont
				{Yu}}, \bibinfo {author} {\bibfnamefont {Dumitru}\ \bibnamefont
				{C{\u{a}}lug{\u{a}}ru}}, \bibinfo {author} {\bibfnamefont {Haoyu}\
				\bibnamefont {Hu}}, \bibinfo {author} {\bibfnamefont {Nicolas}\ \bibnamefont
				{Regnault}}, \bibinfo {author} {\bibfnamefont {Chaoxing}\ \bibnamefont
				{Liu}}, \bibinfo {author} {\bibfnamefont {Oskar}\ \bibnamefont {Vafek}},
			\bibinfo {author} {\bibfnamefont {Piers}\ \bibnamefont {Coleman}}, \bibinfo
			{author} {\bibfnamefont {Alexei}\ \bibnamefont {Tsvelik}}, \bibinfo {author}
			{\bibfnamefont {Zhi-da}\ \bibnamefont {Song}},  \emph {et~al.},\ }\bibfield
		{title} {\enquote {\bibinfo {title} {Topological heavy fermion principle for
					flat (narrow) bands with concentrated quantum geometry},}\ }\href@noop {}
		{\bibfield  {journal} {\bibinfo  {journal} {arXiv preprint arXiv:2404.07253}\
			} (\bibinfo {year} {2024})}\BibitemShut {NoStop}%
		\bibitem [{\citenamefont {Hu}\ \emph {et~al.}(2023{\natexlab{a}})\citenamefont
			{Hu}, \citenamefont {Rai}, \citenamefont {Crippa}, \citenamefont
			{Herzog-Arbeitman}, \citenamefont {C{\u{a}}lug{\u{a}}ru}, \citenamefont
			{Wehling}, \citenamefont {Sangiovanni}, \citenamefont {Valent{\'\i}},
			\citenamefont {Tsvelik},\ and\ \citenamefont {Bernevig}}]{hu2023symmetric}%
		\BibitemOpen
		\bibfield  {author} {\bibinfo {author} {\bibfnamefont {Haoyu}\ \bibnamefont
				{Hu}}, \bibinfo {author} {\bibfnamefont {Gautam}\ \bibnamefont {Rai}},
			\bibinfo {author} {\bibfnamefont {Lorenzo}\ \bibnamefont {Crippa}}, \bibinfo
			{author} {\bibfnamefont {Jonah}\ \bibnamefont {Herzog-Arbeitman}}, \bibinfo
			{author} {\bibfnamefont {Dumitru}\ \bibnamefont {C{\u{a}}lug{\u{a}}ru}},
			\bibinfo {author} {\bibfnamefont {Tim}\ \bibnamefont {Wehling}}, \bibinfo
			{author} {\bibfnamefont {Giorgio}\ \bibnamefont {Sangiovanni}}, \bibinfo
			{author} {\bibfnamefont {Roser}\ \bibnamefont {Valent{\'\i}}}, \bibinfo
			{author} {\bibfnamefont {Alexei~M}\ \bibnamefont {Tsvelik}}, \ and\ \bibinfo
			{author} {\bibfnamefont {B~Andrei}\ \bibnamefont {Bernevig}},\ }\bibfield
		{title} {\enquote {\bibinfo {title} {Symmetric kondo lattice states in doped
					strained twisted bilayer graphene},}\ }\href@noop {} {\bibfield  {journal}
			{\bibinfo  {journal} {Physical review letters}\ }\textbf {\bibinfo {volume}
				{131}},\ \bibinfo {pages} {166501} (\bibinfo {year}
			{2023}{\natexlab{a}})}\BibitemShut {NoStop}%
		\bibitem [{\citenamefont {Zhang}\ \emph {et~al.}(2023)\citenamefont {Zhang},
			\citenamefont {Pan}, \citenamefont {Chen}, \citenamefont {Li}, \citenamefont
			{Sun},\ and\ \citenamefont {Meng}}]{zhang2023polynomial}%
		\BibitemOpen
		\bibfield  {author} {\bibinfo {author} {\bibfnamefont {Xu}~\bibnamefont
				{Zhang}}, \bibinfo {author} {\bibfnamefont {Gaopei}\ \bibnamefont {Pan}},
			\bibinfo {author} {\bibfnamefont {Bin-Bin}\ \bibnamefont {Chen}}, \bibinfo
			{author} {\bibfnamefont {Heqiu}\ \bibnamefont {Li}}, \bibinfo {author}
			{\bibfnamefont {Kai}\ \bibnamefont {Sun}}, \ and\ \bibinfo {author}
			{\bibfnamefont {Zi~Yang}\ \bibnamefont {Meng}},\ }\bibfield  {title}
		{\enquote {\bibinfo {title} {Polynomial sign problem and topological mott
					insulator in twisted bilayer graphene},}\ }\href@noop {} {\bibfield
			{journal} {\bibinfo  {journal} {Physical Review B}\ }\textbf {\bibinfo
				{volume} {107}},\ \bibinfo {pages} {L241105} (\bibinfo {year}
			{2023})}\BibitemShut {NoStop}%
		\bibitem [{\citenamefont {Hu}\ \emph {et~al.}(2023{\natexlab{b}})\citenamefont
			{Hu}, \citenamefont {Bernevig},\ and\ \citenamefont {Tsvelik}}]{hu2023kondo}%
		\BibitemOpen
		\bibfield  {author} {\bibinfo {author} {\bibfnamefont {Haoyu}\ \bibnamefont
				{Hu}}, \bibinfo {author} {\bibfnamefont {B~Andrei}\ \bibnamefont {Bernevig}},
			\ and\ \bibinfo {author} {\bibfnamefont {Alexei~M}\ \bibnamefont {Tsvelik}},\
		}\bibfield  {title} {\enquote {\bibinfo {title} {Kondo lattice model of
					magic-angle twisted-bilayer graphene: Hund’s rule, local-moment
					fluctuations, and low-energy effective theory},}\ }\href@noop {} {\bibfield
			{journal} {\bibinfo  {journal} {Physical review letters}\ }\textbf {\bibinfo
				{volume} {131}},\ \bibinfo {pages} {026502} (\bibinfo {year}
			{2023}{\natexlab{b}})}\BibitemShut {NoStop}%
		\bibitem [{\citenamefont {Kol{\'a}{\v{r}}}\ \emph {et~al.}(2023)\citenamefont
			{Kol{\'a}{\v{r}}}, \citenamefont {Shavit}, \citenamefont {Mora},
			\citenamefont {Oreg},\ and\ \citenamefont {von Oppen}}]{kolavr2023anderson}%
		\BibitemOpen
		\bibfield  {author} {\bibinfo {author} {\bibfnamefont {Kry{\v{s}}tof}\
				\bibnamefont {Kol{\'a}{\v{r}}}}, \bibinfo {author} {\bibfnamefont {Gal}\
				\bibnamefont {Shavit}}, \bibinfo {author} {\bibfnamefont {Christophe}\
				\bibnamefont {Mora}}, \bibinfo {author} {\bibfnamefont {Yuval}\ \bibnamefont
				{Oreg}}, \ and\ \bibinfo {author} {\bibfnamefont {Felix}\ \bibnamefont {von
					Oppen}},\ }\bibfield  {title} {\enquote {\bibinfo {title} {Anderson’s
					theorem for correlated insulating states in twisted bilayer graphene},}\
		}\href@noop {} {\bibfield  {journal} {\bibinfo  {journal} {Physical review
					letters}\ }\textbf {\bibinfo {volume} {130}},\ \bibinfo {pages} {076204}
			(\bibinfo {year} {2023})}\BibitemShut {NoStop}%
		\bibitem [{\citenamefont {Kwan}\ \emph {et~al.}(2021)\citenamefont {Kwan},
			\citenamefont {Wagner}, \citenamefont {Soejima}, \citenamefont {Zaletel},
			\citenamefont {Simon}, \citenamefont {Parameswaran},\ and\ \citenamefont
			{Bultinck}}]{kwan2021kekule}%
		\BibitemOpen
		\bibfield  {author} {\bibinfo {author} {\bibfnamefont {Yves~H}\ \bibnamefont
				{Kwan}}, \bibinfo {author} {\bibfnamefont {Glenn}\ \bibnamefont {Wagner}},
			\bibinfo {author} {\bibfnamefont {Tomohiro}\ \bibnamefont {Soejima}},
			\bibinfo {author} {\bibfnamefont {Michael~P}\ \bibnamefont {Zaletel}},
			\bibinfo {author} {\bibfnamefont {Steven~H}\ \bibnamefont {Simon}}, \bibinfo
			{author} {\bibfnamefont {Siddharth~A}\ \bibnamefont {Parameswaran}}, \ and\
			\bibinfo {author} {\bibfnamefont {Nick}\ \bibnamefont {Bultinck}},\
		}\bibfield  {title} {\enquote {\bibinfo {title} {Kekul{\'e} spiral order at
					all nonzero integer fillings in twisted bilayer graphene},}\ }\href@noop {}
		{\bibfield  {journal} {\bibinfo  {journal} {Physical Review X}\ }\textbf
			{\bibinfo {volume} {11}},\ \bibinfo {pages} {041063} (\bibinfo {year}
			{2021})}\BibitemShut {NoStop}%
		\bibitem [{\citenamefont {Bistritzer}\ and\ \citenamefont
			{MacDonald}(2011)}]{bistritzer2011moire}%
		\BibitemOpen
		\bibfield  {author} {\bibinfo {author} {\bibfnamefont {Rafi}\ \bibnamefont
				{Bistritzer}}\ and\ \bibinfo {author} {\bibfnamefont {Allan~H}\ \bibnamefont
				{MacDonald}},\ }\bibfield  {title} {\enquote {\bibinfo {title} {Moir{\'e}
					bands in twisted double-layer graphene},}\ }\href@noop {} {\bibfield
			{journal} {\bibinfo  {journal} {Proceedings of the National Academy of
					Sciences}\ }\textbf {\bibinfo {volume} {108}},\ \bibinfo {pages}
			{12233--12237} (\bibinfo {year} {2011})}\BibitemShut {NoStop}%
		\bibitem [{\citenamefont {Mera}\ and\ \citenamefont
			{Mitscherling}(2022)}]{mera2022nontrivial}%
		\BibitemOpen
		\bibfield  {author} {\bibinfo {author} {\bibfnamefont {Bruno}\ \bibnamefont
				{Mera}}\ and\ \bibinfo {author} {\bibfnamefont {Johannes}\ \bibnamefont
				{Mitscherling}},\ }\bibfield  {title} {\enquote {\bibinfo {title} {Nontrivial
					quantum geometry of degenerate flat bands},}\ }\href@noop {} {\bibfield
			{journal} {\bibinfo  {journal} {Physical Review B}\ }\textbf {\bibinfo
				{volume} {106}},\ \bibinfo {pages} {165133} (\bibinfo {year}
			{2022})}\BibitemShut {NoStop}%
		\bibitem [{\citenamefont {Herzog-Arbeitman}\ \emph
			{et~al.}(2022{\natexlab{b}})\citenamefont {Herzog-Arbeitman}, \citenamefont
			{Chew}, \citenamefont {Huhtinen}, \citenamefont {T{\"o}rm{\"a}},\ and\
			\citenamefont {Bernevig}}]{herzog2022many}%
		\BibitemOpen
		\bibfield  {author} {\bibinfo {author} {\bibfnamefont {Jonah}\ \bibnamefont
				{Herzog-Arbeitman}}, \bibinfo {author} {\bibfnamefont {Aaron}\ \bibnamefont
				{Chew}}, \bibinfo {author} {\bibfnamefont {Kukka-Emilia}\ \bibnamefont
				{Huhtinen}}, \bibinfo {author} {\bibfnamefont {P{\"a}ivi}\ \bibnamefont
				{T{\"o}rm{\"a}}}, \ and\ \bibinfo {author} {\bibfnamefont {B~Andrei}\
				\bibnamefont {Bernevig}},\ }\bibfield  {title} {\enquote {\bibinfo {title}
				{Many-body superconductivity in topological flat bands},}\ }\href@noop {}
		{\bibfield  {journal} {\bibinfo  {journal} {arXiv preprint arXiv:2209.00007}\
			} (\bibinfo {year} {2022}{\natexlab{b}})}\BibitemShut {NoStop}%
	\end{thebibliography}

\begin{thebibliography}{5}%
	\section*{Supplementary References}
	\makeatletter
	\providecommand \@ifxundefined [1]{%
		\@ifx{#1\undefined}
	}%
	\providecommand \@ifnum [1]{%
		\ifnum #1\expandafter \@firstoftwo
		\else \expandafter \@secondoftwo
		\fi
	}%
	\providecommand \@ifx [1]{%
		\ifx #1\expandafter \@firstoftwo
		\else \expandafter \@secondoftwo
		\fi
	}%
	\providecommand \natexlab [1]{#1}%
	\providecommand \enquote  [1]{``#1''}%
	\providecommand \bibnamefont  [1]{#1}%
	\providecommand \bibfnamefont [1]{#1}%
	\providecommand \citenamefont [1]{#1}%
	\providecommand \href@noop [0]{\@secondoftwo}%
	\providecommand \href [0]{\begingroup \@sanitize@url \@href}%
	\providecommand \@href[1]{\@@startlink{#1}\@@href}%
	\providecommand \@@href[1]{\endgroup#1\@@endlink}%
	\providecommand \@sanitize@url [0]{\catcode `\\12\catcode `\$12\catcode
		`\&12\catcode `\#12\catcode `\^12\catcode `\_12\catcode `\%12\relax}%
	\providecommand \@@startlink[1]{}%
	\providecommand \@@endlink[0]{}%
	\providecommand \url  [0]{\begingroup\@sanitize@url \@url }%
	\providecommand \@url [1]{\endgroup\@href {#1}{\urlprefix }}%
	\providecommand \urlprefix  [0]{URL }%
	\providecommand \Eprint [0]{\href }%
	\providecommand \doibase [0]{http://dx.doi.org/}%
	\providecommand \selectlanguage [0]{\@gobble}%
	\providecommand \bibinfo  [0]{\@secondoftwo}%
	\providecommand \bibfield  [0]{\@secondoftwo}%
	\providecommand \translation [1]{[#1]}%
	\providecommand \BibitemOpen [0]{}%
	\providecommand \bibitemStop [0]{}%
	\providecommand \bibitemNoStop [0]{.\EOS\space}%
	\providecommand \EOS [0]{\spacefactor3000\relax}%
	\providecommand \BibitemShut  [1]{\csname bibitem#1\endcsname}%
	\let\auto@bib@innerbib\@empty
	%</preamble>
	\bibitem [{\citenamefont {{Huhtinen}}\ \emph {et~al.}(2022)\citenamefont
		{{Huhtinen}}, \citenamefont {{Herzog-Arbeitman}}, \citenamefont {{Chew}},
		\citenamefont {{Bernevig}},\ and\ \citenamefont
		{{T{\"o}rm{\"a}}}}]{huhtinen2022revisiting}%
	\BibitemOpen
	\bibfield  {author} {\bibinfo {author} {\bibfnamefont {Kukka-Emilia}\
			\bibnamefont {{Huhtinen}}}, \bibinfo {author} {\bibfnamefont {Jonah}\
			\bibnamefont {{Herzog-Arbeitman}}}, \bibinfo {author} {\bibfnamefont {Aaron}\
			\bibnamefont {{Chew}}}, \bibinfo {author} {\bibfnamefont {Bogdan~A.}\
			\bibnamefont {{Bernevig}}}, \ and\ \bibinfo {author} {\bibfnamefont
			{P{\"a}ivi}\ \bibnamefont {{T{\"o}rm{\"a}}}},\ }\bibfield  {title} {\enquote
		{\bibinfo {title} {{Revisiting flat band superconductivity: Dependence on
					minimal quantum metric and band touchings}},}\ }\href {\doibase
		10.1103/PhysRevB.106.014518} {\bibfield  {journal} {\bibinfo  {journal}
			{\prb}\ }\textbf {\bibinfo {volume} {106}},\ \bibinfo {eid} {014518}
		(\bibinfo {year} {2022})},\ \Eprint {http://arxiv.org/abs/2203.11133}
	{arXiv:2203.11133 [cond-mat.supr-con]} \BibitemShut {NoStop}%
	\bibitem [{\citenamefont {T{\"o}rm{\"a}}\ \emph {et~al.}(2018)\citenamefont
		{T{\"o}rm{\"a}}, \citenamefont {Liang},\ and\ \citenamefont
		{Peotta}}]{torma2018quantum}%
	\BibitemOpen
	\bibfield  {author} {\bibinfo {author} {\bibfnamefont {P{\"a}ivi}\
			\bibnamefont {T{\"o}rm{\"a}}}, \bibinfo {author} {\bibfnamefont {Long}\
			\bibnamefont {Liang}}, \ and\ \bibinfo {author} {\bibfnamefont {Sebastiano}\
			\bibnamefont {Peotta}},\ }\bibfield  {title} {\enquote {\bibinfo {title}
			{Quantum metric and effective mass of a two-body bound state in a flat
				band},}\ }\href@noop {} {\bibfield  {journal} {\bibinfo  {journal} {Physical
				Review B}\ }\textbf {\bibinfo {volume} {98}},\ \bibinfo {pages} {220511}
		(\bibinfo {year} {2018})}\BibitemShut {NoStop}%
	\bibitem [{\citenamefont {Hofmann}\ \emph {et~al.}(2023)\citenamefont
		{Hofmann}, \citenamefont {Berg},\ and\ \citenamefont
		{Chowdhury}}]{hofmann2023superconductivity}%
	\BibitemOpen
	\bibfield  {author} {\bibinfo {author} {\bibfnamefont {Johannes~S}\
			\bibnamefont {Hofmann}}, \bibinfo {author} {\bibfnamefont {Erez}\
			\bibnamefont {Berg}}, \ and\ \bibinfo {author} {\bibfnamefont {Debanjan}\
			\bibnamefont {Chowdhury}},\ }\bibfield  {title} {\enquote {\bibinfo {title}
			{Superconductivity, charge density wave, and supersolidity in flat bands with
				a tunable quantum metric},}\ }\href@noop {} {\bibfield  {journal} {\bibinfo
			{journal} {Physical review letters}\ }\textbf {\bibinfo {volume} {130}},\
		\bibinfo {pages} {226001} (\bibinfo {year} {2023})}\BibitemShut {NoStop}%
	\bibitem [{\citenamefont {{Koshino}}\ \emph {et~al.}(2018)\citenamefont
		{{Koshino}}, \citenamefont {{Yuan}}, \citenamefont {{Koretsune}},
		\citenamefont {{Ochi}}, \citenamefont {{Kuroki}},\ and\ \citenamefont
		{{Fu}}}]{2018PhRvX...8c1087K}%
	\BibitemOpen
	\bibfield  {author} {\bibinfo {author} {\bibfnamefont {Mikito}\ \bibnamefont
			{{Koshino}}}, \bibinfo {author} {\bibfnamefont {Noah F.~Q.}\ \bibnamefont
			{{Yuan}}}, \bibinfo {author} {\bibfnamefont {Takashi}\ \bibnamefont
			{{Koretsune}}}, \bibinfo {author} {\bibfnamefont {Masayuki}\ \bibnamefont
			{{Ochi}}}, \bibinfo {author} {\bibfnamefont {Kazuhiko}\ \bibnamefont
			{{Kuroki}}}, \ and\ \bibinfo {author} {\bibfnamefont {Liang}\ \bibnamefont
			{{Fu}}},\ }\bibfield  {title} {\enquote {\bibinfo {title} {{Maximally
					Localized Wannier Orbitals and the Extended Hubbard Model for Twisted Bilayer
					Graphene}},}\ }\href {\doibase 10.1103/PhysRevX.8.031087} {\bibfield
		{journal} {\bibinfo  {journal} {Physical Review X}\ }\textbf {\bibinfo
			{volume} {8}},\ \bibinfo {eid} {031087} (\bibinfo {year} {2018})},\ \Eprint
	{http://arxiv.org/abs/1805.06819} {arXiv:1805.06819 [cond-mat.mes-hall]}
	\BibitemShut {NoStop}%
	\bibitem [{\citenamefont {{Lee}}\ \emph {et~al.}(2019)\citenamefont {{Lee}},
		\citenamefont {{Khalaf}}, \citenamefont {{Liu}}, \citenamefont {{Liu}},
		\citenamefont {{Hao}}, \citenamefont {{Kim}},\ and\ \citenamefont
		{{Vishwanath}}}]{2019NatCo..10.5333L}%
	\BibitemOpen
	\bibfield  {author} {\bibinfo {author} {\bibfnamefont {Jong~Yeon}\
			\bibnamefont {{Lee}}}, \bibinfo {author} {\bibfnamefont {Eslam}\ \bibnamefont
			{{Khalaf}}}, \bibinfo {author} {\bibfnamefont {Shang}\ \bibnamefont {{Liu}}},
		\bibinfo {author} {\bibfnamefont {Xiaomeng}\ \bibnamefont {{Liu}}}, \bibinfo
		{author} {\bibfnamefont {Zeyu}\ \bibnamefont {{Hao}}}, \bibinfo {author}
		{\bibfnamefont {Philip}\ \bibnamefont {{Kim}}}, \ and\ \bibinfo {author}
		{\bibfnamefont {Ashvin}\ \bibnamefont {{Vishwanath}}},\ }\bibfield  {title}
	{\enquote {\bibinfo {title} {{Theory of correlated insulating behaviour and
					spin-triplet superconductivity in twisted double bilayer graphene}},}\ }\href
	{\doibase 10.1038/s41467-019-12981-1} {\bibfield  {journal} {\bibinfo
			{journal} {Nature Communications}\ }\textbf {\bibinfo {volume} {10}},\
		\bibinfo {eid} {5333} (\bibinfo {year} {2019})},\ \Eprint
	{http://arxiv.org/abs/1903.08685} {arXiv:1903.08685 [cond-mat.str-el]}
	\BibitemShut {NoStop}%
\end{thebibliography}

%
\end{document}